\documentclass[aps,superscriptaddress,onecolumn,10pt,prx,longbibliography]{revtex4-2}

\usepackage{tikz}
\usepackage{amsmath,mathtools,amsthm,amssymb}
\usepackage{tabularray}
\usepackage[caption=false]{subfig}
\usepackage{color}
\usepackage{xcolor}
\usepackage{bbold}
\usepackage{enumitem}
\usepackage{physics}
\usepackage{comment}
\usepackage{array}
\usepackage[hypertexnames=false,colorlinks,linkcolor=blue,urlcolor=blue,citecolor=blue]{hyperref}
\usetikzlibrary{calc,positioning}
\usepackage{graphics}
\usepackage{wrapfig}
\usepackage{biolinum}
\usepackage{bbm}
\usepackage{dsfont}
\usepackage{mathrsfs}
\usepackage{mathdots}
\usepackage{pifont}
\definecolor{myrefcolor}{rgb}{0.067,0.5,0.5}
\definecolor{myurlcolor}{rgb}{0.1,0,0.9}
\newcommand{\rrangle}{\rangle\hspace{-0.8mm}\rangle}
\newcommand{\llangle}{\langle\hspace{-0.8mm}\langle}

\newcommand{\bbraket}[2]{\llangle {#1} | {#2} \rrangle}

\def\urho{\check\rho}
\newcommand{\id}{\mathds{1}}
\newcommand{\average}[1]{\left\langle #1 \right\rangle}
\newcommand{\kket}[1]
{|{#1}\rrangle }
\newcommand{\bbra}[1]
{\llangle{#1}| }
\usepackage{orcidlink}
\usepackage{bibunits}
\defaultbibliographystyle{apsrev4-2}
\defaultbibliography{mainNotes,bib}

\usepackage{etoolbox}
\makeatletter
\patchcmd{\NAT@bibsetnum}{\ref{LastBibItem}}
  {\ref{LastBibItem-\@bibunitname}}{}{}
\patchcmd{\endthebibliography}{\label{LastBibItem}}
  {\label{LastBibItem-\@bibunitname}}{}{}
\let\bu@natlinkstart\hyper@natlinkstart
\let\bu@natlinkbreak\hyper@natlinkbreak
\let\bu@natanchorstart\hyper@natanchorstart
\def\hyper@natlinkstart#1{\bu@natlinkstart{\@bibunitname.#1}}
\def\hyper@natlinkbreak#1#2{\bu@natlinkbreak{#1}{\@bibunitname.#2}}
\def\hyper@natanchorstart#1{\bu@natanchorstart{\@bibunitname.#1}}
\AtBeginDocument{%
  \let\auto@bib\@empty
  \let\auto@bib@innerbib\@empty
}
\makeatother

\makeatletter
\newcommand{\aveE}[2][]{%
  \mathbb{E}%
  \if\relax\detokenize{#1}\relax
  \else_{#1}\fi
  \!\left[#2\right]%
}
\makeatother

\newcommand{\youngcell}{0.7}

\newcommand{\Young}[4]{%
  \begin{scope}[shift={(#1,#2)}]
    \foreach \r [count=\i] in {#3} {
      \foreach \j in {1,...,\r} {
        \draw (\j*\youngcell, -\i*\youngcell)
              rectangle ++(\youngcell,\youngcell);
      }
    }
    \node[anchor=west] at (0.1,0.8*\youngcell) {#4};
  \end{scope}
}

\newcommand{\paircell}[1]{\makebox[1.6em][c]{\ensuremath{#1}}}

\newcommand{\pairinginSn}[2][0.85]{%
\begin{tikzpicture}[baseline=-0.6ex]
  \def\xsep{#1}
  \def\dy{0.19}

  \foreach \i/\lab in {
    1/{1^{\phantom{*}}},
    2/{2^{\phantom{*}}},
    3/{1^*},
    4/{2^*}
  }{
    \node[inner sep=0pt, outer sep=0pt] at ({(\i-1)*\xsep}, \dy) {\paircell{\lab}};
  }

  \foreach \lab [count=\i from 1] in {#2} {
    \node[inner sep=0pt, outer sep=0pt] at ({(\i-1)*\xsep}, -\dy) {\paircell{\lab}};
  }
\end{tikzpicture}%
}

\newcommand{\pairingdiagramtworows}[4][0.9pt]{%
\begin{tikzpicture}[baseline=(#2-base)]
  \def\xsep{0.65}
  \def\dy{0.19}
  \coordinate (#2-base) at (0,0);

  \foreach \i/\lab in {1/{$1$},2/{$2$},3/{$1^*$},4/{$2^*$}}{
    \node[inner sep=1pt] (#2-T\i) at ({(\i-1)*\xsep}, \dy) {\lab};
  }
  \foreach \i/\lab in {1/{$1$},2/{$2$},3/{$1^*$},4/{$2^*$}}{
    \node[inner sep=1pt] (#2-B\i) at ({(\i-1)*\xsep}, -\dy) {\lab};
  }

  \foreach \a/\b in #3 {
    \draw[line width=#1]
      (#2-T\a) to[bend left=55] (#2-T\b);
  }

  \foreach \a/\b in #4 {
    \draw[line width=#1]
      (#2-B\a) to[bend right=55] (#2-B\b);
  }
\end{tikzpicture}%
}

\newcommand{\thickcircle}[2][0.9pt]{%
\begin{tikzpicture}[baseline=-0.55ex]
  \draw[line width=#1] (0,0) circle (#2);
\end{tikzpicture}%
}

\def\Tr{\operatorname{Tr}}

\def\Kr{K}

\def\Pr{\mbox{Prob}}
\def\PP{\mathcal{P}}
\def\PPave{\mathcal{P}}
\def\Pa{\mathsf{P}}
\def\SS{\mathsf{S}}
\def\ggg{\mathsf{G}}

\def\GG{\mathcal{G}}
\def\UU{\mathcal{U}}
\def\OO{\mathcal{O}}
\def\td{\mathsf{t}}
\def\tc{t}

\def\idp{\mathbf{1}}

\def\End{\operatorname{End}}
\def\aa{\mathbf{a}}

\def\tp{\tc_{\rm P}}

\def\HNN{(\mathcal{H}\otimes\mathcal{H}^{\ast})^{\otimes N}}

\begin{document}

\def\titleinfo{Universal purification dynamics in real non-unitary quantum processes}
\title{\titleinfo} 

\author{Federico Gerbino~\orcidlink{0009-0008-1485-3764}}
\email{federico.gerbino@universite-paris-saclay.fr}
\affiliation{Laboratoire de Physique Théorique et Modèles Statistiques, Université Paris-Saclay, CNRS, 91405 Orsay, France}

\author{Donghoon Kim~\orcidlink{0009-0002-8358-5253}}
\affiliation{Analytical Quantum Complexity RIKEN Hakubi Research Team, RIKEN Center for Quantum Computing (RQC), Wako, Saitama 351-0198, Japan}

\author{Guido Giachetti~\orcidlink{0000-0002-4928-7693}}
\affiliation{Laboratoire de Physique de l’École Normale Supérieure, CNRS, ENS and PSL University, Sorbonne Université, Université Paris Cité, 75005 Paris, France}

\author{Andrea De Luca~\orcidlink{0000-0003-0272-5083}}
\affiliation{Laboratoire de Physique de l’École Normale Supérieure, CNRS, ENS and PSL University, Sorbonne Université, Université Paris Cité, 75005 Paris, France}

\author{Xhek Turkeshi~\orcidlink{0000-0003-1093-3771}}
\affiliation{Institut für Theoretische Physik, Universität zu Köln, Zülpicher Strasse 77, 50937 Köln, Germany}

\begin{abstract}
We study purification dynamics in monitored quantum processes governed by ensembles of quantum circuits in different random-matrix symmetry classes. We analyze the universal aspects that emerge away from the measurement-induced phase transition and inside the volume/weak measurement phase and in the scaling limit of large time and Hilbert space dimension.
We present two toy models that reveal two complementary visions and provide quantitative access to universal scaling:
i) a discrete-time dynamic in which each time step corresponds to multiplication by a Gaussian random matrix; ii) weak continuous-time monitoring that induces a Dyson Brownian motion of the eigenvalues of the density matrix. The first approach provides an algebraic characterization based on rotational invariance emerging in Kraus's operator space, focusing in particular on the unitary and orthogonal cases, respectively $\beta=2$ and $\beta=1$, with $\beta$ the Dyson random-matrix index. The second approach, on the other hand, allows for a unified treatment for any $\beta$, thanks to the mapping of the Fokker-Planck evolution of eigenvalues onto the Calogero-Sutherland integrable Hamiltonian diagonalized in terms of Jack polynomials. 
We provide explicit expressions for the universal decrease of R\'enyi entropies. We show that, approaching the universal scaling limit, numerical simulations of different models agree with each other and with our theoretical predictions. 
Our results clarify the existence of different classes of universality for the purification process in hybrid quantum systems, accessible in random circuit architectures and weak measurement protocols.
\end{abstract}

\maketitle

\begin{bibunit}

\section*{Introduction}

Purification is the process by which a quantum system is driven from a mixed state toward a pure one. 
It plays a central role in quantum information science, from quantum sensing and metrology to quantum simulation and quantum computation, where the ability to prepare and stabilize low-entropy quantum states is a fundamental resource~\cite{terhal2002dissipative, verstraete2009quantum, degen2017quantum, barreiro2011open}. 
Beyond its practical utility, purification addresses deep conceptual questions regarding irreversibility and the dynamical nature of entropy in quantum mechanics~\cite{nielsen2002quantum, fidkowski2021dynamical}.

Monitored quantum systems, arising in the quantum trajectory descriptions of open quantum systems~\cite{caves1987quantum, gisin1992quantum, diosi1998non, kurt2006straightforward,daley2014quantum,hamazaki2026introduction}, provide a natural and minimal framework to study purification. 
In this setting, two antagonistic mechanisms coexist: unitary dynamics scrambles quantum information and generates entanglement~\cite{nahum2017quantum, vonkeyserlingk2018operator,sierant2023}, while measurements extract information and drive the system toward purity. 
This competition engenders rich dynamical phenomena distinct from those found in purely coherent or purely dissipative settings~\cite{fisher2023}.
A major advance in this field was the discovery of measurement-induced phase transitions (MIPTs)~\cite{lunt2022quantumsimulationusing,potter2022entanglementdynamicsin,rossini2021coherentanddissipative}. These transitions were first identified in hybrid random quantum circuits, where local unitary gates generate entanglement while repeated measurements suppress it~\cite{chan2019unitaryprojective,li2019measurementdrivenentanglement,skinner2019measurementinducedphase}, and mark a sharp boundary between a weak-measurement phase, where scrambling dominates and the system remains highly entangled (volume-law), and a strong-measurement phase, where the system is rapidly purified into a disentangled state (area-law).
Subsequent studies of random circuits developed this picture in several complementary directions. Numerical works established the robustness of the transition under different measurement protocols and circuit architectures, and investigated its critical exponents and universality properties~\cite{szyniszewski2019entanglementtransitionfrom,szyniszewski2020universalityofentanglement,zabalo2020criticalpropertiesof,sierant2022measurementinducedphase,sharma2022measurementinducedcriticality,liu2022measurementinducedentanglement,PhysRevLett.132.140401}. Other works characterized the critical point through different observables, including operator scaling dimensions, entanglement spectra, entanglement negativity, and multipartite correlations~\cite{zabalo2022operatorscalingdimensions,sierant2022universalbehaviorbeyond,zhang2020nonuniversalentanglementlevel,sang2021entanglementnegativityat,shi2020entanglementnegativityat,weinstein2022measurementinducedpower,lira}. In parallel, analytical mappings to classical statistical-mechanics models and spacetime-dual formulations clarified the mechanism underlying the transition and its relation to quantum error correction, information scrambling, and decoding~\cite{li2021statisticalmechanicsmodel,li2021entanglementdomainwalls,li2021statisticalmechanicsof,lu2021spacetimeduality,ippoliti2021entanglementphasetransitions,ippoliti2022fractallogarithmicand,ippoliti2021postselectionfreeentanglement,li2021robustdecodingin,kelly2022coherencerequirementsfor,klocke2022topologicalorderand,weinstein2022scramblngtransitionin}.
The phenomenon was subsequently shown to extend well beyond random circuits. Measurement-induced criticality has been investigated in Hamiltonian systems under continuous monitoring~\cite{fuji2020measurementinducedquantum,rossini2020measurementinduceddynamics,lunt2020measurementinducedentanglement,tang2020measurementinducedphase,turkeshi2021measurementinduced,turkeshi2022entanglement,xing,Paviglianiti2024enhanced,Li2025,gal} and within field-theoretical descriptions~\cite{minoguchi2022continuousgaussianmeasurements,altland2022dynamicsofmeasured,zhang2021emergentreplica,zhang2022universalentanglementtransitions,zhou2021nonunitaryentanglementdynamics,buchhold2021,fava2023,poboiko2023}. Related transitions were also uncovered in Brownian circuits and random tensor networks~\cite{bentsen2021measurementinducedpurification,yang2022entanglementphasetransitions,medina2021entanglementtransitionsfrom,
han2022measurementinducedcriticality}, and in various other settings investigating memory effects~\cite{chiria}, transport properties~\cite{piro,nava2026informationtransporttransportinducedentanglement} and non-homogeneous effects~\cite{muzzi}.
Further generalizations include circuits with conservation laws or symmetries, where measurement-induced entanglement transitions can coexist with charge-sharpening and infinite-randomness phenomena~\cite{lavasani2021measurement,feng2022measurementinducedphase,barratt2022transitions,agrawal2022entanglmentandchargesharpening,oshima2023,zabalo2022infiniterandomnesscriticality}, as well as systems with feedback, where measurement outcomes are actively used to steer the dynamics~\cite{iadecola2022dynamicalentanglementtransition,odea2022entanglementandabsorbing,ravindranath2022entanglementsteeringin,piroli2022trivialityofquantum,sierant2022controllingentanglementat,Sierant_2023,iade}.
Furthermore, recent experiments on superconducting and trapped-ion platforms have begun to observe these transitions in laboratory settings~\cite{noel2022measurement,koh2023measurement, google2021exponential,Feng2026}. 
While early studies primarily focused on the scaling of entanglement entropy, it has become increasingly clear that purification itself defines an independent and highly nontrivial dynamical 
problem~\cite{gullans2020dynamicalpurificationphase, bentsen2021measurementinducedpurification,loio2023,merritt2022entanglement, Tiutiakina2025fieldtheory}.

In the weak-measurement phase, purification is a gradual process: the entropy of an initially mixed state begins to decrease already at times of order one, and continues to decrease --- while remaining large compared to one --- over a broad window of intermediate timescales~\cite{gullans2020dynamicalpurificationphase,fidkowski2021dynamical}. An $O(1)$ entropy, however, is attained only on a timescale $\tp(L)$ that grows exponentially with the system size $L$, 
\begin{equation}
\label{eq:tpL}
\tp(L) \sim e^{\alpha L},
\end{equation}
in stark contrast with the logarithmic or polynomial purification times characteristic of the strong-measurement phase. A useful way to rationalize the exponential growth of the purification time in the weak-measurement phase is to interpret measurement-induced phase transitions as genuine second-order transitions in an enlarged space of replicated degrees of freedom. 
Within this perspective, the competition between measurements and coherent dynamics can be viewed as a phenomenon of symmetry breaking in the group of permutations that exchange replicas of the system. 
The strong-measurement phase corresponds to a symmetric (disordered) phase, in which measurements efficiently couple different replicas and rapidly drive the system toward a pure state. 
By contrast, in the weak-measurement phase coherent dynamics dominates, leading to a symmetry-broken phase in replica space, where different replicas effectively decouple.
In such a broken-symmetry phase, purification requires rare collective fluctuations that restore the symmetry by coupling replicas over the entire system. 
These processes play a role analogous to instanton events connecting distinct vacua in symmetry-broken phases, and their probability is therefore exponentially suppressed in the system size. 
As a consequence, the characteristic purification time \eqref{eq:tpL} grows exponentially with $L$, reflecting the nonperturbative nature of entropy reduction in the weak-measurement regime.
This exponential separation of timescales has been analyzed from several complementary viewpoints, including quantum trajectory approaches, statistical-mechanics mappings, and information-theoretic arguments \cite{gullans2020dynamicalpurificationphase,gullans2020scalable, giachetti2023elusive}. 
Physically, it reflects the extreme inefficiency of measurements in extracting global information when unitary scrambling remains effective over long times.

The existence of such a long purification timescale suggests that, in the weak-measurement phase, purification becomes the slowest relevant dynamical process in the system. After a possibly model-dependent initial transient, all microscopic relaxation and scrambling mechanisms occur on parametrically shorter timescales, and the system remains mixed for a time of order $\tp(L)\sim e^{\alpha L}$. We stress that the state remains macroscopically mixed --- i.e., its entropy is extensive --- at times much shorter than $\tp(L)$. At times of order $\tp(L)$, the residual entropy is of order one, and it eventually vanishes as the system approaches a pure state: this late stage of the purification dynamics is the one described by the universal scaling functions derived in this work. The earlier regimes, in which purification also takes place while the entropy is still large compared to one, are not expected to be universal and are not addressed here; in the scaling limit considered below, they are compressed to $x\to 0$, where the universal scaling functions diverge logarithmically, matching the large entropies of the preceding stages.
This observation naturally leads to the expectation that purification dynamics in this regime should exhibit a form of universality, becoming insensitive to microscopic details such as lattice geometry, spatial dimensionality, or the precise structure of local interactions. 
Indeed, recent work has argued that, upon rescaling time by the purification time $\tp(L)$, suitable statistical probes of the purification dynamics collapse onto universal scaling functions that depend only on global features of the dynamics~\cite{deluca2025universality}. A physical interpretation of this universality is that, because measurements are too weak or too rare to localize information efficiently, the unitary dynamics has sufficient time to fully scramble all degrees of freedom between successive significant purification events. Here and below, ``significant purification events'' refer to coarse-grained global entropy-reducing fluctuations, not to individual measurement processes, which may occur continuously or at an extensive microscopic rate. The universal regime considered below is the joint limit $t,L\to\infty$ at fixed $x=t/\tp(L)$; microscopic or polynomial-time transients are therefore compressed to $x=0$. As a result, the many-body system effectively behaves as a single macroscopic quantum dot, whose dynamics can be captured by an appropriate random-matrix description \cite{bulchandani_random-matrix_2023, gerbino2024dyson}. 
This effective reduction from a spatially extended many-body problem to a $0+1$-dimensional stochastic process provides a powerful and unifying perspective on slow purification dynamics.

A central message of this work is that the universal features of purification dynamics are controlled by an emergent symmetry group associated with this effective random-matrix description. 
In the most generic situation, coherent dynamics explores the full unitary group acting on the many-body Hilbert space, leading to an effective symmetry group $\GG = \UU(q)$, the unitary group over the whole $q$-dimensional Hilbert space. 
This case was analyzed in detail in previous work, where the corresponding universality class was characterized in terms of scaling functions governing spectral observables of the density matrix \cite{deluca2025universality}. 
However, many physically relevant systems possess additional constraints, such as time-reversal symmetry or {purely imaginary Hamiltonians (i.e., real evolution operators)}, which restrict the accessible dynamics to a subgroup of the unitary group. 
In such cases, the effective symmetry group is reduced to the orthogonal group $\GG = \OO(q)$~\cite{hunterjones2018operatorgrowthrandomquantum,grevink2025glueshortdepthdesignsunitary,Kalsi2022,Braccia_2024,west2025realclassicalshadows,khanna2025randomquantumcircuitstimereversal}.

The primary goal of the present work is to show that this reduction of symmetry gives rise to a distinct universality class of purification dynamics. 
We demonstrate that purification processes invariant under orthogonal transformations exhibit scaling behavior that is qualitatively different from that of the unitary class, despite sharing the same microscopic phenomenology of exponentially slow purification. 
These differences are directly visible in the scaling behavior of spectral observables, such as the Rényi entropies. 
As a representative example, we find that in the scaling regime the average Rényi entropies behave at short rescaled times $x = \tc/\tp$ as
\begin{equation}
S_n(x) \sim -\log x +
\begin{cases}
O(x^2), & \text{unitary symmetry } (\beta = 2), \\
O(x),   & \text{orthogonal symmetry } (\beta = 1),
\end{cases}
\end{equation}
revealing a clear and robust distinction between the two universality classes. 
These differences ultimately originate from the symmetry constraints imposed on the effective stochastic dynamics. A concrete instance of such a physical process is a monitored brickwork circuit whose nearest-neighbor gates are sampled from $\OO(q^2)$, together with local measurements in a real basis, for example the computational basis. Standard real gates such as $R_y(\theta)$, Hadamard, CNOT, controlled-(Z), and SWAP provide elementary examples. 

To establish these results, we develop two complementary theoretical descriptions of universal purification dynamics. 
The first is based on a discrete-time random-matrix formulation, in which the effect of monitoring is encoded in random Kraus operators drawn from appropriate matrix ensembles. 
In this approach, universal scaling functions emerge from the algebraic and combinatorial structure of the effective interactions between replicated degrees of freedom, with the nature of the symmetry group determining the relevant mathematical framework. 
The second description relies on a continuous-time weak-measurement limit, in which monitoring induces an infinitesimal stochastic evolution of the density matrix. 
In this limit, the dynamics of the density matrix eigenvalues is governed by a stochastic process closely related to Dyson Brownian motion, whose associated Fokker--Planck operator maps onto the Hamiltonian of the Calogero--Sutherland model. 
This formulation naturally encompasses different symmetry classes through the Dyson index $\beta$, with the cases $\beta=1,2$ corresponding to orthogonal and unitary symmetry, respectively \cite{mehta2004random}.

In addition to developing the theoretical framework, we provide analytical and numerical evidence that purification dynamics with real-valued monitoring protocols converges to the same universal scaling functions across different microscopic models, confirming the existence of a well-defined orthogonal universality class. 
The remainder of the paper is organized as follows. 
In the next section, we introduce the general framework and observables used to characterize purification dynamics and formulate the scaling limit underlying universality. 
Subsequent sections develop the discrete-time and continuous-time descriptions in detail and present explicit results for the universal scaling functions.

\section{Framework, averages, and scaling limit
\label{sec:framework}
}

We study the purification dynamics of a finite-dimensional quantum system undergoing monitored evolution. 
The state of the system is described by a density matrix $\rho$, initially taken to be highly mixed. 
Throughout this work we focus on systems without conservation laws, for which the natural initial condition is the maximally mixed state $\rho(0)=\mathbb{1}/q$, where $q$ denotes the dimension of the many-body Hilbert space $\mathcal{H}$, exponentially large in the system size $L$. (We refer to 1D systems for simplicity's sake, but the discussion is not limited to them.) 
Purification is then defined as the stochastic process by which $\rho$ evolves toward a rank-one projector at long times.

At the microscopic level, the evolution over a finite time-step $\Delta t$ is described by a quantum channel $\mathcal C$, which can always be written in terms of a set of Kraus operators $\{\hat K_a\}_a$ as
\begin{equation}
\label{eq:quantumchannel}
\mathcal C_{\Delta \tc}[\rho] = \sum_a \hat K_a \rho \hat K_a^\dagger ,
\qquad
\sum_a \hat K_a^\dagger \hat K_a = \mathbb{1}.
\end{equation}
Here the index $a$ labels the outcomes of generalized measurements, as well as possible sources of classical randomness or noise. 
The specific nature of these outcomes will not be important in the following, and the sum over $a$ should be understood in a generalized sense, possibly including integrations over a continuous probability space, as we will see in practical cases.

Conditioning on a given outcome $a$ leads to an unraveling of the quantum channel, in which the density matrix undergoes the stochastic update
\begin{equation}
\label{eq:rhoevol}
\rho \;\longrightarrow\; \rho_a = \frac{\check\rho_a}{\Tr(\check\rho_a)},
\qquad
\check\rho_a = \hat K_a \rho \hat K_a^\dagger,
\end{equation}
where $\check\rho$ denotes the unnormalized density matrix and the probability $\Pr(a)=\Tr(\check\rho_a)$, in accordance with Born’s rule. 
By iterating this evolution for several time-steps, we obtain a quantum trajectory conditioned on the time series of outcomes $\aa = (a_1,a_2,\ldots, a_\td)$ up to time $\tc = \td \Delta \tc$, with $\td$ the (integer) number of time-steps, whose unnormalized density matrix
\begin{equation}
\label{eq:qtraj}
\check{\rho}(t = 
\td \Delta \tc) := \check{\rho}_{\aa_\td} = \Kr_{a_\td}\ldots  \Kr_{a_2}\Kr_{a_1}\rho(t=0)
\Kr_{a_1}^{\dag}\Kr_{a_2}^{\dag}\ldots  \Kr_{a_\td}^{\dag}
\end{equation}
from which $\rho_{\aa_\td} = \check{\rho}_{\aa_\td}/\Tr[\check{\rho}_{\aa_\td}]$ and the probability of the whole trajectory is $\Pr(\aa_\td) = \Tr[\check\rho_{\aa_\td}]$.

To quantify the purity of the state along the quantum trajectory, one can look at the spectrum of $\rho_{\aa_\td}$. A general set of indicators consists of the moments 
\begin{equation}
\label{eq:momdef}
\PP_{\mathbf{r}}[\rho] :=   \Tr[\rho]^{r_1}
        \Tr[\rho^2]^{r_2} \ldots = \prod_{j=2}^\infty \Tr[\rho^j]^{r_j}
\end{equation}
where we introduce the notation for integer partitions~\cite{macdonald1998symmetric}
\begin{equation}
\label{eq:rdef}
 \mathbf{r} = \{r_1,r_2,\ldots\} = 1^{r_1} 2^{r_2} \ldots , \quad \mbox{with } r_j \in \mathbb{N} \;, \quad |\mathbf{r}| := \sum_j j r_j \;.
\end{equation}
In the last equality of \eqref{eq:momdef}, we used that $r_1$ is irrelevant for normalized density matrices with $\Tr[\rho] = 1$. The moment $\PP_{\mathbf{r}}[\rho]$ is a homogeneous polynomial of degree $|\mathbf{r}| := \sum_{j} j r_j$ in the entries of $\rho$. 
These moments provide a
complete description of the joint distribution of the spectrum of $\rho$; for instance, 
it is possible to extract the entire distribution of Rényi entropies. Indeed, from the definition of the $n$th Rényi entropy of a density matrix $S_n[\rho] = (1-n)^{-1} \log \Tr[\rho^n]$, we deduce
\begin{equation}
\label{eq:renyidef}
    e^{(1-n) k S_n[\rho]} {=} \Tr[\rho^n]^k = \PP_{n^k}[\rho]
\end{equation}
where, in agreement with (\ref{eq:momdef},\ref{eq:rdef}), the subscript $n^k$ denotes a sequence $\mathbf{r}$ whose only non-zero element is $r_n = k$. 
Consistently, the von Neumann entropy can be interpreted as the formal limit $S_1[\rho] \equiv \lim_{n \to 1} S_n[\rho]$.  
One is interested in analyzing the behavior of the moments $\PP_{\mathbf{r}}[\rho]$ on a typical quantum trajectory. Quite generally, we can say that for a highly mixed state, $\PP_{\mathbf{r}}[\rho]$ is exponentially small in the system size, e.g. for the maximally mixed state $\PP_{\mathbf{r}}[\mathbb{1}/q] = q^{\sum_j r_j (1  - j)}$. In contrast, for a pure state $\PP_{\mathbf{r}}[\ket{\Psi}\bra{\Psi}] = 1$.
During the non-unitary stochastic dynamics \eqref{eq:qtraj}, the density matrix evolves from a mixed state to a pure one and the purification dynamics can be characterized by the growth in time of the moments $\PP_\mathbf{r}[\rho]$ and $\tp$ represents the characteristic time scale for such a process. 
The details of this process of growth of moments (and therefore of purification itself) depend on how the different realizations of trajectories $\aa$  are weighted. Two natural choices are
\begin{subequations}
\label{eq:Mave}
\begin{align}
&\average{\PP_{\mathbf{r}}(t)}_{\rm BR} := \sum_{\aa_\td} \PP_{\mathbf{r}}[\rho_{\aa_\td}] \Pr(\aa_\td) \;, && \mbox{Born's rule (BR)}
\label{eq:MaveBR}
\\ 
&\average{\PP_{\mathbf{r}}(t)}_{\rm FM} := \frac{1}{Z_{\rm FM}}\sum_{\aa_\td} \PP_{\mathbf{r}}[\rho_{\aa_\td}]  \;, && \mbox{Forced measurements (FM)}
\label{eq:MaveFM}
\end{align}
\end{subequations}
where $Z_{\rm FM}$ is chosen so that $\average{1}_{\rm FM} = 1$. In practice, in Eq.~\eqref{eq:MaveBR}, measurement outcomes are sampled according to Born's rule, while in Eq.~\eqref{eq:MaveFM} all outcomes are given equal probability. It should be noted that the basic rules of quantum mechanics predict that through several repetitions of this protocol, it is possible to 
realize the random process \eqref{eq:qtraj}, i.e. to
sample the density matrix $\rho_{\aa_\td}$ precisely according to the distribution $\Pr(\aa_\td)$. Therefore, Eq.~\eqref{eq:MaveBR} might look like the most natural choice in this context. However, in order to experimentally access  nonlinear quantities in the density matrix, such as moments whenever $|\mathbf{r}| > 1$, it becomes necessary to obtain the same quantum state multiple times, which requires some form of post-selection or of an auxiliary classical simulation~\cite{PhysRevLett.130.220404,Feng2026}. In this case, the uniform sampling of Eq.~\eqref{eq:MaveFM} also becomes relevant and we consider it here. 

From a more abstract point of view of statistical physics, the averages on $\aa$ can be seen as an average over disorder. In this perspective, Eq.~\eqref{eq:MaveFM} corresponds to the usual quenched average of disordered systems, where different realizations of disorder (labeled by the index $\aa$) are uncorrelated. Conversely, Eq.~\eqref{eq:MaveBR} is the direct consequence of Born's rule, in which the outcomes of measurements are correlated with each other as they depend on the state itself.
A unified description of both cases is obtained from the generalization
\begin{equation}
\label{eq:MaveN}
\average{\PP_{\mathbf{r}}(\tc)}^{(N)} := \frac{1}{Z^{(N)}}\sum_{\aa_{\td}} \PP_{\mathbf{r}}[\rho_{\aa_{\td}}] \Pr(\aa_{\td})^N = \frac{1}{Z^{(N)}}\sum_{\aa_{\td}} \PP_{\mathbf{r}}[\check\rho_{\aa_{\td}}] \Tr[\check \rho_{\aa_{\td}}]^{N - |\mathbf{r}|}  
\end{equation}
where $Z^{(N)} = \sum_{\aa_{\td}} \Tr[\urho_{\aa_\td}]^N$ again enforces normalization $\langle 1 \rangle^{(N)} = 1$; 
in the last equality we used homogeneity of the moments \eqref{eq:momdef}. Clearly, the cases $N \to 0,1$ reduce respectively to the {FM/BR} averages in Eq.~\eqref{eq:Mave}, with $Z^{(N=1)} = 1$ and $Z^{(N=0)} = Z_{\rm FM}$.

As explained in the introduction, 
recently, in~\cite{deluca2025universality}, it was argued that in the weak measurement phase and in the scaling limit $\tc, L\to \infty$, the moments collapse onto universal curves of $x = t/\tp(L)$, independent of microscopic details and even spatial dimensionality
\begin{equation}
    \label{eq:scalinglim}
\lim_{\substack{t, L \to \infty\\ x = t/\tp(L)}} \average{\PP_{\mathbf{r}}(t)}^{(N)}   =: \PPave_{\mathbf{r}}^{(N)}(x) \;.
\end{equation}
Thus, the set of these scaling functions identifies a universality class and can be used to express directly the average of the Rényi entropies in the scaling limit
\begin{equation}
\label{eq:scalinglimRenyi}
\mathcal{S}_n^{(N)}(x) := \lim_{\substack{t, L \to \infty\\ x = t/\tp(L)}} \average{S_{n}(t)}^{(N)} = \frac{\left. \partial_k \PPave_{r_n = k}^{(N)}(x) \right|_{k = 0}}{1-n} \;.
\end{equation}

A crucial assumption underlying this framework is the persistence of strong scrambling throughout the purification process. 
Because measurements are inefficient in the weak-measurement phase, unitary dynamics thoroughly scrambles the system between successive significant purification events. 
More specifically, let us assume that the time window $\Delta \tc$ in \eqref{eq:quantumchannel} is such that $\tc_{\rm micro}(L)\ll \Delta \tc \ll \tp(L)$, where $t_{\rm micro}$ refers broadly to a microscopic scale of scrambling and relaxation. For example, estimating $\tc_{\rm micro}$ with the saturation of entanglement entropies, we obtain $\tc_{\rm micro} = O(L)$, which, given the exponential scaling of $\tp(L)$, therefore provides a wide range for the choice of $\Delta \tc$. In this sense, scrambling translates into the fact that coherent dynamics can be approximated by a uniformly distributed random matrix in the reference group $\GG$: in the most general case, $\GG=\UU(q)$, the entire unitary group of transformations of the global Hilbert space; but there are several interesting situations in which the group $\GG$ is reduced to just a subgroup of $\UU(q)$. A reference case that we will consider here for its physical and mathematical interest is the orthogonal group $\OO(q)$: it is obtained with a purely imaginary Hamiltonian, or for a circuit, if each gate is a purely real matrix.  In general, the natural implication of complete scrambling within the group $\GG$ is to assume that the purification dynamics are invariant under the transformation $K_a \to U K_a V$, where $U, V$ are elements chosen with the Haar measure of $\GG$. As we will see, the purification dynamics, within the universal scaling limit described by Eq.~\eqref{eq:scalinglim}, depends exclusively on the group $\GG$ and the calculation of the scaling functions $\PPave_{\mathbf{r}}^{(N)}(x)$ provides an interesting mathematical challenge~\cite{deluca2025universality}.
 
This symmetry structure admits two complementary theoretical descriptions that are well-captured by the toy models we will introduce in the next sections.

\section{Discrete-time model: products of independent matrices from the Ginibre ensemble} \label{sec:discrete_model}

We consider here a model of monitored dynamics for a quantum dot. Let $\mathcal{H}$ denote a $q$-dimensional Hilbert space, and assume, as anticipated, the system is initially in the maximally mixed state $\rho(t=0)= q^{-1} \id$. 
We evolve the system in discrete steps by applying Kraus operator unraveling, as described in Eq.~\eqref{eq:quantumchannel}. But in this model, we consider a continuous set of Kraus operators so that the index $a$ indexes an ensemble of Gaussian random matrices. Formally, for a given $a$, the entries $[\Kr_a]_{\alpha,\beta}$ are chosen independently as complex ($\beta = 2$) / real ($\beta = 1$) gaussian random variables with variance $1/q$, often known as $\beta$-Ginibre random matrix ensembles. In practice, we can replace
the sum over $a$ with the gaussian average
\begin{equation}
\label{eq:sumatoave}
    \sum_a (\ldots) \to \aveE{\ldots}
\end{equation}
defined for each matrix $\Kr_a$ (we omit the index $a$ to lighten the notation) by the covariance 
\begin{equation}
\label{eq:covK}
    \aveE{\Kr_{\mu\nu}^\ast \Kr_{\gamma\delta} } = \frac 1  q \delta_{\mu \gamma} \delta_{\nu \delta} \;, \qquad  \forall \ \mu,\nu,\gamma,\delta = 1,\dots,q \,.
\end{equation}
where $\ast$ denotes the complex conjugation. By summing over {$\mu = \gamma$}, one easily verifies that this choice enforces the proper normalization \eqref{eq:quantumchannel} of a quantum channel. 
According to Eq.~\eqref{eq:qtraj}, we can write 
\begin{equation}
\label{eq:evolve_onediscrete}
    \urho(\tc = \td \Delta t) = {q^{-1}} M_{\td} M^\dag_{\td} \;, \quad M_{\td} = \Kr_\td \Kr_{\td-1} \ldots \Kr_{1}
\end{equation}
where the $K_j$ are matrices independently sampled from the $\beta$--Ginibre ensemble. In this way, we see that the spectrum of $\check \rho(t)$ corresponds to the singular values of a product of random matrices, a problem very much studied in both the physics and mathematics literature~\cite{furstenberg_noncommuting_1963,furstenberg1960products,bellman1954limit,oseledec_multiplicative_1968, bouchard_rigorous_1986}. For products of Ginibre matrices, exact finite-dimensional results are known for the joint distributions and correlation functions of the eigenvalues and singular values, together with their determinantal or Pfaffian structures in the different symmetry classes; their macroscopic densities, microscopic bulk and edge statistics, and several asymptotic regimes have also been investigated extensively. In particular, in the limit of a large number of factors at fixed matrix dimension, the logarithms of the singular values are governed by Lyapunov exponents and become asymptotically Gaussian and increasingly separated. We refer to Ref.~\cite{akemann_ipsen_review} for a review of these developments. Here, rather than studying the unnormalized singular values or their local correlations, we consider nonlinear observables of the trace-normalized spectrum.

We deal with it with the convenient framework of replicas. First, it is useful to consider the vectorization 
\begin{equation}
\label{eq:vecdef}
X \in \End(\mathcal{H})\to \kket{X} = \sum_{\alpha,\beta} X_{\alpha,\beta} \kket{\alpha,\beta} \in \mathcal{H}\otimes \mathcal{H}^\ast    
\end{equation}
where $X_{\alpha,\beta} $ is the matrix representation of $X$ (the specific basis being irrelevant here). Applying this to
\eqref{eq:evolve_onediscrete}, we have
\begin{align}
\urho(t) \to    \kket{\urho(t)} := \frac 1 q M_{\td}\otimes M_{\td}^\ast \kket{\id} \ , 
\end{align}
where the vectorization of the identity is simply $\kket{\id} = \sum_{\alpha=1}^q\kket{\alpha,\alpha}$. 
Within this framework, we will focus on the calculation of averages in Eq.~\eqref{eq:MaveN}. As explained before, the main idea is to treat $N$ as an integer and introduce the averaged replicated density matrix in $\HNN$, via \eqref{eq:sumatoave}
\begin{equation}
    \kket{\urho^{(N)}(t)} := \aveE{\kket{\urho(t)}^{\otimes N}} \;.
\end{equation}
It is convenient to denote the different Hilbert spaces in $\HNN$ as the set of $2N$ elements 
\begin{equation}
\label{eq:inddef}
    I = \{1, 1^\ast, 2, 2^\ast, \ldots, N, N^\ast\}
\end{equation}
where unstarred/starred integers corresponds to copies of $\mathcal{H}/\mathcal{H}^\ast$.
The possible ways to connect unstarred indices with starred ones can be seen as permutations $\sigma \in \SS_N$, and to each of these we associate the state in $\HNN$, defined as
\begin{equation}
\label{eq:sigmadef}
    \kket{\sigma} = \sum_{\{\alpha\}} \kket{\alpha_1, \alpha_{\sigma_1}}\kket{\alpha_2, \alpha_{\sigma_2}}\ldots\kket{\alpha_N, \alpha_{\sigma_N}} \;.
\end{equation}
Note that the density matrix at initial time $\kket{\check{\rho}^{(N)}(t = 0)} = q^{-N} \kket{\mathbb{1}}^{\otimes N} = q^{-N} \kket{\idp}$, where $\idp$ denotes the identity element in $\SS_N$. 
It is well known that elements of $\SS_N$ can be organized in conjugacy classes $\mathcal{C}_{\mathbf{r}^{(N)}}$ identified with integer partitions of $N$, or more explicitly sequences $\mathbf{r}^{(N)} = \{r_1^{(N)}, r_2^{(N)},\ldots\}$, where {$r_j^{(N)}$} $\in \mathbb{N}$ denotes the number of cycles of size $j$~\cite{fulton2013representation} and such that $\sum_{j} j r_j^{(N)} = N$. 
Now, given an arbitrary sequence $\mathbf{r}$ of {non-negative} integers associated with the moment $\PP_{\mathbf{r}}$ in Eq.~\eqref{eq:momdef}, we can always extend it to a conjugacy class in $\SS_N$ for $N \geq |\mathbf{r}|$, setting {(as $r_1$ is irrelevant for normalized states, we take $r_1 = 0$ in $\mathbf{r}$)}
\begin{equation}
    \mathbf{r}^{(N)} = \{ r_1^{(N)} = N - | \mathbf{r}|, r_2, r_3, \ldots \} \;. 
\end{equation}
Then, one simply has
\begin{equation}
\label{eq:mombraket}
\bbraket{\sigma(\mathbf{r}^{(N)})}{ \urho^{(N)}(t)} = \aveE{\PP_{\mathbf{r}^{(N)}}[\urho]} =
    \aveE{\PP_{\mathbf{r}}[\urho] \Tr[\urho]^{N - |\mathbf{r}|}}
    \;,  
\end{equation}
where with a slight abuse of notation, we indicate with $\sigma(\mathbf{r}^{(N)})$ any element of the conjugacy class $\mathcal{C}_{\mathbf{r}^{(N)}}$. We see that, via the identification \eqref{eq:sumatoave}, 
apart from the normalization $Z^{(N)}$,
Eq.~\eqref{eq:mombraket} reproduces  Eq.~\eqref{eq:MaveN}. Thus, the calculation of the moments \eqref{eq:momdef} reduces to the calculation of overlaps between the evolved replicated density matrix and permutation states. 

Note that, since the average in \eqref{eq:sumatoave} amounts to choose independent matrices at different timesteps, we have
\begin{equation}
\label{eq:Tdef}
    \kket{\urho^{(N)}(t = \td \Delta \tc)} = \underbrace{\aveE{(K_{\td}\otimes K_{\td}^\ast)^{\otimes N}}}_{=: T} \kket{\urho^{(N)}((\td -1)\Delta \tc)}= T^\td \kket{\urho^{(N)}(0)}\,. 
\end{equation}
where $T \in \operatorname{End}(\HNN)$ can be interpreted as a transfer matrix. In principle, the transfer matrix $T$ has huge size $q^{2N} \times q^{2N}$. However, the random matrix structure of this problem allows us to highlight that the effective size of the matrix $T$ is actually much smaller. The details of this reduction depend on the ensemble of random matrices under consideration, so we will consider separately the fundamental case $\beta = 2$ (already analyzed in Ref.~\cite{deluca2025universality}) and then the case $\beta = 1$.

\subsection{Complex Ginibre ensemble $(\beta = 2)$} \label{subsec:complexginibre}
\begin{figure}
    \centering
    \includegraphics[width=\linewidth]{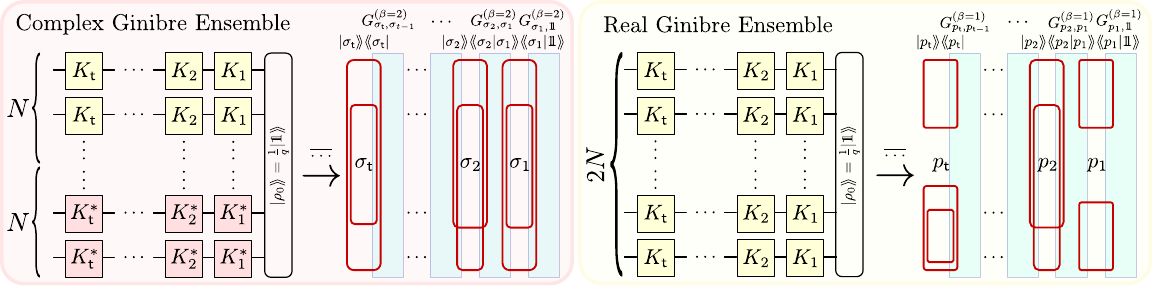}
    \caption{Products of independent random matrices from the Ginibre ensemble. \textit{Left:} complex Ginibre matrices. By Wick's theorem, averaging at each time-step $i$ yields a sum over the permutations $\sigma_i\in \SS_N$ of the $2N$ indices. Permutations are indeed depicted as loops connecting one index from the first batch of $N$ matrices to one from the second batch. Blueish rectangles represent the overlap matrix $G^{(\beta=2)}_{\sigma,\sigma'}$ Eq.~\eqref{eq:grambeta2}. 
    \textit{Right:} real Ginibre matrices. Averaging at each time-step $i$ yields a sum over pairings $p_i\in \Pa_N$ of the $2N$ matrix indices. Pairings are represented by loops connecting couples of indices in the batch of $2N$, while the Gram matrix elements $G^{(\beta=1)}_{p,p'}$ \eqref{eq:grambeta1} are represented by blueish rectangles.  }
    \label{fig:replicas}
\end{figure}

In this section, we consider the case where all entries of $\Kr$ are complex numbers and the only non-vanishing correlations can be read in Eq.~\eqref{eq:covK}. Thus, the matrix elements of the transfer matrix $T$ can be computed by a simple application of the Wick's theorem: this amounts to all possible ways to contract one entry from the set of $N$ $K$'s with one entry from the set of $N$ $K^\ast$,
leading to
\begin{equation}
\label{eq:Tbeta2}
   T := \aveE{(K_{\td}\otimes K_{\td}^\ast)^{\otimes N}} = q^{-N} \sum_{\sigma \in \SS_N} \kket{\sigma}\bbra{\sigma}
\end{equation}
where $\kket{\sigma}$ labels the permutation states defined in Eq.~\eqref{eq:sigmadef}. We see that the transfer matrix effectively spans a set of $|\SS_N| = N!$ states. However, a simple inspection shows that these states are not orthogonal and give rise to the $N! \times N!$ Gram matrix
\begin{equation}
\label{eq:grambeta2}
    G_{\sigma, \sigma'}^{(\beta=2)} := \bbraket{\sigma}{\sigma'} = q^{N - d_N(\sigma, \sigma')} \;, \qquad \sigma, \sigma' \in \SS_N
\end{equation}
where $d_N(\sigma, \sigma')$ is the transposition distance, i.e. the minimal number of transpositions connecting $\sigma, \sigma' \in \SS_N$. In the following we will omit the superscript $\beta=2$ when there is no ambiguity. 
In Fig.~\ref{fig:replicas} we graphically represent the evolution of a single instance $\kket{\urho_t^{\otimes N}}$, and its average $\kket{\urho_t^{(N)}}$, leading to the transfer matrix $T$ Eq.~\eqref{eq:Tbeta2} and the Gram matrix elements $G^{(\beta=2)}_{\sigma,\sigma'}$ \eqref{eq:grambeta2} for the symmetry class $\beta=2$.

Inserting Eqs.~\eqref{eq:Tdef} and \eqref{eq:Tbeta2} in Eq.~\eqref{eq:mombraket}, we see that the calculation of the moments reduces to
\begin{equation}
\label{eq:momfromG}
\aveE{\PP_{\mathbf{r}}[\urho] \Tr[\urho]^{N - |\mathbf{r}|}} = q^{-N (\td+1)}\sum_{\sigma_1,\ldots, \sigma_{\td}} G_{\sigma(\mathbf{r}^{(N)}), \sigma_{\td}} \ldots G_{\sigma_2, \sigma_1}G_{\sigma_1, \sigma_0 = \idp} = q^{-N (\td+1)} [G^{\td+1}]_{\sigma(\mathbf{r}^{(N)}), \idp} \;,
\end{equation}
We thus see that the calculation of the moments has been reduced to the calculation of the matrix elements of the Gram matrix of reduced (and $q$-independent) size $N! \times N!$.  The emergence of permutation states from the average on an ensemble of random matrices is a crucial mechanism underlying scrambling in ergodic quantum dynamics. In this case, there is a second fundamental ingredient towards universality: we are interested in studying purification dynamics in the scaling limit \eqref{eq:scalinglim} where both the size of the system and time are large. From Eq.~\eqref{eq:grambeta2} we have the expansion
\begin{equation}
\label{eq:Glargeqbeta2}
    G_{\sigma, \sigma'}^{(\beta=2)} = q^{N}\left(\delta_{\sigma, \sigma'} + \frac{1}{q} A_{\sigma, \sigma'}^{(\beta=2)} + O(q^{-2})\right)
\end{equation}
where $A$ denotes the adjacency matrix of the transposition graph, i.e.
\begin{equation}
\label{eq:Adefbeta2}
    A_{\sigma, \sigma'}^{(\beta=2)} = \begin{cases}
        1 & d_N(\sigma, \sigma') = 1 \;,\\
        0 & \mbox{otherwise}
    \end{cases} \;.
\end{equation}
We define the purification time scale as the cost associated to the elementary step from $\idp$ to the transposition $(12)$, in units of $\Delta t$, thus 
\begin{equation}
\label{eq:tpdefbeta2}
    \frac{\Delta t}{\tp} = \frac{G_{(12), \idp}^{(\beta=2)}}{G_{\idp, \idp}^{(\beta=2)}} = q^{-1} \;.
\end{equation}
Therefore, in the scaling limit \eqref{eq:scalinglim}, $x = \tc/\tp = \td/q$ and accounting for the normalization factor, we obtain the expression
\begin{equation}
\label{eq:PPaveexpbeta2}
\PPave_{\mathbf{r}}^{(N)}(x; \beta = 2) = \frac{[e^{x A^{(\beta=2)}}]_{\sigma(\mathbf{r}^{(N)}),\idp}}{[e^{x A^{(\beta=2)}}]_{\idp,\idp}} 
\;, \qquad \sigma \in \mathcal{C}_{\mathbf{r}^{(N)}} 
\end{equation}
where $[e^{x A^{(\beta=2)}}]_{\idp,\idp}$ is a normalization corresponding to the $N$-th moment of the trace.
We can therefore see that the calculation of moments is reduced to the one of the elements of an $N! \times N!$ matrix. In this way, the combinatorial interpretation becomes clear: expanding in powers of $x$, we obtain terms of the type $[A^\ell]_{\sigma(\mathbf{r}^{(N)}),\idp}$, with $\ell \in \mathbb{N}$; since $A$ is the adjacency matrix of the transposition graph, these terms are clearly identifiable as the number of paths of length $\ell$ connecting two specific nodes, a quantity also known as Hurwitz multiplicity~\cite{stanley_enumerative_1999, STANLEY1981255}. 

Eq.~\eqref{eq:PPaveexpbeta2} provides an operational way to calculate moments in generalized ensembles defined by the index $N$, whenever $N {\geq} |\mathbf{r}|$, in terms of finite-dimensional matrix traces. However, the most physically interesting cases are the specific cases corresponding to Born's rule ($N \to 1$) and forced measurements ($N \to 0$). Extracting these behaviors from the integer values calculable via Eq.~\eqref{eq:PPaveexpbeta2} is a non-rigorous and often heuristic procedure called the replica trick. 
As explained in Ref.~\cite{deluca2025universality}, there are various ways to obtain predictions in these regimes: one possibility is to consider the perturbative expansion at small $x$ in terms of Hurwitz multiplicities. As we will see, in some cases, the latter admit a simple functional expression in terms of the parameter $N$ (e.g., a polynomial~\cite{stanley2009combinatorial, olshanski2009plancherel}) for which it is straightforward to take the limit $N \to 0$ or $N \to 1$. An alternative approach is to consider appropriate generating functions based on the summation of all integer values of $N$. This approach has the advantage of providing a prediction for arbitrary values of the scaling parameter $x$, but it is technically more difficult. An example of its use is provided in \cite{deluca2025universality} for the calculation of the von Neumann entropy in the Born's rule case, using that, in that case, both the Rényi index $n$ and the replica index $N$ need to be sent to $1$, so one can consider $\lim_{N \to 1} \mathcal S_N^{(N)}$ in Eq.~\eqref{eq:scalinglimRenyi}.

\subsection{Real Ginibre ensemble $(\beta = 1)$} \label{subsec:real}
We will now extend the previous discussion to the case where the matrix elements of each matrix $K$ are real numbers, still assuming that Eq.~\eqref{eq:covK} holds for the covariance between the matrix elements. However, since the elements are real, $\Kr^\ast = \Kr$ and thus we have that
\begin{equation}
\label{eq:covKbeta1}
\aveE{\Kr_{\mu\nu} \Kr_{\gamma\delta} } = \aveE{\Kr_{\mu\nu}^\ast \Kr_{\gamma\delta} } = \frac 1  q \delta_{\mu \gamma} \delta_{\nu \delta} \;, \qquad  \forall \ \mu, \nu,\gamma,\delta = 1,\dots,q \,.
\end{equation}
The effect of this additional non-zero term manifests itself in the calculation of the transfer matrix $T$ using Wick's theorem: while for $\beta = 2$ in Eq.~\eqref{eq:Tbeta2}, the only relevant contractions were those between a matrix $K$ and its complex conjugate $K^\ast$, in this case, all possible contractions between the $2N$ matrices must be considered.
Applying Eq.~\eqref{eq:covKbeta1}, one has
\begin{equation}
\label{eq:Tbeta1}
T := \aveE{(\Kr_{\td}\otimes \Kr_{\td}^\ast)^{\otimes N}}  = 
\aveE{\Kr_\td^{\otimes 2N}} = q^{-N}\sum_{p\in \Pa_N} \kket{p} \bbra{p} \,,
\end{equation}
where we introduced the pairing states
\begin{equation}
    \kket{p} =\sum_{a_1,\dots,a_{2N}=1}^q  \left( \prod_{(i,j)\in p} \delta_{a_i,a_j} \right)\ket{a_1,\dots, a_{2N}} \quad \in \quad  \HNN \,,
\end{equation}
where $p = \{(i_1,i_2), \dots,(i_{2N-1},i_{2N})\}$ is an element of the set of pairings $\Pa_{N}$ of $I$ in \eqref{eq:inddef}. 
There are two crucial observations about the set of pairings. First, the set of permutation states $\kket{\sigma}$ in \eqref{eq:sigmadef} is a subset of the set of pairings $\kket{p}$, corresponding to pairs $(i, i')$ only joining starred and unstarred elements in $I$, so formally $\SS_N \subset {\Pa_N}$. Secondly, the representation of a pairing $p = \{(i_1, i_2),\ldots\}$ can be interpreted as the decomposition into cycles of a permutation of $\SS_{2N}$ composed exclusively of $2-$cycles. In other words, this implies
\begin{equation}
\label{eq:pairingsize}
    \Pa_{N} = \mathcal{C}_{2^{N}} \subset \SS_{2N} \quad \to \quad | \Pa_N | = \frac{(2N)!}{N! 2^N}\;.
\end{equation}
where $|\ldots|$ denotes the cardinality of a finite set and $\mathcal{C}_{\mathbf{r}}$ denotes the conjugacy class in the permutation group with cycle structure $\mathbf{r}$.

\begin{figure}
\begin{center}
\def\LW{0.9pt}
\subfloat{
\begin{minipage}{0.48\columnwidth}
\centering
\text{Pairings of $\Pa_2$ as permutations in $\SS_4$:}

\vspace{0.8em}

\begin{tikzpicture}[every node/.style={inner sep=0pt, outer sep=0pt}]
  \node (L1) {$p_\idp = \{(1,1^*)(2,2^*)\} = \pairinginSn[0.3]{1^*,2^*,1,2}$};

  \node[below=1em of L1] (L2)
    {$p_- = \{(1,2^*)(2,1^*)\} = \pairinginSn[0.3]{2^*,1^*,2,1}$};

  \node[below=1em of L2] (L3)
    {$p_+ = \{(1,2)(1^*,2^*)\} = \pairinginSn[0.3]{2,1,2^*,1^*}$};

  \draw[->, line width=\LW]
    ([xshift=1.2em]L1.east) to[bend left=35]
    node[midway, right=0.8em] {\small two swaps}
    ([xshift=1.2em, yshift=0.2em]L2.east);

  \draw[->, line width=\LW]
    ([xshift=1.2em]L2.east) to[bend left=35]
    node[midway, right=0.8em] {\small two swaps}
    ([xshift=1.2em, yshift=0.2em]L3.east);

  \draw[->, line width=\LW]
  ([xshift=-1.2em, yshift=0.3em]L3.west)
  to[bend left=25]
  node[midway, left=0.8em] {\small two swaps}
  ([xshift=-1.2em,yshift=-0.3em]L1.west);

\end{tikzpicture}
\end{minipage}
}
\hfill
\subfloat{
\begin{minipage}{0.48\columnwidth}
\centering
$G_{p_\idp,p_\idp}^{(\beta=1)}$
\;=\;
\pairingdiagramtworows[\LW]{diagA}{{1/3,2/4}}{{1/3,2/4}}
\;=\;
\thickcircle[\LW]{0.3}
\thickcircle[\LW]{0.3}
\;=\;
$q^2$ 

\vspace{0.3em}
$G_{p_\idp,p_-}^{(\beta=1)}$
\;=\;
\pairingdiagramtworows[\LW]{diagA}{{1/3,2/4}}{{1/4,2/3}}
\;=\;
\thickcircle[\LW]{0.5}
\;=\;
$q^1$ 

\vspace{0.1em}
$G_{p_\idp,p_+}^{(\beta=1)}$
\;=\;
\pairingdiagramtworows[\LW]{diagB}{{1/3,2/4}}{{1/2,3/4}}
\;=\;
\thickcircle[\LW]{0.5}
\;=\;
$q^1$
\end{minipage}
}
\end{center}
\caption{Graphical explanation for the overlap of pairing states, in the case of $|\Pa_2|=3$ pairings of $2N=4$ objects, dubbed $p_\idp,p_-,p_+$. \textit{Left:} When interpreted as permutations in $\SS_4$, the three considered pairings are 2 swaps apart from each other, from which $d_4(p,p')=2$. \textit{Right:} A graphical justification of the Gram matrix elements $G_{p,p'}^{(\beta=1)}$ shows Eq.~\eqref{eq:grambeta1} for the case $N=2$. 
\label{fig:overlap}}
\end{figure}
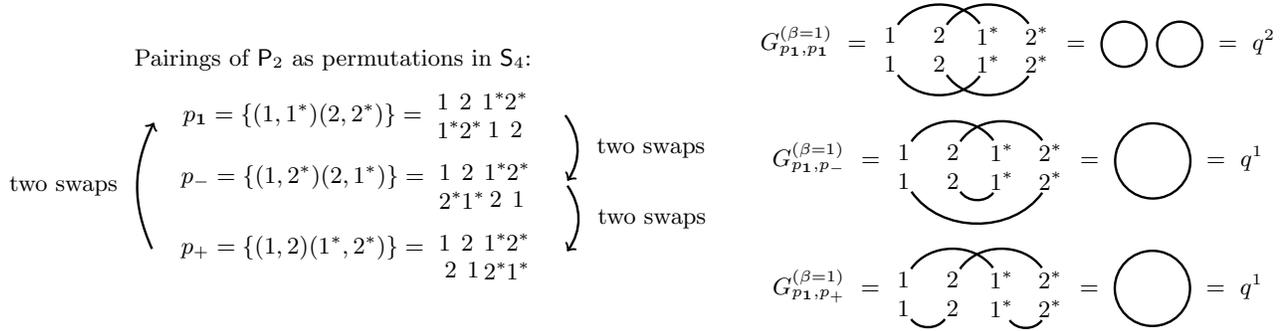

Repeating the derivation in the previous section, we see that the moments can still be expressed by Eq.~\eqref{eq:momfromG}, but with a larger Gram matrix whose expression takes the form
\begin{equation}
\label{eq:grambeta1}
     G_{p, p'}^{(\beta=1)} = \bbraket{p}{p'} = q^{N - d_{2N}(p, p')/2} \;, \qquad p,p' \in \Pa_N  \,.
\end{equation}
where the pairings $p,p'$ are interpreted as elements in $\SS_{2N}$ and 
$d_{2N}(p,p')$ is the corresponding transposition distance.
To better clarify the origin of this equation, a graphical analysis of the case $N=2$ is depicted in Fig.~\ref{fig:overlap}. 
Note that Eq.~\eqref{eq:grambeta1} reduces to Eq.~\eqref{eq:grambeta2} when the pairings $p=\sigma, p'=\sigma'$ are permutations in $\SS_N$, so that $G^{(\beta = 2)}$ is a submatrix of $G^{(\beta=1)}$.
In Fig.~\ref{fig:replicas}, moreover, we represent the transfer matrix $T$ Eq.~\eqref{eq:Tbeta1} and the Gram matrix elements $G_{p,p'}^{(\beta=1)}$ for the real case $\beta=1$, as compared to the complex case $\beta=2$ previously described.
Then, the calculation of the moments can once again be performed in terms of matrix elements of powers of $G$, using Eq.~\eqref{eq:momfromG}.

Similarly to Eq.~\eqref{eq:Glargeqbeta2}, also for $\beta=1$, the Gram matrix admits the large-$q$ expansion
\begin{equation}
\label{eq:Glargeqbeta1}
        G_{p, p'}^{(\beta=1)} = q^{N}\left(\delta_{p, p'} + \frac{1}{q} A_{p, p'}^{(\beta=1)} + O(q^{-2})\right)
\end{equation}
where we introduced the adjacency matrix in the pairing graph
\begin{equation}
\label{eq:Adefbeta1}
    A_{p, p'}^{(\beta=1)} = \begin{cases}
        1 & d_{2N}(p, p') = 2 \;,\\
        0 & \mbox{otherwise}
    \end{cases} \;.
\end{equation}
Note that conservation of parity in $\SS_{2N}$ imposes that the distance between pairings $p,p' \in \Pa_N \subset \SS_{2N}$ is necessarily even; correspondingly, 
the matrix $G^{(\beta=1)}$ only depends on integer powers of $q$ and $A^{(\beta=1)}$ connects elements with the minimal non-vanishing distance, being $2$. As a consequence, consistently with Eq.~\eqref{eq:tpdefbeta2}, even in this case we can identify the purification time $\tp = q \Delta \tc$ and obtain in the scaling limit of fixed $x = \tc/\tp$ an expression analogous to Eq.~\eqref{eq:PPaveexpbeta2}
\begin{equation}
\label{eq:PPaveexpbeta1}
\PPave_{\mathbf{r}}^{(N)}(x; \beta = 1) = \frac{[e^{x A^{(\beta=1)}}]_{\sigma(\mathbf{r}^{(N)}),\idp}}{[e^{x A^{(\beta=1)}}]_{\idp,\idp}} \;, \qquad \sigma \in \mathcal{C}_{\mathbf{r}^{(N)}}
\end{equation}
where $\sigma(\mathbf{r}^{(N)})$ and $\idp$ are elements of $\SS_N \subset {\Pa_N}$.

\section{General $\GG$-invariant ensembles and the commutant structure}
\label{sec:general_G_invariant}
We now present a unified formulation that explains the origin of the effective transfer matrices derived in the previous sections and further clarifies why their structure is universal. The style and content of this section are more mathematical, and readers who are less interested in these aspects may choose to skip reading it.
The key ingredient is invariance of the Kraus ensemble under a compact subgroup $\GG$ of $\UU(q)$. Once again, we assume to be in the volume phase of a measurement-induced phase transition emerging in the limit of large $q$, where the dimension of the group $\GG$ becomes large. We will keep the discussion about a generic group $\GG$, specifying how it specializes to the situations treated in the cases $\GG=\UU(q)$ and $\GG=\OO(q)$. We assume that each timestep is described by a  set $\{K_a\}$ of Kraus operators, with the crucial property, justified in the previous sections, that the ensemble is invariant under the action of a compact group $\GG$ sampled with Haar measure:
\begin{equation}
\{K_a\}\;\overset{d}{=}\;\{U K_a V\},
\qquad U,V\sim \text{Haar}(\GG).
\label{eq:G_invariant_ensemble}
\end{equation}
where the equality holds in the sense of distributions. 
Then, one can once again compute the moments from the powers of the transfer matrix
\begin{equation}
\label{eq:mombraketGeneric}
\aveE{\PP_{\mathbf{r}^{(N)}}[\urho]}  = {q^{-N}}\bbraket{\sigma(\mathbf{r}^{(N)})| T^{{\td}}}{ \idp} \;,  \qquad T=\aveE{(K\otimes K^{{\ast}})^{\otimes N}}  \;.
\end{equation}
We introduce the \emph{commutant algebra}
\begin{equation}
\mathcal A_{\GG}^{(N)} :=  \{ X\in \End(\mathcal{H}^{\otimes N})\;:\; [X,U^{\otimes N}]=0\ \ \forall U\in\GG\} \simeq \{ {\kket{X}}\in \HNN \;:\; (U\otimes U^\ast)^{\otimes N}\kket{X} = \kket{X} \ \ \forall U\in\GG\}
\end{equation}
where in the second equality we rewrote the same notion using the vectorized notation \eqref{eq:vecdef} (after replication) to show that $\mathcal{A}_\GG^{(N)}$ can be interpreted as the space left invariant by the replicated action of the group $\GG$. 
{Denoting by $\mathcal{U}_N(U) := (U\otimes U^\ast)^{\otimes N}$ the replicated action of $U \in \GG$ on $\HNN$, Eq.~\eqref{eq:G_invariant_ensemble} implies $T = \mathcal{U}_N(U)\, T\, \mathcal{U}_N(V)$ for all $U,V\in\GG$, i.e.}
\begin{equation}
 {\mathcal{U}_N(U)\, T = T = T\, \mathcal{U}_N(V)  \;, \quad \forall U,V \in \GG \quad \Rightarrow \quad T = P_\GG\, T\, P_\GG
\;\in\; \operatorname{End}(\mathcal{A}_{\GG}^{(N)})}
\end{equation}
{with $P_\GG = \int_\GG dU\, \mathcal{U}_N(U)$ the orthogonal projector onto $\mathcal{A}_{\GG}^{(N)}$.}
This already implies the observed reduction for the action of the transfer matrix from the huge space $\HNN$ to the commutant vector space $\mathcal{A}_{\GG}^{(N)}$.

To proceed further, it should be noted that both the transfer matrix $T$ in Eq.~\eqref{eq:mombraketGeneric} and the symmetry transformations $(U \otimes U^\ast)^{\otimes N}$ have a structure consisting of the repetition of the exact same object across different replicas. This property allows for an additional symmetry consisting of replica exchanges. In the most general case, $K$ and $K^\ast$ (and analogously for $U$ and $U^\ast$) are different from each other, but it is still possible to freely permute the replicas corresponding to the starred/unstarred indices in \eqref{eq:inddef}. This identifies a group of discrete symmetries equal to $\SS_N \times \SS_N$. This group of symmetries can be enlarged in certain cases, such as in the case of real Kraus matrices $K=K^\ast$, in which case it is possible to permute freely all $2N$ replicas indiscriminately, giving rise to $\SS_{2N}$. In general, we assume that there exists a discrete symmetry group $\ggg_N \supseteq \SS_N \times \SS_N$ that acts on {$\HNN$ by permuting the $2N$ tensor factors labeled by $I$ in \eqref{eq:inddef}, through a
representation $R:  \ggg_N \to \operatorname{End}(\HNN)$ leaving $\mathcal{A}_{\GG}^{(N)}$ invariant,} and such that
\begin{equation}
\label{eq:UUTR}
    [(U\otimes U^\ast)^{\otimes N}, R(g)] = [T, R(g)] = 0 \;, \quad \forall g \in \ggg_N \;, \quad U \in \GG \,.
\end{equation}
A natural consequence of this is that the action of $R(g)$ preserves the invariant states. Starting from the reference state $\kket{\idp}$, one must have $\mathcal{A}_{\GG}^{(N)} \supseteq \operatorname{span}[ R(g) \kket{\idp}, \; g \in \ggg_N]$. We will assume that in practice, the equality holds, i.e. the entire structure of the commutant space can be obtained by using the action of the group $\ggg_N$ on the reference vector $\kket{\idp}$. 
This is a compatibility requirement between the group $\GG$ and the discrete symmetry group among replicas $\ggg_N$, that in the cases $\GG = \UU(q)$ and $\GG = \OO(q)$ follows from Schur-Weyl duality.

In practice, the basis consisting of the states $R(g)\kket{\idp}$ contains redundancies, for example because there may exist $g,g' \in \ggg_N$ such that $R(g)\kket{\idp} = R(g')\kket{\idp}$. To take this into account, we introduce the stabilizer group
\begin{equation}
\mathsf{H}_N :=
\{ h\in \ggg_N \; | \; R(h) \kket{\idp} = \kket{\idp} \}
\end{equation}
We assume that for sufficiently large $q$, there are no other linear dependencies between elements of the type $R(g)\kket{\idp}$. In this case, we can identify
\begin{equation}
\label{eq:commbeta2}
\mathcal{A}_{\GG}^{(N)} \equiv \mathbb{C}[\ggg_N / \mathsf{H}_N
] = \operatorname{Span}[ \kket{g \mathsf{H}_N} \;, g \in \ggg_N]
\end{equation}
i.e. the commutant vector space can be seen as the span of the left cosets of the group over the stabilizer{, with $\kket{g \mathsf{H}_N} := R(g)\kket{\idp}$}. 
For instance, in the case $\GG = \UU(q)$, as we stated above $\ggg_N = \SS_N \times \SS_N$, with the action $R((\sigma, \sigma')) \kket{\idp} = \kket{\sigma \sigma'^{-1}}$; thus $\kket{\idp}$ is left invariant by choosing $\sigma' = \sigma$, with $\mathsf{H}_N = \{(\sigma, \sigma) \in \ggg_N | \sigma \in \SS_N \} \equiv \SS_N$ and for $q \geq N$
\begin{equation}
\label{eq:commAU}
\mathcal{A}_{\UU(q)}^{(N)} = \mathbb{C}[\SS_N \times \SS_N / \SS_N] \equiv  \mathbb{C}[\SS_N]  \simeq \operatorname{Span}[\kket{\sigma} \;, \sigma \in \SS_N ]\;,
\end{equation}
the already introduced vector space of permutations.  Instead, in the case $\GG = \OO(q)$, all replicas can be permuted and $\ggg_N = \SS_{2N}$. In this case, {viewing} a pairing $p \in \mathcal{C}_{2^N} \subset \SS_{2N}$, the group $\SS_{2N}$ acts on the pairings as $R(\sigma) \kket{p} = \kket{\sigma p \sigma^{-1}}$ for $\sigma \in \SS_{2N}$. The whole set of  pairings can be generated acting on the state $\kket{\idp}$, seen as the reference pairing  $p_{\idp} = ((1,1^\ast), (2, 2^\ast), \ldots, (N, N^\ast))$. However the mapping $\sigma \to \sigma {p_{\idp}} \sigma^{-1}$ is not one-to-one due to the existence of a non-trivial stabilizer group, corresponding to the wreath product, i.e., $\mathsf{H}_N = \SS_2^N \rtimes \SS_N$. Intuitively, it corresponds to the exchanges within each pairing, or by permutations of the $N$ pairings.
As a consequence, {for $q \geq N$}
\begin{equation}   
\label{eq:commAbeta1}
\mathcal{A}_{\OO(q)}^{(N)} = \mathbb{C}[\SS_{2N} / \SS_2^N \rtimes \SS_N] \equiv \mathbb{C}[\Pa_N] \equiv  \operatorname{Span}[\kket{p} \;, p \in \Pa_N ]\;,
\end{equation}
also known as the Brauer algebra~\cite{56b5304f-de34-3e73-ad68-4648cd3270d9}. Note that the dimension matches consistently with Eq.~\eqref{eq:pairingsize}. Moreover, as $\GG$ is a subgroup of $\UU(q)$, $\mathcal{A}_{\UU(q)}^{(N)} = \mathbb{C}[\SS_N] \subset \mathcal{A}_{\GG}^{(N)}$ as a subalgebra.

Eq.~\eqref{eq:UUTR} alone allows us to simplify the calculation of the powers of $T$ required to determine the moments in Eq.~\eqref{eq:mombraketGeneric}. Indeed, Schur's lemma ensures that $T$ is {proportional to} the identity in each irreducible representation of $\ggg_N$ in $\mathcal{A}_{\GG}^{(N)}$ {(when it appears with unit multiplicity)}. But we can proceed further as we are interested in exploring the scaling limit where $q$ and $t$ become large, under the assumption of a long purification time. Due to Eq.~\eqref{eq:UUTR}, we know that $T$ {commutes with $R(\ggg_N)$} and therefore it is natural to consider an expansion in terms of the conjugacy classes of $\ggg_N$. Let us consider the conjugacy class in $\ggg_N$ associated with the swap $[(12)] = \{g (12) g^{-1} | g \in \ggg_N\}$ {(for $\ggg_N = \SS_N \times \SS_N$, we take $(12) \equiv ((12), e)$, acting on unstarred replicas only)} and the linear sum over all its elements leads to the operator
\begin{equation}
\label{eq:Mdef}
 M = \sum_{h \in [(12)]} R(h) \;,
\end{equation}
also belonging to the center of {the group algebra of} $\ggg_N$, i.e.
\begin{equation}
\label{eq:Mcentral}
[R(g), M] = 0
\qquad \forall g \in \ggg_N \;.
\end{equation}
We assume that $T = \eta_0 \mathbb{1} + \eta_1 M + \ldots$, where the neglected terms correspond to more complex transitions suppressed within the scaling limit. 
The structure of the operator $M$ can be further clarified using the coset description 
$\mathcal{A}_{\GG}^{(N)} \simeq \mathbb{C}[\ggg_N / \mathsf{H}_N]$. 
In the natural basis $\{ \kket{g \mathsf{H}_N} \}$, the matrix elements of $M$ are given by
\begin{equation}
M_{g'\mathsf{H}_N,\, g\mathsf{H}_N}
=
\#\big( [(12)] \cap g' \mathsf{H}_N g^{-1} \big),
\end{equation}
where $\#$ denotes the cardinality of the finite set. So, all the entries are integers and in particular, the diagonal contribution is independent of the coset and equals
\begin{equation}
\label{eq:mvaldef}
m = |[(12)] \cap \mathsf{H}_N|,
\end{equation}
which yields the decomposition
\begin{equation}
\label{eq:AfromM}
M = m\,\mathbb{1} + {c}\, A,
\end{equation}
where $A$ has vanishing diagonal and {entries in $\{0,1\}$, and $c$ is the number of elements of $[(12)]$ mapping a coset into an adjacent one (assumed independent of the pair of cosets; $c=1$ for $\GG = \UU(q)$ and $c=2$ for $\GG = \OO(q)$).}
The matrix $A$ admits a natural interpretation as the adjacency matrix of the 
Schreier graph~\cite{lubotzky1994discrete} associated with the action of $\ggg_N$ on the coset space 
$\ggg_N / \mathsf{H}_N$, with generating set given by the conjugacy class $[(12)]$. 
Vertices correspond to cosets, and two vertices are connected {if}
{there exist} elements of the class mapping one coset into the other. It provides a natural generalization of the matrices introduced in Eqs.~(\ref{eq:Adefbeta2}, \ref{eq:Adefbeta1}).

Therefore, within the scaling regime where higher conjugacy classes are suppressed, 
the transfer matrix takes the form consistent with Eqs.~(\ref{eq:Glargeqbeta2}, \ref{eq:Glargeqbeta1})
\begin{equation}
T
=
\eta_0 \mathbb{1}
+
\eta_1 M + \ldots
=
(\eta_0 + m \eta_1)\,\mathbb{1}
+
{c}\eta_1 A + \dots \;.
\end{equation}
Consistently with \eqref{eq:tpdefbeta2}, we can define the time scale for purification from the (normalized) matrix element
\begin{equation}
\frac{T_{(12),\idp}}{T_{\idp,\idp}} \simeq {\frac{c\eta_1}{\eta_0 + m \eta_1}}
=  \frac{\Delta t}{\tp}
\end{equation}
In the scaling limit of interest this ratio is expected to become exponentially small in the system size. For the moments, we therefore obtain 
\begin{equation}
\label{eq:PPaveexpgeneralG}
\PPave_{\mathbf{r}}^{(N)}(x; \GG) = \frac{[e^{x A}]_{\sigma(\mathbf{r}^{(N)}),\idp}}{[e^{x A}]_{\idp,\idp}} \;, \qquad \sigma \in \mathcal{C}_{\mathbf{r}^{(N)}}
\end{equation}
formally analogous to Eqs.~(\ref{eq:PPaveexpbeta2},\ref{eq:PPaveexpbeta1}), which therefore justifies their validity and generality.

\subsection{Spectrum of the adjacency matrix}
In general, replica limits require analytical treatment for generic $N$ and therefore benefit from the possibility of diagonalizing the adjacency matrix $A$ and thus simplifying the calculation of the matrix elements in Eq.~\eqref{eq:PPaveexpgeneralG}. This can be done on the basis of a purely algebraic general treatment based on the considerations of the previous section.

We now give a general and abstract discussion of the diagonalization of the 
adjacency operator $A$, valid for any discrete symmetry group 
$\ggg_N$ acting transitively on the invariant space 
$\mathcal A_{\GG}^{(N)} \simeq \mathbb C[\ggg_N / \mathsf H_N]$. It is convenient to consider directly the operator \eqref{eq:Mdef}, trivially related to $A$ via \eqref{eq:AfromM}.

By Schur's lemma, Eq.~\eqref{eq:Mcentral} implies that on each irreducible block of $\ggg_N$ inside $\mathcal{A}_{\GG}^{(N)}$, $M$ is a multiple of the identity. 
We write the decomposition into irreducible representations of $\ggg_N$ as
\begin{equation}
\mathbb C[\ggg_N/\mathsf H_N]
=
\bigoplus_{\Lambda \in \operatorname{Irreps}(\ggg_N)}
m_\Lambda \, V_\Lambda,
\end{equation}
where $V_\Lambda$ runs over irreducible representations of $\ggg_N$
and $m_\Lambda$
is the corresponding multiplicity.
Let $P_\Lambda$ denote the projector onto the $\Lambda$-isotypic component, i.e. onto the $m_\Lambda$ copies  of $V_\Lambda$ corresponding to the same irrep $\Lambda$.
Then one has the spectral decomposition
\begin{equation}
\label{eq:Airreps}
A = \sum_{\Lambda \in \operatorname{Irreps}(\ggg_N)} \nu(\Lambda)\, P_\Lambda,
\qquad
e^{xA}
=
\sum_{\Lambda \in \operatorname{Irreps}(\ggg_N)} e^{x\nu(\Lambda)} P_\Lambda \;,
\end{equation}
and the eigenvalue $\nu(\Lambda)$ then appears with multiplicity
$m_\Lambda d^\Lambda$ and $d^\Lambda = \dim V_\Lambda$, the dimension of each irreducible representation $\Lambda$ of $\ggg_N$.

The eigenvalues can be expressed in terms of characters, by simply considering $\Tr[M P_\Lambda]$.
Let
$\chi^\Lambda(g)$ be the character of $V_\Lambda$.
Denoting by $d_{[(12)]}$ the size of the conjugacy class $[(12)]$, one finds
\begin{equation}
\label{eq:nulambdagen}
\nu(\Lambda)
=
{\frac{1}{c}\Big(}\frac{d_{[(12)]} \chi^\Lambda(g = (12))}{d^\Lambda}
-
m {\Big)}
\end{equation}
where the integer $m$ is defined in Eq.~\eqref{eq:mvaldef} {and $c$ below Eq.~\eqref{eq:AfromM}}.
This formula provides the spectrum of $A$ in complete generality.

In order to compute matrix elements of functions of $A$, such as \eqref{eq:PPaveexpgeneralG}, it is convenient to describe the spectral projectors  $P_\Lambda$. They  can be written explicitly in terms of
characters $\chi^\Lambda(g)$ of $\ggg_N$, relative to $\Lambda \in \operatorname{Irreps}(\ggg_N)$ on the element $g \in \ggg_N$.  
Using the orthogonality of characters, one has
\begin{equation}
\label{eq:Pfromchar}
P_\Lambda
=
\frac{d^\Lambda}{|\ggg_N|}
\sum_{g \in \ggg_N}
\chi^\Lambda(g)^\ast\, R(g).
\end{equation}
Using \eqref{eq:Airreps}, this expression can be used to compute the matrix elements of $e^{xA}$ required for the moments. 

A particularly relevant case is the one where $(\ggg_N, \mathsf H_N)$ forms a Gelfand pair~\cite{macdonald1998symmetric}, i.e. if the multiplicities
${m_\Lambda} \in \{0,1\}$. We will see that this is true in the unitary/orthogonal cases considered here. In this case,
for those $\Lambda$ such that $m_\Lambda = 1$,
the projector $P_\Lambda$ reduces to the projector onto the single irrep $V_\Lambda$. 
As a consequence, one can introduce the spherical functions~\cite{macdonald1998symmetric}
\begin{equation} \label{eq:spherical}
    \omega^\Lambda(g) := \frac{1}{|\mathsf{H}_N|}
\sum_{h \in \mathsf H_N}
\chi^\Lambda(gh^{-1})^\ast
\end{equation}
and obtain finally
\begin{equation}
 [e^{x A}]_{\sigma(\mathbf{r}^{(N)}),\idp} =\frac{|\mathsf{H}_N|}{|\ggg_N|}\sum_{\Lambda \in \operatorname{Irreps}(\ggg_N)} e^{x\nu(\Lambda)} d^\Lambda \omega^\Lambda(\sigma) \;.
\end{equation}

\subsection{Specialization to the $\beta=1,2$ cases
\label{ref:specialbeta12}}
We conclude this section by seeing how this discussion specializes in the cases of interest in this article, thus also providing a summary of the relevant formulas. Before proceeding, it is useful to recall that the irreducible representations of $\SS_N$ correspond to the integer partitions of the integer $N$, already introduced in \eqref{eq:rdef}. Here, however, following the standard literature, we go from the notation via multiplicities $\mathbf r= \{r_1, r_2,\dots\}$ to the one in terms of a Young diagram $\lambda = \{\lambda_1,...,\lambda_{\ell}\}$, with the conventions $\lambda_i \geq \lambda_{i+1}$ and $\lambda_{j > \ell} = 0$, and the relations
\begin{equation} \label{eq:partitionYoung}
|\lambda|:=\sum_{j=1}^\ell\lambda_j=\sum_{j} j r_j = |\mathbf{r}| \;, \quad {\ell(\lambda)} = \sum_{j\geq 1} r_j \;,
\quad  \mbox{e.g.: }\quad  \lambda = (3,3,1) = 
\begin{tikzpicture}[baseline=-5.0ex, scale=0.65, every node/.style={font=\small}]
  \Young{0}{0}{3,3,1}{}
\end{tikzpicture}
\end{equation}
so that $r_j$ denotes the number of $\lambda$'s equal to $j$. We write $\lambda \vdash N$ to say that $\lambda$ is an integer partition of $N$, i.e. $|\lambda| = N$.
In what follows, with a slight abuse of notation, we may sometimes switch, when there is no ambiguity, between notations for integer partitions, depending on whether it is more convenient to indicate the multiplicities $r_j$ or the components $\lambda_i$.
We will denote by $\mathcal{R}_\lambda$ the irreducible representation of $\SS_N$ associated with the integer partition $\lambda$, with dimension $d^\lambda=\dim(\mathcal{R}_\lambda)$. 
Explicitly, the dimension is given in terms of the Hook-length formula 
\begin{equation}
\label{eq:hooklengthformula}
d^\lambda \;=\; \frac{N!}{\displaystyle\prod_{(i,j)\in Y(\lambda)} h_\lambda(i,j)} \,,
\qquad 
h_\lambda(i,j)=\lambda_i-j+\lambda'_j-i+1,
\end{equation}
where $(i,j)$ labels the squares of the Young diagram (from $(1,1)$ at the top/left) and
we introduced the dual partition $\lambda'$ associated to $\lambda$ by exchanging row/columns of the corresponding Young diagram, as
\begin{equation} \label{eq:conjPart}
    \lambda'_i = \# \{ \lambda_j\; |\; \lambda_j \geq i\}  \;.
\end{equation}
For instance, one has
\begin{equation}
\begin{tikzpicture}[scale=0.65, every node/.style={font=\small}]
  \Young{0}{0}{3,3,1}{$\lambda=(3,3,1)$}

  \node at (4.6, -0.8) {$\xrightarrow{\ \text{transposition}\ }$};

  \Young{6.2}{0}{3,2,2}{$\lambda'=(3,2,2)$}
\end{tikzpicture} \;.
\end{equation}
To avoid complicating the notation, we use $\chi^\lambda(g)$ as the notation for the characters of $\SS_{|\lambda|}$ related to the irreducible representation labeled by the partition $\lambda$, distinguishing them from the characters for generic groups $\chi^\Lambda$ introduced in Eq.~\eqref{eq:Pfromchar}.

Then, we have
\begin{itemize}
    \item in the case $\GG = \UU(q)$, we recall from Eq.~\eqref{eq:commAU} that $\ggg_N = \SS_N \times \SS_N$, with {$\mathcal{A}_{\UU(q)}^{(N)}$} $= \mathbb{C}[\SS_N]$, with the left/right multiplicative action $R((\sigma_1, \sigma_2))\kket{\sigma} = \kket{\sigma_1 \sigma \sigma_2^{-1}}$. The irreps $\Lambda = (\lambda_1, \lambda_2)$ of $\ggg_N$ are then in correspondence with $V_\Lambda  \equiv \mathcal{R}_{\lambda_1} \otimes \mathcal{R}_{\lambda_2}$, with $\lambda_1, \lambda_2 \vdash N$, and $\mathcal{R}_\lambda$ labelling the irreps of $\SS_N$. Then, one can verify that the only irreps having a non-vanishing (and unit) multiplicity require $\lambda_1 = \lambda_2$ and thus
    \begin{equation}
    {\mathcal{A}_{\UU(q)}^{(N)}} \equiv \mathbb{C}[\SS_N]
=
\bigoplus_{\lambda \vdash N}
\mathcal{R}_\lambda \otimes \mathcal{R}_\lambda,
\end{equation}
Additionally, one can employ Eq.~\eqref{eq:nulambdagen} where one finds $m = 0${, $c=1$} and the explicit form
\begin{equation} \label{eq:nu2}
    \nu_{\beta=2}(\lambda) = \frac 1 2 \sum_{j} \lambda_j^2 - \frac{1}{2} \sum_j (\lambda_j')^2 \,.
\end{equation} 
Finally, as $|\ggg_N| = N!^2$, $|\mathsf{H}_N| = N!$ and noting that for the characters of $\ggg_N = \SS_N \times \SS_N$, one has $\chi^{\Lambda}(\sigma_1, \sigma_2) = \chi^{\lambda_1}(\sigma_1) \chi^{\lambda_2}(\sigma_2^{-1})$ and that $d^\Lambda = d^{\lambda_1} d^{\lambda_2}$, we have
\begin{equation} \label{eq:sphericalbeta2}
    \omega^{\Lambda = (\lambda, \lambda)}(g = (\sigma_1, \sigma_2)) := \frac{1}{N!}
\sum_{\sigma \in \SS_N}
\chi^\lambda(\sigma_1 \sigma^{-1})^\ast \chi^\lambda(\sigma_2 \sigma^{-1}) = \frac{1}{d^\lambda}\chi^{\lambda} (\sigma_1 \sigma_2^{-1})
\end{equation}
where in the last equality we used the orthogonality relation for the characters of $\SS_N$. This leads to the final expression
\begin{equation} \label{eq:momentsbeta2}
    [e^{x A}]_{\sigma(\mathbf{r}^{(N)}),\idp} = \frac{1}{N!} \sum_{\lambda \vdash N} d^\lambda \chi^\lambda(
    \sigma(\mathbf{r}^{(N)})) \ e^{x \nu_{\beta=2}(\lambda)} \;.
\end{equation}
This expression coincides with what was obtained in~\cite{deluca2025universality} (see Eq.~(99) therein) but places its origin in a more general framework.
 \item in the case $\GG = \OO(q)$, we recall from {Eq.~\eqref{eq:commAbeta1}} that the replica permutation symmetry enlarges to $\ggg_N = \SS_{2N}$ and that the commutant space can be identified with the vector space of pairings,
    \begin{equation}
        {\mathcal{A}_{\OO(q)}^{(N)}} \equiv \mathbb{C}[\SS_{2N}/\mathsf H_N] \equiv \mathbb{C}[\Pa_N],
        \qquad
        \mathsf H_N = \SS_2^N \rtimes \SS_N,
    \end{equation}
    where $\mathsf H_N$ is the stabilizer of the reference pairing ${p_{\idp}}=((1,1^\ast),(2,2^\ast),\ldots,(N,N^\ast))$. The irreps $\Lambda$ of $\ggg_N=\SS_{2N}$ are labelled by partitions $\Lambda \vdash 2N$, with $V_\Lambda \equiv \mathcal{R}_\Lambda$ and character $\chi^\Lambda$. Also, we are dealing with a Gelfand pair $(\SS_{2N}, \SS_2^N\rtimes \SS_N)$
    and only partitions with even parts appear with multiplicities $m_\Lambda = 1$, i.e.
    \begin{equation}
        {\mathcal{A}_{\OO(q)}^{(N)}} \equiv \mathbb{C}[\Pa_N]
        =
        \bigoplus_{\lambda \vdash N} \mathcal{R}_{2\lambda},
    \end{equation}
    where $2\lambda$ denotes the partition of $2N$ obtained by doubling each part of $\lambda$. 
    Simple counting shows that the integer coefficient $m$ in \eqref{eq:mvaldef} is given by 
    \begin{equation}
        m=\#\big([(12)]\cap \mathsf H_N\big)=N,
    \end{equation}
    corresponding to the $N$ transpositions that swap the two elements within any fixed pairing. {Moreover, $c = 2$, as each pair of adjacent pairings is connected by two distinct transpositions, e.g. $(13)$ and $(24)$ both map $(12)(34)$ into $(14)(23)$.} Using Eq.~\eqref{eq:nulambdagen} and the explicit form of characters over transpositions, one obtains for the eigenvalues
\begin{equation} \label{eq:nubeta1}
        \nu_{\beta=1}(\lambda)={\frac{1}{2}\Big(}\frac{d_{[(12)]}\,\chi^{2\lambda}_{[(12)]}}{d_{2\lambda}}-m{\Big)}
        \;=\sum_j\lambda_j^2 - \frac{1}{2}\sum_j (\lambda_j')^2 - \frac N 2 \,,
    \end{equation}
which provides the explicit diagonalization in the orthogonal case. The functions \eqref{eq:spherical} in this case are known as zonal spherical functions~\cite{macdonald1998symmetric}.
One arrives at the expression    \begin{equation} \label{eq:momentsbeta1}
        \big[e^{xA}\big]_{\sigma(\mathbf{r}^{(N)}),\idp}
        =
        \frac{1}{| \Pa_N|}\sum_{\lambda\vdash N} d^{2\lambda}\,
        \omega^{2\lambda}(\sigma(\mathbf{r}^{(N)}))\, e^{x\nu_{\beta=1}(\lambda)}.
    \end{equation}
\end{itemize}


\section{Continuous-time model: Dyson Brownian Motion and Calogero-Sutherland mapping} \label{sec:continuous_model}
We now turn to a complementary description of purification dynamics in continuous time. This approach provides a powerful analytical framework by mapping the purification process onto the dynamics of an integrable many-body system. 
While the discrete-time model highlighted the combinatorial structure of the commutant algebra acting on replicated spaces, the continuous-time limit reveals the connection with Dyson Brownian motion and the Calogero-Sutherland model. 
The formulation is analogous to the one introduced in Refs. \cite{gerbino2024dyson,bulchandani_random-matrix_2023} for the unitary class, but we will extend it in a way that applies to a generic Dyson index $\beta$, thus encompassing the $\beta=2$-unitary case $\GG = \UU(q)$ and the $\beta=1$-orthogonal case $\GG = \OO(q)$~\footnote{Our discussion embraces naturally also the symplectic $\beta=4$ case. However, that does not have a simple interpretation as a many-body monitored dynamics. So we do not analyze it in detail in this work.}.

\subsection{Stochastic evolution and eigenvalue dynamics} \label{subsec:weakmeastoeigen}
We consider a model of monitoring dynamics in continuous time, as before for a macroscopic quantum dot whose Hilbert space has dimension $q$. 
Let us start by briefly discussing the generalization of projective measurements to weak measurements. One way to introduce them is to consider coupling the system to an auxiliary system (ancilla) and performing projective measurements exclusively on the latter. This allows, by means of an experimentally relevant protocol, the effect of the measurement on the system to be tunable and arbitrarily small~\cite{SM}. By choosing the ancilla as a spin $1/2$ and considering a generic observable $\hat O$, labeling with $s=\pm 1$ the measurement outcome on the ancilla, we have the hermitian Kraus operators
\begin{equation}
\label{eq:KaFull}
    \Kr_{s}(\hat O; \kappa, \Delta t)  =  \cos \left(  \kappa \hat O \Delta t + s \frac{\pi}{4} \right) \,,
\end{equation}
satisfying $\sum_{s = \pm 1} K_s(\hat O; \kappa, \Delta t)^2 = \id$. In the Equation above, $\Delta t$ is the time interval during which the ancilla and the system interact, and $\kappa$ is a coupling constant. 
The famous stochastic Schrödinger equation~\cite{caves1987quantum, diosi1998non, gisin1992quantum, cao2019} is obtained by iterating 
this procedure several times, leading to a quantum trajectory (see \eqref{eq:rhoevol}), with the continuous-time limit emerging by taking $\Delta t \to 0$, $\kappa \to \infty$ at fixed rate $\Gamma = \kappa^2 \Delta t$; then, the sequence of measurement outcomes along the trajectory 
\begin{equation}
\label{eq:meastowien}
    \sqrt{\Delta t}\sum_{\td' <\td} s_{\td'} \to W(t = \td \Delta t) \;, \qquad \kappa = \sqrt{\frac{\Gamma}{\Delta t}}
\end{equation} 
converges to a continuous Wiener process $W(t)$, with a measurement rate $\Gamma$. In the absence of further effects, this dynamics leads for $\tc \to \infty$ to the projection onto one of the eigenstates of $\hat O$. In our approach, however, we assume that at \emph{each time-step}, the choice of operator $\hat O$ is also randomized. Setting $\hat O= \sum_{\alpha,\beta=1}^q o_{\alpha\beta} \ket{\alpha}\bra{\beta}$, written in terms of an orthonormal basis of states $\{\ket{\alpha}\}_{\alpha=1}^q$, we consider $o_{\alpha\beta}$ a $q\times q$ random matrix with Gaussian entries, and with given symmetry parameterized by a Dyson index $\beta$. Here, Gaussianity is not crucial as long as rotational invariance is maintained, but we assume it for simplicity of the derivation.
Then, the probability distribution for a matrix $o_{\alpha\beta}$ drawn from a Gaussian Ensemble at given $\beta$, and the corresponding correlations, are 
\begin{equation}
\label{eq:Podef}
    P(o) = \frac{1}{z} e^{-\frac{q \beta}{4}\Tr (o o^\dagger)} \,,\qquad 
    \aveE{o_{\alpha \beta} o_{\gamma\delta}^*} = \frac{1}{q} \left( \delta_{\alpha\gamma}\delta_{\beta\delta} + \frac{2-\beta}{\beta} \delta_{\alpha\delta}\delta_{\beta\gamma} \right) \,.
\end{equation}
Crucially, this distribution is invariant under the symmetry transformation $o \to U^\dag o U$, with $U$ an element of the unitary ($\beta=2$) or orthogonal ($\beta = 1$) group. From the combined choice $a = (s, o)$, we obtain a continuous family of Kraus operators. Formally, with respect to the notation of Sec.~\ref{sec:framework}, one has to interpret the sum over the generalized measurements' outcomes as
\begin{equation}
     \sum_a (\ldots) \longrightarrow \int do \; P(o) \sum_{s = \pm 1} (\dots ) \;.
\end{equation}
We can now consider the limit $\Delta t \to 0$ as in Eq.~\eqref{eq:meastowien}; in this case it is convenient to introduce the matrix-valued
Wiener process $do_{\alpha\beta} = o_{\alpha\beta}(t)\ dW$, satisfying $\aveE{do_{\alpha\beta} do_{\gamma\delta}^*}=\frac{dt}{q}\left( \delta_{\alpha\gamma}\delta_{\beta\delta} + \frac{2-\beta}{\beta} \delta_{\alpha\delta}\delta_{\beta\gamma} \right)$. 
The notation $do_{\alpha\beta} = o_{\alpha\beta}(t)\, dW$ can be regarded as a polar (direction–magnitude) decomposition of an isotropic multidimensional Wiener increment in the continuous-time limit, where only the second moments are retained.
Accordingly, we can define $d\hat O=\sum_{\alpha,\beta=1}^q do_{\alpha\beta}\ket{\alpha}\bra{\beta}$. Then, the evolution equation \eqref{eq:rhoevol} can be rewritten as a stochastic equation in the continuous time limit, which reads
\begin{equation} \label{eq:VNstochastic}
    d \urho = e^{\sqrt{\Gamma} dO - \Gamma dO^2  } \urho \, e^{\sqrt{\Gamma} dO - \Gamma dO^2} - \urho =
    - \Gamma dt \left( \urho - \frac{\Tr \urho}{q} \id\right) + \sqrt{\Gamma} \left\{ dO, \urho \right\}\,,
\end{equation} 
where the Itô convention is assumed, and where $\urho= \urho(t)$ hereafter for simplicity.
This dynamics can be extended by considering in each $\Delta \tc$ also a unitary evolution (deterministic or stochastic), by simply replacing $d\urho \to d\urho {-} i[H, {\urho}] dt$. However, provided that $e^{-i H dt} \in \GG$ (holding for any hermitian Hamiltonian for $\beta = 2$ and any purely imaginary one for $\beta =1$), such an evolution does not modify the spectrum of the density matrix but only acts as a symmetry transformation of its eigenvectors. A simple example of a Hamiltonian term satisfying these constraints is the Pauli matrix $\hat H = \sigma_y$, for $q=2$. 
As we consider the infinite-temperature initial state $\urho(t=0) = \rho(t=0) = q^{-1}\id$ and 
since our ensemble of Kraus operators generated by the measurements is automatically invariant under $\GG$, we can omit a Hamiltonian term without loss of generality~(see also Ref.~\cite{gerbino2024dyson} and~\cite{SM}). Also, being the initial state and the distribution of observables \eqref{eq:Podef} invariant under $\GG$, the distribution of $\urho$ will be itself invariant under $\urho \to U \urho U^{-1}$ for any $U \in \GG$. This point is fundamental, inasmuch as it allows us to decouple the dynamics of the eigenvalues $\{y_\alpha\}_{\alpha=1}^q$ of $\urho$ from the one of the eigenvectors.
Following second-order perturbation theory~\cite{dyson1962brownian}, from \eqref{eq:VNstochastic} we can  write the evolution equation for the eigenvalues:
\begin{equation}
    dy_\alpha = d\urho_{\alpha\alpha} + \sum_{\alpha'\neq \alpha}\frac{|d\urho_{\alpha\alpha'}|^2}{y_\alpha-y_{\alpha'}} \,.
\end{equation}
Then we can insert the infinitesimal variations $d\urho_{\alpha\beta}$ as determined via Eq.~\eqref{eq:VNstochastic}. 
Moreover, we can exploit the rotational invariance in Hilbert space by the elements of $\GG$, so that $\urho(t)$ is assumed to be diagonal in an appropriate orthonormal basis 
i.e., $\urho_{\alpha\beta} = y_\alpha \delta_{\alpha\beta}$, yielding
\begin{equation}
\begin{split}
    & d\urho_{\alpha\alpha} = -\frac{\Gamma dt}{q} \left(q y_\alpha - \sum_{\alpha'} y_{\alpha'} \right) + 2 \sqrt{\Gamma} dO_{\alpha\alpha} y_{\alpha} \,, \quad  d\urho_{\alpha \neq \alpha'} = \sqrt{\Gamma}  dO_{\alpha \alpha'} (y_\alpha + y_{\alpha'} ) \,.
\end{split}
\end{equation}
Finally, by a rescaling of the diagonal elements of $d\hat{O}$ in terms of standard Wiener increments, we write 
$dO_{\alpha\alpha} = dB_\alpha \sqrt{2/(\beta q)}$, with $dB_\alpha dB_{\alpha'} = \delta_{\alpha\alpha'} \,dt$, so that one has 
a multiplicative Dyson Brownian motion for the eigenvalues of $\urho$, namely
\begin{equation}
\label{eq:dystoch}
    dy_\alpha = \frac{4\Gamma}{q} dt \sum_{\alpha'\neq \alpha} \frac{y_\alpha y_{\alpha'}}{y_\alpha -y_{\alpha'}} + \sqrt{\frac{8\Gamma}{\beta q}} dB_\alpha\,y_\alpha \,.
\end{equation}

It is useful to connect this formalism with the moments introduced in \eqref{eq:MaveN}. First, one has $\PP_{\mathbf{r}}[\urho] = \prod_{j} (\sum_{\alpha} y_\alpha^j)^{r_j}$ and thus
\begin{equation}
\label{eq:momfromys}
\average{\PP_{\mathbf{r}}(\tc)}^{(N)} = \frac{\aveE[t]{p_{\mathbf{r}^{(N)}}}}{\aveE[t]{p_{1^N}}} \;, \qquad p_{\mathbf{r}} := \prod_j \bigl(\sum_\alpha y_\alpha^j\bigr)^{r_j}
\end{equation}
where $\aveE[t]{\ldots}$ represents the average over $\{y_\alpha\}$ solutions of Eq.~\eqref{eq:dystoch} at time $t$, equivalent to the sum over the generalized measurement outcomes $\aa$; in the second equality, we introduced the power-sum symmetric polynomials $p_{\mathbf{r}}(y)$ associated to the partition $\mathbf{r}$.

Summing over $\alpha$, one clearly sees that 
$\aveE{\sum_\alpha y_\alpha}$ is kept constant by the dynamics \eqref{eq:dystoch}, ultimately a consequence of the preservation of total probability $\sum_{\aa} \operatorname{Prob}(\aa)$ by each  quantum channel \eqref{eq:quantumchannel}. So in the case $N=1$, we can get rid of the denominator by simply normalizing the initial condition $y_{\alpha}(t = 0) = 1/q$.
From a qualitative point of view, in \eqref{eq:dystoch}, purification results from the combined action of multiplicative noise, which tends to increase eigenvalues exponentially, together with level repulsion, which keeps them well separated from each other.
Indeed, a pure state corresponds to a single maximal eigenvalue being much larger than all the others. 
To handle the multiplicative noise and identify the stationary properties, it is convenient to switch to logarithmic variables
$w_\alpha = \log y_\alpha {+} x \left[ 1/\beta + (q-1)/2 \right]$ 
with a suitable constant drift accounting for center-of-mass motion, and where we introduced the rescaled time $x=4\Gamma t/q$. It will be clear later that $\tp=q/(4\Gamma)$ is the purification time for the weak measurement protocol in continuous time. As the dynamics emerges from a multiplicative random matrix process, the variables $\{w_\alpha\}$ can be interpreted as Lyapunov exponents~\cite{GoldsheidMargulis1989,GuivarcRaugi1989,akemann_universality_2020,Liu2023,mochizuki2025measurement}.
Using Itô's lemma, the evolution maps onto a gradient flow in a potential
\begin{equation} \label{eq:FP_notrace}
    dw_\alpha = - \frac{2}{\beta} \frac{\partial V^0_\beta}{\partial w_{\alpha}} dx + \sqrt{\frac 2 \beta} \ dB_\alpha \,.
\end{equation}
The potential $V^0_\beta$ governing this Dyson Brownian motion is given by
\begin{equation}
    V^0_\beta(\vec w) = - \frac{\beta}{4} \sum_{\alpha \neq \gamma} \log \sinh \left|\frac{w_\alpha - w_\gamma}{2}\right| \,.
\end{equation}
It represents a pairwise logarithmic repulsion typical of random matrix ensembles, but with a hyperbolic deformation ($\sinh$) arising from the multiplicative nature of the noise; with standard methods, one can write down the associated Fokker-Planck equation for the probability distribution of the Lyapunov exponents $P(\vec w, x)$ 
\begin{equation} \label{eq:FP_Ham}
    \beta \ \partial_x P(\vec w, x) = - H^{\rm FP}_\beta P(\vec{w}, x) \,,\qquad
    H_{\beta}^{\rm{FP}} = - \sum_\alpha \partial_\alpha \left(\partial_\alpha + 2\partial_\alpha V_\beta^0 \right) \,.
\end{equation}
For clarity, we stress that the \emph{Hamiltonian} $H$  we introduced here is an effective one that emerges in the description of the joint pdf of the singular values through the Fokker-Planck. So it should not be confused with the quantum dynamics of the original quantum system in \eqref{eq:VNstochastic}.

\subsection{Mapping to the Calogero-Sutherland model}

The Fokker-Planck operator $H_{\beta}^{\rm{FP}}$ is non-Hermitian. However, it can be mapped onto a well-known Hermitian Hamiltonian through a similarity transformation. 
First, one observes that the stationary solution of \eqref{eq:FP_Ham} {is $\Psi_0^2$, where $\Psi_0$} admits the simple exponential form
\begin{equation}
    \Psi_0(\vec{w}) = e^{-V_\beta^0(\vec{w})} = (\tilde \Delta(\vec{w}))^{\beta/2} \;, \qquad \tilde \Delta(\vec{w}) = \prod_{\alpha<\alpha'} \sinh \frac{w_\alpha- w_{\alpha'}}{2} \;,
\end{equation}
where $\tilde \Delta$ plays the role of a generalized Vandermonde determinant.
By performing the transformation $P(\vec{ w}, x) = \Psi_0(\vec{w}) \psi(\vec{w}, x)$, the evolution  takes the form of an  imaginary-time Schrödinger equation
\begin{equation}
\label{eq:FPtoCS}
    -\beta \ \partial_x \psi = \left( \Psi_0^{-1} H_{\beta}^{\rm{FP}} \Psi_0 \right) \psi = \left( H_{\beta}^{\rm CS} + E^0_{q,\beta} \right) \psi \,.
\end{equation}
where the shift with $E^0_{q,\beta} = \beta^2 q(q^2-1)/48$ enforces the zero-energy ground state and, remarkably, the resulting operator is the Hamiltonian of the hyperbolic Calogero-Sutherland (CS) model
\begin{equation} \label{eq:calogero}
    H_{\beta}^{\rm CS} = - \sum_\alpha \partial_\alpha^2 + \frac{\beta}{4}\left(\frac{\beta}{2}-1\right) \sum_{\alpha<\alpha'} \frac{1}{\sinh^{2}\left(\frac{w_\alpha - w_{\alpha'}}{2} \right)} \,.
\end{equation}
The CS Hamiltonian emerges naturally in the study of Dyson Brownian motion for the eigenvalues of random matrices~\cite{potters2020first}. However, while the standard case of additive noise is usually turned into the rational variant of the CS model, with interaction $\sim {(w_\alpha-w_\gamma)^{-2}}$, in our case the noise is multiplicative. This generates the non-rational structure of the interaction. Moreover, the fact that $\rho$ is multiplied by hermitian matrices $O$ in Eq.~\eqref{eq:VNstochastic} ensures that the dynamics of eigenvalues remains constrained on $\mathbb R^+$. This has to be compared to the case of the unitary Brownian motion, with the eigenvalues taking complex values and the exponents $\theta_\alpha = -i \ln y_\alpha \in [0,2\pi)$ representing phases, which generates a trigonometric version of the CS interaction.

This mapping establishes a profound connection: the universal purification dynamics of a monitored quantum system is governed by the same integrable Hamiltonian describing interacting particles in one dimension, and this mapping becomes exact for weak measurement in continuous time. The symmetry class of the monitoring enters simply as the coupling constant $\beta$ in the CS model. Note that for $\beta=2$, the interaction term in Eq.~\eqref{eq:calogero} vanishes, recovering free diffusion for the eigenvalues~\cite{gerbino2024dyson, bulchandani_random-matrix_2023}, which allows us to express the all-$x$ joint distribution of Lyapunov exponents in terms of a determinantal point process.
For $\beta=1$, however, the non-trivial interaction persists, leading to the distinct universality class studied here. This suggests that, as often happens in random-matrix ensembles by the replacement $\beta=2\to\beta=1$, the distribution of the $w_\alpha$'s could be expressed for any $x$ via a Pfaffian point process; however we do not explore this possibility here.

We can now further exploit the integrability of the CS model~\cite{pasquier, Estienne_2012} to derive exact expressions for the finite-time value of the moments \eqref{eq:momfromys}. The eigenvectors of $H_{\beta}^{\rm CS}$ can be put in correspondence with integer partitions, already introduced in Eq.~\eqref{eq:rdef} and in the discussion at the beginning of Sec.~\ref{ref:specialbeta12}. 
Then, the eigenvector  $\Psi_{\lambda,\beta}$ associated to a given partition $\lambda$ is known to be linked to the so-called Jack polynomial $J_\lambda^{(\alpha)}$, indexed by a parameter $\alpha$, through the equation of eigenvalues. 
\begin{equation}
 \Psi_{\lambda, \beta} (w) := \tilde \Delta^{\beta/2}(\vec w) \ J_\lambda^{(2/\beta)}(e^w) \;, \quad
({(H_{\beta}^{\rm CS} + E^0_{q,\beta})} \cdot  \Psi_{\lambda, \beta})(w) = \epsilon^\lambda_{q,\beta} \Psi_{\lambda,\beta}(w) \,,\quad 
    \epsilon^\lambda_{q,\beta} = - \sum_{j=1}^q \lambda_j \left( \lambda_j + \frac{\beta}{2} (q+1-2j)\right) \,,
\end{equation}
where $\epsilon^\lambda_{q,\beta}$ denotes the corresponding eigenvalue. 
We refer to Sec.~2 of the Supplemental Material~\cite{SM} for the definition and a summary of the properties of Jack polynomials. Here, it suffices to note that they form a complete basis for symmetric polynomials. Therefore, it is possible to derive linear relationships with another basis, that of power-sum  symmetric  polynomials
\begin{equation} \label{eq:changeofbasis}
    J_\lambda^{(\alpha)} = \sum_{\mu\vdash N} \theta_\mu^\lambda(\alpha) p_\mu \,,\qquad 
    p_\mu = \sum_{\lambda\vdash N} \gamma_\mu^\lambda(\alpha) J^{(\alpha)}_\lambda \,.
\end{equation}
For clarity, we stress again that we are using a different notation for integer partitions with respect to Eq.~\eqref{eq:momfromys}, where
\begin{equation}
\label{eq:plambdadef}
    p_\lambda(y) = \prod_{j = 1}^\ell \left(\sum_\alpha {y_\alpha^{\lambda_j}}\right) \;.
\end{equation}
The evaluation of the change-of-basis  coefficients $\theta_\mu^\lambda(\alpha)$, $\gamma_\mu^\lambda(\alpha)$ can be done iteratively but as analytical expressions are not always available, it represents a technical challenge~\cite{SM}. 

We recall that our aim is to evaluate averages of the moments $\mathcal P_\mathbf{r}$ expressed in terms of power-sum symmetric polynomials of the eigenvalues $y$'s. Crucially, we have converted the averages over the realization of the Wiener processes $dB_\alpha
$ in terms of the probability density $P(\vec w,x)$, which we now know to be evolved by $H^{\rm CS}_\beta$. Exploiting the hermiticity of $H^{\rm CS}_\beta$, we can
switch the time evolution $e^{-H^{\rm CS}_\beta \ x/\beta}$ from the probability distribution to the power-sum symmetric polynomials 
\begin{equation}
\aveE[t]{p_{\mu}} := \int_{[0,\infty)^q} d^q y \ p_{\mu}(y) \ P(y,t) = e^{-\left[ N \left( \frac{\beta}{2}(q-1) + 1 \right) + E^0_{q,\beta} 
    \right] \frac{x}{\beta}} \int_{\mathbb R^q} d^q w\, \tilde \Delta^{-\beta/2} P(\vec{w}, 0) \left( e^{-H^{\rm CS}_\beta \ \frac{x}{\beta}} \right) \left(  p_{\mu}(e^{w})  \,  \tilde \Delta^{\beta/2} \right) \,,
\end{equation}
where $t =   q x/(4 \Gamma)$, we chose $\mu \vdash N$ corresponding to $\mathbf{r}^{(N)}$ (see Eq.~\eqref{eq:partitionYoung}) 
and 
we restored explicitly the shift of the center-of-mass thanks to the homogeneity $p_{\mu}(k y) = k^{|\mu|} p_{\mu}(y) = k^{N} p_{\mu}(y)$. We also used the transformation 
\eqref{eq:FPtoCS} and 
set the initial condition for the probability distribution of the $w$'s variables $P(\vec w, 0) = \prod_\alpha \delta(w_\alpha +\log q)$.
Crucially, we can use Eq.~\eqref{eq:changeofbasis} to expand $p_\mu$ on the basis of the Jack's polynomials $J^{(2/\beta)}_\lambda$, thus obtaining a spectral decomposition for the time-evolution operator
\begin{equation} \label{eq:pmu}
    \aveE[t]{p_{\mu}}
    =  
    \sum_{\lambda\vdash N} \gamma_\mu^\lambda(2/\beta) \, e^{-\left[ N \left( \frac{\beta}{2}(q-1) + 1 \right) + \epsilon_{q,\beta}^\lambda
    \right] \frac{x}{\beta}} \, J^{(2/\beta)} _\lambda(q^{-1}) \,.
\end{equation}
In this last expression, we
performed the integral over the variables $w$'s using the initial condition, which leads to coinciding points inside the Jack's polynomials. 

We observe that for this continuous-time evolution model, thanks to the integrability of the evolution of the eigenvalues of the density matrix, we have obtained an explicit and compact expression for the moments that holds for finite time $t$ and finite matrix size $q$. In the next section, we will discuss how this expression can be further simplified in the scaling limit where $q$ and $t$ diverge simultaneously.

\subsection{Scaling limit} \label{subsec:SL}

We now analyze the behavior of Eq.~\eqref{eq:pmu} in the large $q$ limit, showing how the universal form for the moments emerges in the scaling limit. First of all,  we recall that $J_\lambda^{(2/\beta)}(q^{-1})$ denotes the Jack's polynomial involving $q$ variables all evaluated at the same coinciding value $q^{-1}$. At large $q$, its value takes a particularly simple asymptotic form. To see this, we conveniently move to the basis of power-sum symmetric polynomials via the basis change matrix $\theta_\mu^\lambda$ and use that, from the definition \eqref{eq:plambdadef}, $p_{\mu}(q^{-1}) = q^{\ell(\mu) - N}$. 
Thus, in the limit of large $q$, the sum over $\mu$ is dominated by $\mu = (1,1,\ldots, 1) \equiv (1^N)$, which maximizes $\ell(\mu \equiv 1^N) = N$, so that explicitly
\begin{equation}
    J_\lambda^{(2/\beta)}(q^{-1})  = \frac{1}{q^N}\sum_\mu \theta^{\lambda}_{\mu}(2/\beta) \,p_\mu(1) = \sum_\mu \theta^\lambda_\mu(2/\beta) \,q^{\ell(\mu)-N} = \theta^\lambda_{(1^N)}(2/\beta) + O(q^{-1}) = 1+O(q^{-1}) \,,
\end{equation}
where we used the identity $\theta^\lambda_{(1^N)}(\alpha)=1$ independently of $\alpha$ and $\lambda$ (see the normalization chosen in Sec.~2 of the Supplemental Material~\cite{SM}, see also standard references \cite{macdonald1998symmetric, stanley1989some}).
Additionally, we can simplify the exponent in \eqref{eq:pmu} defining 
\begin{equation} \label{eq:eigen_new}
    \nu_\beta(\lambda) = \frac{1}{\beta} \left[ 
    - N \left( \frac{\beta}{2}(q-1) + 1 \right) - \epsilon_{q,\beta}^\lambda \right]= \sum_j \left( \frac{1}{\beta} \lambda_j^2 - \frac 1 2 (\lambda'_j)^2 \right) + \left(\frac 1 2 - \frac 1 \beta \right) N \,,
\end{equation}
where the last equality is based on $\sum_j(\lambda_j')^2=2\sum_j j\lambda_j - N$ (compare with Eq.~(74) of \cite{gerbino2024dyson}). 
Notice that the values $\nu_{\beta=1,2}(\lambda)$ correspond exactly to the eigenvalues of the adjacency matrix of transpositions $A^{(\beta=1,2)}$ as given in Eqs.~\eqref{eq:nu2} and \eqref{eq:nubeta1}.

Finally, we consider the scaling limit. In this case, the appropriate scaling $t,q \to \infty$, with the scaling variable $x = 4\Gamma t/q$ finite.
Using the definition \eqref{eq:scalinglim} and Eq.~\eqref{eq:momfromys} we obtain 
\begin{equation} \label{eq:pmu_SL}
\lim_{q\to \infty} \aveE[t]{p_\mu}  =  \sum_{\lambda\vdash N} \gamma_\mu^\lambda(2/\beta) \, e^{x \,  \nu_\beta(\lambda)} \,.
\end{equation}
This equation provides a nice generalization of Eqs.~(\ref{eq:momentsbeta2}, \ref{eq:momentsbeta1}) for arbitrary values of $\beta$, provided $N\in \mathbb N$. 
Indeed, for the case $\beta=2$, the Jack polynomials $J_\lambda^{(1)}$ are proportional to the well-known Schur polynomials $s_\lambda$ (see Refs.~\cite{macdonald1998symmetric,stanley1989some,SM}); consequently, the change-of-basis coefficients $\gamma_\mu^\lambda(2/\beta)$ and $\theta^{\lambda}_\mu(2/\beta)$ are simply related to the characters of the irreducible representations of $\SS_N$~\cite{SM} and Eq.~\eqref{eq:pmu_SL} reduces to Eq.~\eqref{eq:momentsbeta2}, obtained through a purely algebraic construction based on the Gelfand pair $(\SS_N \times \SS_N, \SS_N)$. In the case $\beta = 1$, the Jack polynomials coincide with the so-called Zonal polynomials, $Z_\lambda := J^{(2)}_\lambda$, and the coefficients for the change of basis \eqref{eq:changeofbasis} are in this case algebraically more complex. Nonetheless, one verifies that the coefficients {$\gamma_\mu^\lambda (2)$} coincide{, up to the factor $d^{2\lambda}/|\Pa_N|$,}
with the zonal spherical functions $\omega^{\Lambda=2\lambda}(\sigma(\mathbf{r}^{(N)}))$, thus recovering the expression in~\eqref{eq:momentsbeta1}. Furthermore, the choice $\tp = q/(4\Gamma)$ as the purification time for the weak measurement protocol in continuous time is now clear~\cite{SM}.

The fact that these two approaches based on discrete Gaussian models and on a continuous model of Dyson Brownian motion, through mapping onto the Calogero-Sutherland model, provide identical results in the scaling limit is a direct manifestation of the universality we discuss in this paper. We stress that this universal description can be accessed via the formulation at integer number $N\geq 2$ of replicas, and that, while Eq.~\eqref{eq:pmu_SL} is well-defined for arbitrary $\beta$ at integer $N$, there is no direct interpolation for real $\beta$ once the replica limit $N \to 0,1$ is taken. 

\section{Small-$x$ expansion of Entropies in the replica limit $N \to 0,1$ for $\beta = 1$} \label{sec:small_x}

In this section, we analyze in detail the calculation of scaling functions as a perturbative series at small $x=t/\tp$. We base our discussion on the combinatorial analysis of paths in the graph characterized by the adjacency matrix $A$. We show that it is possible to compute systematically the first few orders in the expansion in powers of $x$.  
We focus in particular on the case $\beta=1$, since the case $\beta=2$ has already been considered in \cite{deluca2025universality, gerbino2024dyson}. However, some aspects become clearer by showing the similarities and differences between the two cases. Moreover, while Eq.~\eqref{eq:pmu_SL} is valid for arbitrary real $\beta$ at integer $N$, obtaining the explicit expressions for the entropies requires an exact resummation at integer $N$ and the replica limit $N\to 0,1$. This is not possible for arbitrary $\beta$. Therefore, for generic real $\beta$, our result~Eq.~\eqref{eq:pmu_SL} should be understood as an exact interpolation of the integer-$N$ moments, rather than as a fully controlled interpolation of the entropies between $\beta=1$ and $\beta=2$.

We specialize the calculation of moments to the interesting case of Rényi entropies $\mathcal S_n^{(N)}(x)$, both for their physical relevance and because, beyond the analytic continuation in $N$ for the limits \eqref{eq:Mave}, they also require one in the number of cycles, i.e., $k$ in Eq.~\eqref{eq:scalinglimRenyi}.
Thus, using Eq.~\eqref{eq:renyidef}, we need to consider $\mathbf{r} = n^k$ and so $\mathbf r^{(N)} = \{r_1=N-nk,r_n=k\}$, which consists of $k$ cycles of length $n$ and $N-nk$ fixed points. 
Expanding Eq.~\eqref{eq:momentsbeta1}
perturbatively at small $x$, we have
\begin{equation} \label{eq:expansion}
\left[e^{x A}\right]_{\sigma(r_n=k,r_1=N-nk),\idp}  = \sum_{\ell=0}^\infty \frac{x^\ell}{\ell!}   \left[ A^\ell\right]_{\sigma(r_n=k,r_1=N-nk),\idp}  \,.
\end{equation}
Therefore, our goal reduces to computing the number of paths of length $\ell$ along the graph with adjacency matrix $A$, connecting
the identity  with $\sigma=\sigma(r_n=k,r_1=N-nk)$, a representative of the conjugacy class $\mathcal C_{\mathbf r^{(N)}}$. Let us start by summarizing the case $\beta=2$. In this case, a path is equivalent to decomposing the permutation in terms of a product of transpositions
\begin{equation}
    \sigma(r_n=k,r_1=N-nk) = \tau_1 \tau_2 \ldots \tau_\ell \;, \qquad (\beta = 2)
\end{equation}
It is well-known that the minimal number of transpositions to decompose an $n$-cycle is $n-1$. Also, there are precisely $n^{n-2}$ distinct decompositions of such a cycle~\cite{stanley_enumerative_1999}. Since the final state consists of $k$ of these $n$-cycles, it is clear that for $\ell<(n-1)k$, there are no possible paths. For $\ell=(n-1)k$, however, considering one of these $n^{n-2}$ decompositions for each of the $k$ $n$--cycles, there are $[(n-1)k]!/[(n-1)!^k]$ non-equivalent ways to order them. 
We therefore obtain
\begin{equation}
\label{eq:beta2expansion}
    \left[e^{x A^{(\beta = 2)}}\right]_{\sigma(r_n=k,r_1=N-nk),\idp}  = \left(\frac{n^{n-2}}{(n-1)!}\right)^k x^{(n-1) k} \left(1 +  \frac{a_{N, n,k, 1} }{(n +1)} x^2 + 
    O(x^4)\right) \;.
\end{equation}
The next order corresponds to decompositions longer than the minimal ones and, due to parity conservation, this necessarily requires an even number of additional transpositions. The coefficient $a_{N, n,k, 1}$ can be determined by considering separately the cases in which the two extra transpositions i) remain within one of the $k$ $n$-cycles; ii) combine $2$ of the $k$ $n$--cycles; iii) combine one of the $n$--cycles with one of the $N-nk$ fixed points;
iv) combine two of the $N-nk$ fixed points. We refer directly to Ref.~\cite{deluca2025universality} where the calculation for $n=2$ is carried out in great {detail} (see Eq. 76 and B7 therein) which we report for convenience
\begin{equation} \label{eq:aUnitary}
    \frac{a_{N, n,k, 1}^{(\beta = 2)}}{n+1} = \frac{1}{2} \left[\binom{N-n k}{2} + \frac{2n}{n+1} n k (N-n k) + 
    n^3 \binom{k}{2} + \frac{n^2(n-1)}{12} k \right]\,,
\end{equation}
The crucial aspect is the resulting purely polynomial dependence on the parameters $k$ and $N$, which therefore allows for formal limits and derivatives. 

We now move on to the case $\beta=1$. Here, the non-trivial structure of the coset \eqref{eq:commAbeta1} requires a more complex analysis. Consider $ \sigma(\mathbf{r}^{(N)}) \in \SS_N$ as an element in $\SS_{2N}$ that only permutes the starred indices in Eq.~\eqref{eq:inddef}. To shorten the notation, we indicate it simply as $\sigma\equiv\sigma(\mathbf{r}^{(N)})$. Then, we can see the elements $\idp$ and $\sigma(\mathbf{r}^{(N)})$ as elements of $\mathcal{C}_{2^N} \simeq \Pa_{N}$ (the conjugacy class of the two cycles) in $\SS_{2N}$. Explicitly
\begin{equation}
    \idp \to p_{\idp}=\{(1,1^*),\ldots, (N,N^*)\} \;, \qquad \sigma \to p_\sigma \equiv \sigma p_\idp \sigma^{-1} = \{(1,\sigma(1^*))\dots (N,\sigma(N^*))\} \,.
\end{equation}
The adjacency matrix $A^{(\beta=1)}$ acts on the {Schreier} graph (the flip graph) of $\mathcal C_{2^N}=\Pa_N \subset \SS_{2N}$ where two distinct pairings $p, p'$ are connected if there exists a transposition $\tau \in \SS_{2N}$ such that $p' = \tau p \tau$.
Equivalently, two pairings on the flip graph are adjacent if one is obtained from the other by choosing two pairs $(a_i,a_j)$, $(a_m,a_n)$ with $1\leq i,j,m,n \leq 2N$ and replacing them by one of the two alternative pairings on the same four vertices:
\begin{equation}
(a_1,a_2)(a_3,a_4) \quad \longrightarrow \quad
(a_1,a_3)(a_2,a_4)\quad\text{or}\quad (a_1,a_4)(a_2,a_3) \,.
\end{equation}
By this property, for every pairing $p$ we have ${\rm d} = N(N-1)$ possible choices. In other words, the graph of $\Pa_N$ is regular of degree $\rm d$, meaning that each vertex has exactly $\rm d$ neighbors.
Accordingly, the matrix element $[(A^{(\beta=1)})^\ell]_{\sigma,\idp}$ counts the number of sequences of $\ell$  transpositions {(each defined up to the twofold redundancy $\tau p \tau = \tau' p \tau'$ discussed in Sec.~\ref{sec:general_G_invariant})} such that
\begin{equation} \label{eq:pRule}
    p_\sigma = \sigma p_\idp \sigma^{-1} = \tau_1 \tau_2\ldots \tau_\ell p_{\idp}
    \tau_\ell \ldots \tau_2 \tau_1 \,.
\end{equation}
Note, however, that this equation does not necessarily imply that $\tau_1 \ldots \tau_\ell = \sigma$. In fact, due to the structure in Eq.~\eqref{eq:commAbeta1}, we can rather write that
\begin{equation}
\label{eq:decomptaubeta1}
    \tau_1\ldots \tau_{\ell} h = \sigma
\end{equation}
where $h\in \mathsf{H}_N = \SS_2^N \rtimes \SS_N$. Therefore, in principle, elements of the stabilizer $\mathsf{H}_N$ can be used to complete the sequence of transpositions. However, since $\sigma$ is a permutation that involves only starred elements, while the elements of the stabilizer necessarily swap either pairs with each other or elements within a pair, it is easy to see that at the leading order, the stabilizer does not allow any shortcuts. Consequently, the count of possible decompositions of $\sigma$ in terms of transpositions becomes equivalent to what was done in Eq.~\eqref{eq:beta2expansion} for the case $\beta=2$. To account for the higher orders in $x$, it is convenient to organize the expansion again as
\begin{equation}
\label{eq:beta1expansion}
    \left[e^{x A^{(\beta = 1)}}\right]_{\sigma(r_n=k,r_1=N-nk),\idp}  = \left(\frac{n^{n-2}}{(n-1)!}\right)^k x^{(n-1) k} \left(1 +  \frac{a_{N, n,k, 1}^{(\beta = 1)}}{(n +1)}x +
    O(x^2)\right) \;.
\end{equation}
Note that in this case there is no parity constraint for non-minimal paths, as the sign can always be incorporated into the stabilizer $h$ in Eq.~\eqref{eq:decomptaubeta1}. Therefore, the first correction is linear in $x$.
The coefficient $a_{N, n,k, 1}$ can be simplified by noting that with a single extra transposition, $k-1$ cycles must necessarily follow a decomposition of minimal length {$n-1$} and only one a quasi-minimal decomposition of length {$n$}. Explicitly, 
\begin{equation} \label{eq:quasiminimal}
     \left[ \left(A^{(\beta=1)}\right)^{k(n-1)+1}\right]_{\sigma,\idp} = k (n^{n-2})^{k-1}
     \frac{[k(n-1)+1]!}{(n-1)!^{k-1} \ n!} \ m_n^{(\beta=1,n)}  \,, \quad
     m_n^{(\beta=1,\ell)}  := [(A^{(\beta = 1)})^{\ell}]_{n^1, 1^n} \,,
\end{equation}
where the factor $k$ accounts for the possible ways of choosing the cycle out of the $k$ available, $(n^{n-2})^{k-1}$ the number of minimal decompositions for the remaining $k-1$ cycles, and finally the last combinatorial factor accounts for the possible ways to order the decompositions of the $k$ cycles. We denote by $m_n^{(\beta=1,\ell)}$ the matrix element counting the number of decompositions of length $\ell$ of a single $n$-cycle in $\Pa_n$. Accordingly, $m_n^{(\beta=1,n-1)}=n^{n-2}$, and in the specific case of quasi-minimal decompositions we set $\ell=n$.
To explicitly calculate this coefficient, one can exploit a one-to-one correspondence between the decompositions of an $n$-cycle into $n-1$ transpositions and the spanning trees on $n$ elements, as is done for minimal decompositions.
For quasi-minimal paths, instead, we turn to a resummation of the spectral representation Eq.~\eqref{eq:momentsbeta1}, which is shown in Sec.~3 of the Supplemental Material~\cite{SM}, leading to a closed expression in terms of the incomplete $\Gamma$ function:
\begin{equation} \label{eq:coeff1}
     m_n^{(\beta=1,n)}  = \frac{n(n-1)}{2} \ e^n \ \Gamma (n-1,n) \quad\Rightarrow\quad
     \frac{a_{N,n,k,1}}{n+1} = \frac{k}{2}\frac{n-1}{n^{n - 2}} \ e^n \  \Gamma(n - 1, n) \,.
\end{equation}
For higher orders, explicit combinatorial calculation quickly becomes complex. However, from these arguments, the polynomial dependence on $N$ and $k$ can be easily verified. For instance, for the second-order term (with the superscript $(\beta=1)$ implied), one can prove the expansion
\begin{equation} \label{eq:coeff2}
    \frac{a_{N,n,k,2}}{(n+1)^2} = \binom{N-nk}{2} + \frac{2n}{n+1} nk (N -nk) + \frac{1}{n^{2(n - 2)}} \left[\frac{2 n}{2^{2n-1}} m_{2n}^{(2n-1)} + \frac{1}{n^2} \left(m_n^{(n)}\right)^2 \right] \binom{k}{2} + \frac{m_n^{(n+1)}}{n(n+1) n^{n-2}} \ k \,,
\end{equation}
where the dependence on $N$ comes from coupling the $N-nk$ fixed points among themselves, or to the $n$-cycles via non-minimal decompositions. The remaining terms can be understood as follows. For the part $\propto \binom{k}{2}$, we can: $(i)$ join two of the $n$-cycles into a new $2n$-cycle, and then use the remaining $2n-1$ steps to decompose it minimally, from which the multiplicity coefficient $m_{2n}^{(\beta=1,2n-1)}=(2n)^{2n-2}$; $(ii)$
choose quasi-minimal decompositions for two chosen $n$-cycles, each with multiplicity given by the matrix element $m_n^{(n)}$. The remaining factors in $n$ come from factorizing the multinomial coefficient and the multiplicity $n^{k(n-2)}$ in front, as in Eq.~\eqref{eq:beta1expansion}. For the term $\propto k$, we consider 2-quasi-minimal decompositions of a chosen $n$-cycle, which are two steps longer than the minimal one. The corresponding matrix element $m_n^{(\beta=1,n+1)}$ can be resummed exactly~\cite{SM}:
\begin{equation} \label{eq:2quasiminimal}
    m_n^{(\beta=1,n+1)} = n^{n-2} \frac{n^3 (n^2 - 1)}{3} \left(\frac{7}{8} - \frac{e^n \ \Gamma(n - 1, n)}{n^{n - 1}} \right) \,.
\end{equation}
Again, the remaining $n$-dependence is a consequence of convenient factorizations.
This justifies
the form, as in Ref.~\cite{deluca2025universality} 
\begin{equation}
\label{eq:acoeffpol}
    a_{N,n,k,\ell} = \sum_{i,j=0}^\ell c_{ij}(n,\ell) \ N^i k^j \,.
\end{equation}
In order to evaluate the $(\ell+1)^2$ coefficients $c_{ij}(n,\ell)$ by polynomial interpolation, we need to know $a_{N,n,k,\ell}$ for $k =0,...,\ell$, $N=n\ell,...,(n+1)\ell$. 
This allows us to have access, at small values of $x$, to an analytic continuation of $a_{N,n,k,\ell}$ in the variables $k$ and $N$, so that the replica limit $N \to 0,1$ can be taken explicitly term by term.

In practice, to compute the coefficients $a_{N,n,k,\ell}$ for several values of $N$ and $k$ we find it convenient to employ the expressions in terms of Jack polynomials $J^{(2)}_\lambda$ by Gram-Schmidt orthogonalization~\cite{SM}, from which the values of the irrep dimensions $d^{2\lambda}$, the spherical functions $\omega^{2\lambda}$ and the eigenvalues $\nu_{\beta=1}(\lambda)$ can be derived algorithmically. As a benchmark, we verified that Eq.~\eqref{eq:acoeffpol} is consistent with all accessible values of $N$ (up to $N = 19$) and $nk \leq N$, beyond the values explicitly needed for the polynomial interpolation. 
For the interesting case $n=2$, we found closed expressions up to $\ell=6$, in particular we report
\begin{equation}
    a_{N,2,k,1} = \frac{3k}{2} \,,\quad 
    a_{N,2,k,2} = \frac 9 2 k + \frac{297}{4} \binom{k}{2} + 24 k (N-2k) + 9 \binom{N-2k}{2} \,.
\end{equation}

Finally, we have access to R{\'e}nyi entropies via differentiation of the moments, as explained in Eq.~\eqref{eq:scalinglimRenyi}, namely: 
\begin{equation} \label{eq:entropyDef}
    \mathcal{S}_n^{(N)}(x) = \frac{1}{1-n} \left. \partial_k \PPave_{r_n=k}^{(N)}(x;1) \right|_{k=0} =
    \frac{1}{1-n} \frac{\partial_k \left[e^{x A^{(\beta=1)}}\right]_{\sigma,\idp} }{\Omega_N^{(\beta=1)}(x)} \Bigg|_{k=0} \,, \quad 
    \Omega_N^{(\beta=1)}(x) := \left[e^{x A^{(\beta=1)}}\right]_{\idp,\idp}
    \,. 
\end{equation}
Notice that the denominator $\Omega_N^{(\beta=1)}(x)$ provides an $N$-dependent normalization. The evaluation of this term is technically challenging~\cite{SM}. Nonetheless, it is independent of $n$ and $k$, and it trivializes to $1$ in the replica limits $N\to0$ and $N\to1$. Therefore, we leave it unexpressed in the formulas for generic $N$ and can then simply delete it when these limits are taken into account.
Moreover, an interesting consequence of the definition of $\Omega_N^{(\beta)}(x)$ Eq.~\eqref{eq:entropyDef} is that the Born-rule purity $\Pi^{(\beta)}(x) := \average{e^{-S_2(x)}}_{\rm BR}$ in the scaling limit can be written directly in terms of its replica derivative, namely
\begin{equation} \label{eq:trick}
    \Pi^{(\beta)}(x) = \beta\ \partial_x \left[ \lim_{N\to 1} \partial_N \Omega_N^{(\beta)}(x)\right] \,.
\end{equation}
The rationale behind this expression is that, differentiating the spectral representation Eq.~\eqref{eq:pmu_SL} for $\Omega_N^{(\beta)}(x)$ in $x$, the exponent $\nu_\beta(\lambda)$ descends from the exponential for each term, yielding the correct coefficient $\gamma_{(2^11^{N-2})}^\lambda(2/\beta)$ up to a $\beta$- and $N$-dependent normalization. This term simply gives one in the replica limit $N\to1$. We refer to the Supplemental Material for further details~\cite{SM}.

From the expression above and from the knowledge of the coefficients $a_{N,n,k,1}$ \eqref{eq:coeff1} and $a_{N,n,k,2}$ \eqref{eq:coeff2}, one can compute the first- and second-order terms of R{\'e}nyi entropies of arbitrary $n$:
\begin{equation}
\label{eq:Srenyigen}
\begin{split}
    \mathcal{S}_n^{(N)}(x)  =  & -\frac{1}{\left[e^{x A^{(\beta=1)}}\right]_{\idp,\idp}} \Bigg[ \left( 1+\binom{N}{2} x^2 \right)\left( \ln x + \frac{1}{n-1} \ln \frac{n^{n-2}}{(n-1)!} \right)+ {\frac{C_n}{2 n^{n-2}}} \ x +  \\
 & {-} \left( \frac{n}{2}\frac{N-1}{n-1} - \frac{nN}{2}\frac{3n-1}{n^2-1}
 - \frac{n^2}{3}
\left(
\frac{7}{8}
-\frac{C_n}{n^{n-1}}
\right)
+\frac{n^3}{2(n-1)}  
+\frac{n-1}{2} \left(\frac{C_n}{2{n^{n-2}}}  \right)^2
 \right) x^2 \Bigg]  + O(x^3) \,, 
\end{split}
\end{equation}
with $C_n = e^n \Gamma(n-1,n)$,
from which we obtain the von Neumann entropy and the second Rényi entropy explicitly, respectively
\begin{subequations} \label{eq:S_theo}
\begin{equation} \label{eq:S1_theoN}
\begin{split}
\mathcal{S}_{\rm VN}^{(N)}(x) = -\frac{1}{\left[e^{x A^{(\beta=1)}}\right]_{\idp,\idp}} \Bigg[ & \gamma_{\rm E}-1  + \left(1+\binom{N}{2}x^2\right) \ln x +
\frac e 2 \Gamma(0,1) \ x  + \\
& {-} \left( \frac{29}{24} - \frac{N+1}{2} + \binom{N}{2} (1-\gamma_{\rm E}) + \frac{C_1}{3} \right) x^2 \Bigg] + O(x^3) \,, 
\end{split}
\end{equation}
\begin{equation} \label{eq:S2_theo}
    \mathcal{S}_{2}^{(N)}(x)  = -\frac{1}{\left[e^{x A^{(\beta=1)}}\right]_{\idp,\idp}} \left[\left(1+\binom{N}{2}x^2\right)\ln x +
\frac x 2 + \left( \frac{2N}{3} - \frac{21}{8}\right) x^2 \right] + O(x^3) \,,
\end{equation}
\end{subequations}
with $\gamma_{\rm{ E}}$ the Euler-Mascheroni constant and $C_1 = e\ \Gamma(0,1)$. It is evident that the quantitative difference between Born's rule ($N\to1$)/Forced Measurements($N\to0$) in Eq.~\eqref{eq:Mave} 
appears at the order $O(x^2)$, as the $O(x)$ term \eqref{eq:coeff1} is independent of $N$. 
The values of R{\'e}nyi entropies can be read from the square brackets in the Eqs.~(\ref{eq:Srenyigen}, \ref{eq:S_theo}), i.e., replacing $\left[e^{x A^{(\beta=1)}}\right]_{\idp,\idp} \to 1$. The case of the second R{\'e}nyi entropy (and in principle that of all integer low-$n$ entropies with a fixed definite $n$) can be explored further. We find the expressions 
\begin{subequations}
\label{eq:S2_theo_N}
\begin{equation} \label{eq:S2_theo_N1}
    \mathcal{S}_2^{(1)}(x) = -\ln x -\frac x 2  + \frac{47}{24} x^2 + \frac{83}{30}x^3 - \frac{5911}{320}x^4 - \frac{210541}{2520}x^5 + \frac{142480817}{302400}x^6 + O(x^7) \,,
\end{equation}
\begin{equation}\label{eq:S2_theo_N0}
    \mathcal{S}_2^{(0)}(x) = -\ln x -\frac x 2  + \frac{21}{8} x^2 + \frac{18}{5}x^3 - \frac{24149}{960}x^4 - \frac{2206}{21}x^5 + \frac{26532911}{43200}x^6 + O(x^7) \,,
\end{equation}
\end{subequations} 
Importantly, in the expressions Eqs.~(\ref{eq:S_theo}, \ref{eq:S2_theo_N}), the linear term $O(x)$ is a signature of the considered symmetry class $\beta=1$, as compared to the unitary case $\beta=2$ of Ref.~\cite{deluca2025universality}, where the first nontrivial order is $O(x^2)$, namely 
\begin{subequations}
\begin{equation}
    \mathcal{S}_2^{(1)}(x) = -\ln x + x^2 - \frac{949}{180}x^4 + \frac{1900303}{22680}x^6 + O(x^8) \quad  (\beta=2) \,,
\end{equation}
\begin{equation}
    \mathcal{S}_2^{(0)}(x) = -\ln x + \frac{4}{3} x^2 - \frac{637}{90}x^4 + \frac{301328}{2835}x^6 + O(x^8) \quad (\beta=2) \,,
\end{equation}
\end{subequations}
In the $\beta=1$ case, purification is slightly more effective, as confirmed by the linear negative term: heuristically, this is explained by the fact that the space of real matrices is smaller than that of complex matrices. 

We stress that, in contrast to the $\beta=2$ case~\cite{deluca2025universality}, 
the dependence of the coefficients $a_{N,n,k,\ell}$ on the parameter $n$ is not polynomial for $\beta = 1$, as evidenced already at the first order $\ell = 1$ by the expression \eqref{eq:coeff1}. This is a clear manifestation of the more complex combinatorial structure. Therefore, there is no simple algorithmic procedure for extracting the dependence on $n$ of the higher orders. This generally prevents us from having expressions for Rényi entropies for arbitrary $n$ and, in particular, for the von Neumann entropy, which is linked to the formal limit $n\to 1$ in Eq.~\eqref{eq:scalinglimRenyi}. It is worth mentioning that the von Neumann entropy $\mathcal S_{\rm VN}^{(N)}$, in the Born's rule case, can also be treated within a different approach, where, as explained at the end of Subsec.~\ref{subsec:complexginibre}, the Rényi index $n$ is identified with the replica index $N$, $n = N \to 1$, into a single limit. The starting point for this analysis is the equivalent expression
\begin{equation} \label{eq:S1_equiv}
    \mathcal S_{\rm VN}^{(1)}(x) = - \lim_{N\to 1}\partial_N \left(M_N^{(\beta=1)}(x) - \Omega_N^{(\beta=1)}(x)\right) \,,
\end{equation}
with $\Omega_N^{(\beta=1)}(x)$ as written in Eq.~\eqref{eq:entropyDef}, and the definition of the moment
\begin{equation}
    M_N^{(\beta=1)}(x) := \left[e^{x A^{(\beta=1)}}\right]_{\sigma(r_N=1),\idp} \,.
\end{equation}
Notice that the matrix elements $m_n^{(\beta=1,\ell)}$ introduced in Eq.~\eqref{eq:quasiminimal} simply correspond to the $\ell$-th derivatives at $x=0$ of $M_N^{(\beta=1)}(x)$.
This approach has been used successfully in \cite{deluca2025universality} for the case $\beta=2$, providing a semi-analytical method for calculating the von Neumann entropy scaling function in a non-perturbative manner in $x$. In the case $\beta=1$, technical complications arise, and we report 
the expressions for the two moments up to order $O(x^4)$ in Sec.~3 of the Supplemental Material~\cite{SM}. Those {can} be used to compute the small-$x$ expansion of $\mathcal S_{\rm VN}^{(1)}(x)$ to the same order
\begin{equation} \label{eq:S1_theo_N1}
    \mathcal S_{\rm VN}^{(1)}(x) = 1 - \gamma_{\rm E} - \ln  x
   -\frac{C_1}{2} \ x
   + \left( \frac{5}{24}+  \frac{C_1}{3} \right)  x^{2}
   + \left( \frac{23}{90} - \frac{103}{720} C_1 \right) x^3
   + \left(-\frac{57397}{181440} + \frac{71}{1620} C_1 \right) x^4 + O(x^5) \,.
\end{equation}
We can hope to push \eqref{eq:S1_theo_N1} to high order and 
\emph{resum} the expansion. At $\beta=2$, the non-perturbative evaluation of Eq.~\eqref{eq:S1_equiv} rests on two special structures: $M_N^{(\beta=2)}$ is elementary, and the generating function
of $\Omega_N^{(\beta=2)}$ over $N$ collapses into a determinant~\cite{deluca2025universality}.  At $\beta=1$ the natural analogue would be to consider a Pfaffian representation. While it can be written for integer $N$, its replica continuation proves technically arduous, and we do not pursue it here. A closed resummation, however, is not required to obtain qualitative results. It is indeed possible to resum the replica derivative terms $\partial_N M^{(\beta=1)}_N(x)\Big|_{N=1}$ and $\partial_N \Omega^{(\beta=1)}_N(x)\Big|_{N=1}$ approximately, using a procedure based on a Borel-resummation of the Pad\'e approximants of each term. This procedure, though not exact, is able to capture the behavior of the von Neumann entropy, as well as of the purity, at large values of $x$, as we show in Sec.~\ref{sec:numerics} and we discuss in the Supplemental Material~\cite{SM}.

Finally, for comparison with the small-$x$ expansion Eq.~\eqref{eq:S1_theo_N1}, we report the expression of the von Neumann entropy in the case of FM average $N\to 0$ obtained from the second-order expression
Eq.~\eqref{eq:Srenyigen}:
\begin{equation} \label{eq:S1_theo_N0}
    \mathcal S_{\rm VN}^{(0)}(x) = 1 - \gamma_{\rm E} - \ln  x
   -\frac{C_1}{2} \ x
   + \left( \frac{17}{24}+  \frac{C_1}{3} \right)  x^{2}+  O(x^3) \,.
\end{equation}

\begin{figure}
\centering
\subfloat{
\includegraphics[width=0.5\linewidth]{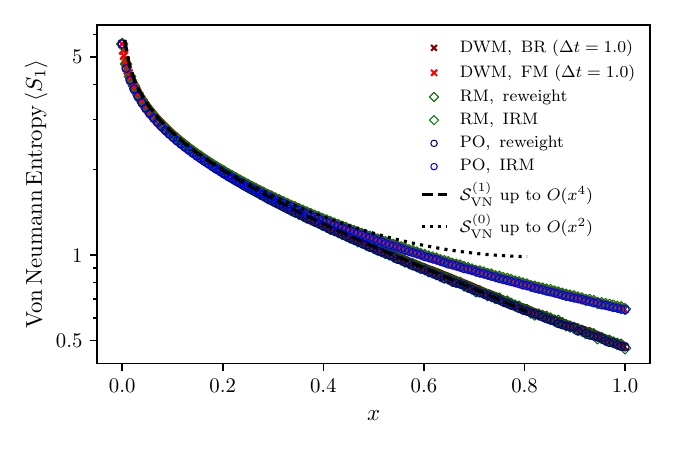}}
\subfloat{
\includegraphics[width=0.5\linewidth]{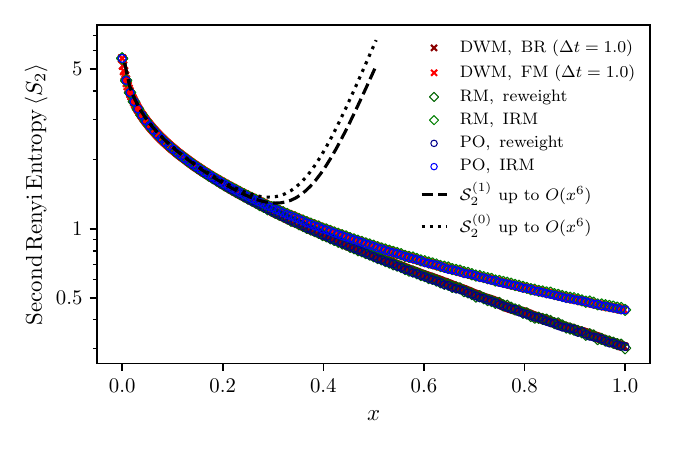}
}

\caption{Comparison of the numerical data for the protocols discussed in Sec.~\ref{sec:numerics}, at $q=256$, with the theoretical scaling functions obtained in Sec.~\ref{sec:small_x}, in the perturbative expansion at small $x$. Both the BR - reweight - $N\to1$ average and the FM - independent random matrices (IRM) - $N\to 0$ averages are displayed.
RM and PO refer to the discrete-time processes $(i)$ and $(ii)$ in Subsec.~\ref{subsec:simdiscrete} respectively, namely, to the multiplication of independent random matrices, and to the protocol alternating projection and multiplication times an orthogonal matrix, while DWM refers to the discrete-time weak-measurement protocol, see Subsec.~\ref{subsection:continuous_num}.
\textit{Left:} Sample-mean of the von Neumann entropy $\average{S_1}$, compared to the $O(x^4)$ prediction \eqref{eq:S1_theo_N1} for $\mathcal S_{\rm VN}^{(1)}$ (dashed) and to the $O(x^2)$ prediction $\mathcal{S}_{\rm VN}^{(0)}$  Eq.~\eqref{eq:S1_theo_N0} (dotted). 
\textit{Right:} Sample mean of the second Rényi entropy $\average{S_2}$, compared to the $O(x^6)$ predictions for $\mathcal{S}_{2}^{(1)}$ (dashed) and $\mathcal{S}_{2}^{(0)}$ (dotted)
Eqs.~\eqref{eq:S2_theo_N}.}
\label{fig:S1_S2_ortho}
\end{figure}

\section{Numerics} \label{sec:numerics}

In this Section, we report the numerical verification of the analytical results discussed in previous Sections and show the emergence of the universal scaling functions relative to orthogonal symmetry class (real Kraus matrices). In particular, the figures presented in this Section illustrate the convergence of the numerical data to the theoretical predictions. 
We focus on two different classes of models: on the one hand, we consider products of random matrices (``RM" protocol), including real Ginibre matrices (directly related to the toy model of Sec.~\ref{subsec:real}), and products of Haar-distributed orthogonal matrices interleaved with measurements in the form of projectors (we dub this protocol ``PO"). 
On the other hand, we consider weak measurements (WM), both in discrete-time (DWM model) and in the continuous-time model (in line with the toy model of Sec.~\ref{sec:continuous_model}). 
These simulation protocols are described in detail in the next Subsecs.~\ref{subsec:simdiscrete} (RM and PO cases) and \ref{subsection:continuous_num} (DWM and WM). All the numerical results presented in this Section, for each of the protocols described, refer to the orthogonal symmetry class. Numerical analyses of the unitary symmetry class can be found in Refs.~\cite{deluca2025universality,gerbino2024dyson}.

As significant probes of our analytical results and of the universal scaling behavior, in all cases we focus our attention on the direct evaluation of the von Neumann entropy $\average{S_{\rm VN}}$ and of the second R{\'e}nyi entropy $\average{S_2}$, which are computed at the trajectory level via the normalized density matrix $\rho \equiv \urho/\Tr \urho$. For both observables, we consider both the BR average \eqref{eq:MaveBR}, equivalent to the limit $N\to 1$ in the replica description, and the FM average \eqref{eq:MaveFM}, corresponding to the limit $N\to 0$. 
We focus on the scaling regime obtained by taking the large-time $\td$, large-$q$ limit with the finite scaling variable
\begin{align}
	x = \frac{\td \ \Delta t}{\tp} \,,
\end{align}
where $\tp$ is the purification time and it depends on the specific model, while $\td$ is the number of discrete time-steps. For the multiplication of random matrices, we have simply $\tp  = q \ \Delta t$. 
As regards the weak measurements model, we note that both for numerical implementation and to add a further way of testing universality, we consider the model obtained by applying Kraus operators in Eq.~\eqref{eq:KaFull}, with finite steps $\Delta t$. In this case, it can be verified that  $\tp = q \ \Delta t/ \mathsf f(\Gamma \Delta t)$, with $\mathsf f$ a (nonuniversal) function of the time-step $\Delta t$ and the measurement rate $\Gamma$ (see Sec.~1 in the Supplemental Material~\cite{SM} for a definition of $\mathsf f$). In the continuous-time limit $\Delta t \to 0$, $\mathsf f(\Gamma \Delta t) \simeq 4\Gamma \Delta t$ so that $\tp = {q/(4\Gamma)}$ as discussed in Subsec.~\ref{subsec:weakmeastoeigen}. 
For all protocols, we consider Hilbert spaces of increasing sizes $q=128$ and $q=256$, averaging over batches of $3\times 10^4$ samples in each case.

\begin{figure}
\centering    \includegraphics[width=0.49\linewidth]{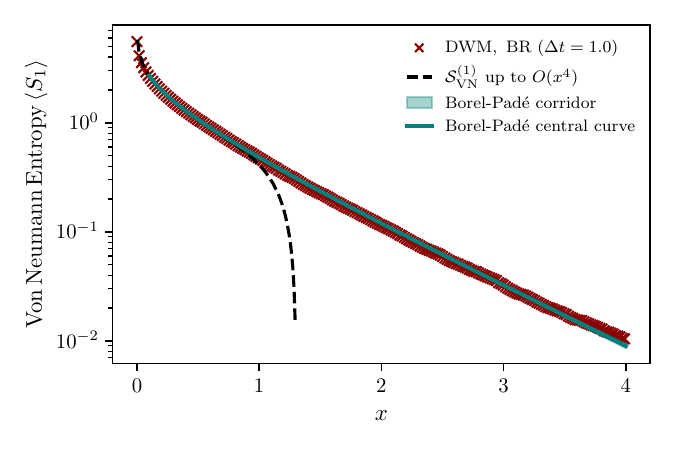}
\hfill
\includegraphics[width=0.49\linewidth]{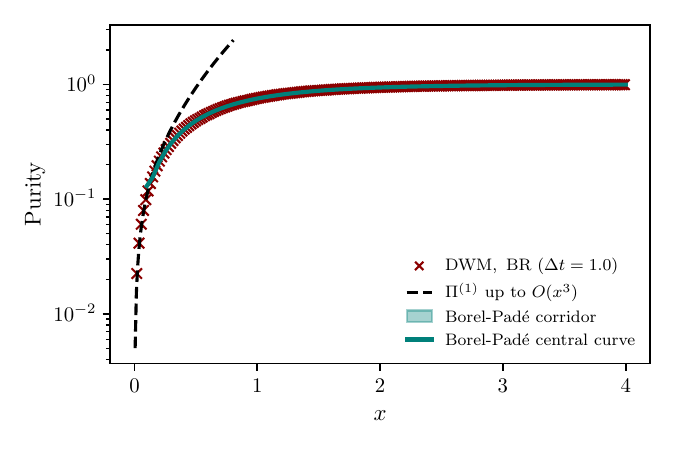}
\caption{Comparison of the numerical data  for the DWM protocol, described in Subsec.~\ref{subsection:continuous_num}, with the approximate Borel-Pad\'e resummation method explained below Eq.~\eqref{eq:S1_theo_N1}. 
\textit{Left:} the BR von Neumann entropy $\average{S_1}$ from data is compared to the small-$x$ expansion Eq.~\eqref{eq:S1_theo_N1} obtained in the scaling limit, and with the approximate resummed expression. While the match with the perturbative series is good for $x<1$ but then gets worse, the approximate resummation provides good quantitative predictions for arbitrary values of $x$. \textit{Right:} the BR purity $\Pi$ from data is compared to the small-$x$ prediction $\Pi^{(\beta=1)}(x) = x+x^2/2+5x^3/2$ obtained through Eq.~\eqref{eq:trick} and the short-time expansion of $\Omega^{(\beta=1)}_{{N}}$, showing good match at small $x$. At larger values of $x$, the resummed expression for the purity provides a good effective prediction, also showing better agreement with data due to lower statistical noise.}
\label{fig:BP}
\end{figure}

The numerical results are displayed in Fig.~\ref{fig:S1_S2_ortho}, where we show the convergence to the predictions Eqs.~(\ref{eq:S1_theo_N1},\ref{eq:S1_theo_N0}) for the von Neumann entropy $\mathcal{S}_{\rm VN}^{(N)}$ and Eq.~\eqref{eq:S2_theo_N} for the Rényi entropy $\mathcal{S}_2^{(N)}$, both in the BR and FM cases.
The differentiation of the numerical data from different models into two well-separated $N$-dependent curves is an evident validation of universality. Moreover, the collapse of those curves to the theoretical small-$x$ predictions is clear.
In all figures in this Section, green tiles and blue circles represent data from RM and from PO protocols, respectively. 
We indicate instead with red crosses the data points from DWM obtained setting $\Delta t=1.0$, and with red solid lines the curves from forward Euler evolution of the Stochastic Schr{\"o}dinger equation for weak measurements, setting $\Delta t= 10^{-5}$. 
We also test the Borel-Pad\'e resummation method explained below Eq.~\eqref{eq:S1_theo_N1}, comparing the resummed expression provided in the Supplemental material~\cite{SM} with numerical data, as shown in Fig.~\ref{fig:BP}.

\begin{figure}
\centering
\subfloat{\includegraphics[width=0.4\linewidth]{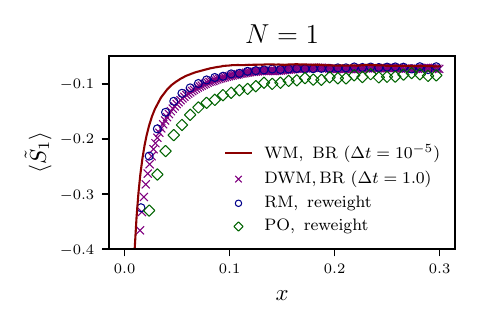}}
\hspace{2pt}
\subfloat{\includegraphics[width=0.4\linewidth]{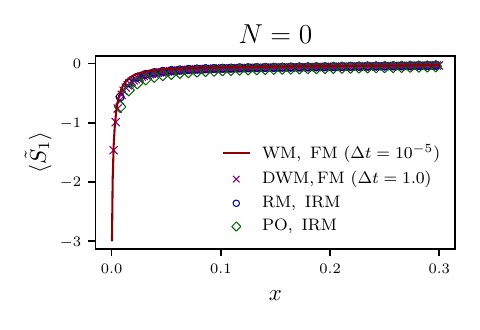}} 

\subfloat{\includegraphics[width=0.4\linewidth]{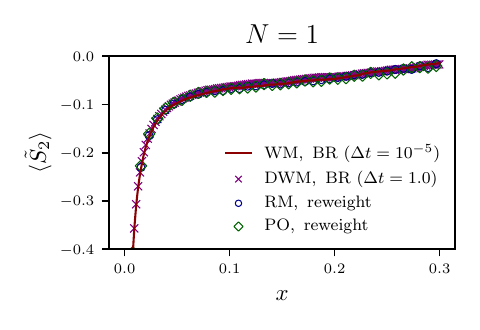}}
\hspace{2pt}
\subfloat{\includegraphics[width=0.4\linewidth]{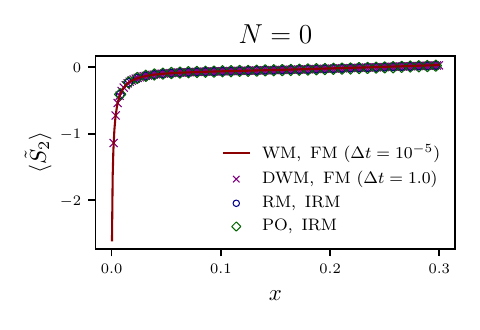}}

\caption{Convergence of the numerical data for various models to universal curves. \textit{Top:} shifted von Neumann entropy $\average{\tilde S_1}$ (see \eqref{eq:shifted}). Curves merge at around {$x \simeq 0.2$} where the effect of model-dependent  and finite-size corrections becomes negligible. 
\textit{Bottom:} shifted R{\'e}nyi entropy $\average{\tilde S_2}$ (see \eqref{eq:shifted}). The juxtaposition of numerical data is excellent already at small $x$. 
\textit{Left:} BR - reweight - $N\to1$ averages.
\textit{Right:} FM - RM - $N\to 0$ averages.
}
\label{fig:convergence}
\end{figure}

As a further proof of universality, we check the collapse of the numerical data beyond the leading order $\sim \ln x$. In Fig.~\ref{fig:convergence}, we plot the shifted sample averages 
\begin{equation} \label{eq:shifted}
    \average{\tilde S_1}=\average{S_1}-(1-\gamma_{\rm E})+\ln x\,,\qquad 
    \average{\tilde S_2}=\average{S_2}+\ln x \,.
\end{equation}
In general, for the von Neumann entropies, the effect of model-dependent and finite-size corrections becomes irrelevant for {$x \gtrsim 0.1$}, where universal behavior is explicit. For the second R{\'e}nyi entropies, instead, curves collapse perfectly for any value of the scaling parameter $x$.

\begin{figure}
    \centering
    \subfloat{
    \includegraphics[width=0.5\linewidth]{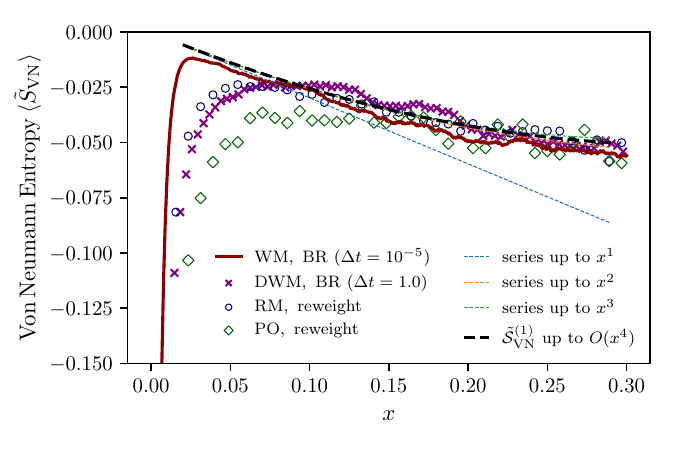}}
    \subfloat{
    \includegraphics[width=0.5\linewidth]{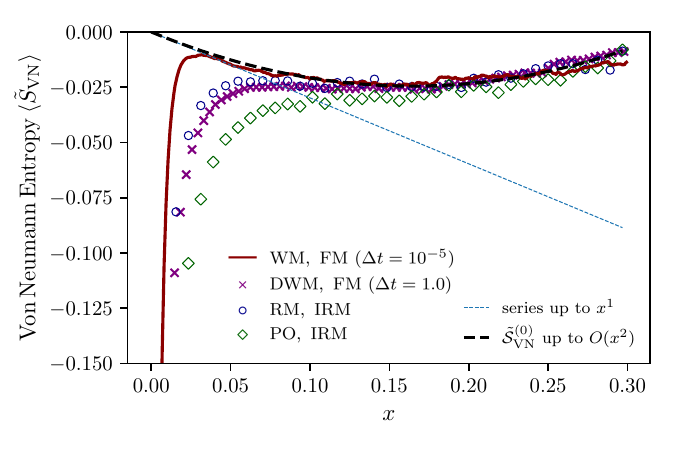}}
    \caption{Next-to-leading-order terms in the von Neumann entropy $\average{\tilde S_1}$ Eq.~\eqref{eq:shifted}, via the extrapolation formula \eqref{eq:extrapolation}. The convergence to the theoretical prediction $\tilde{\mathcal{S}}_{\rm VN}^{(1/0)} = \mathcal{S}_{\rm VN}^{(1/0)} - (1-\gamma_E) + \ln x$ (dashed) is very good for $x \gtrsim 0.1$. 
    Moreover, it is particularly explicit for the continuous-time WM model in the region of $x\approx 0.05$, where the linear prediction alone is enough to capture the empirical behavior.
    \textit{Left:} BR - reweight - $N\to 1$ averages compared to $\tilde{\mathcal{S}}_{\rm VN}^{(1)}$ Eq.~\eqref{eq:S1_theo_N1}. \textit{Right:} FM - RM - $N\to 0$ averages compared to $\tilde{\mathcal{S}}_{\rm VN}^{(0)}$ Eq.~\eqref{eq:S1_theo_N0}
    }
    \label{fig:S1_ortho_corrected}
\end{figure}

\begin{figure}
\centering
\subfloat{
\includegraphics[width=0.5\linewidth]{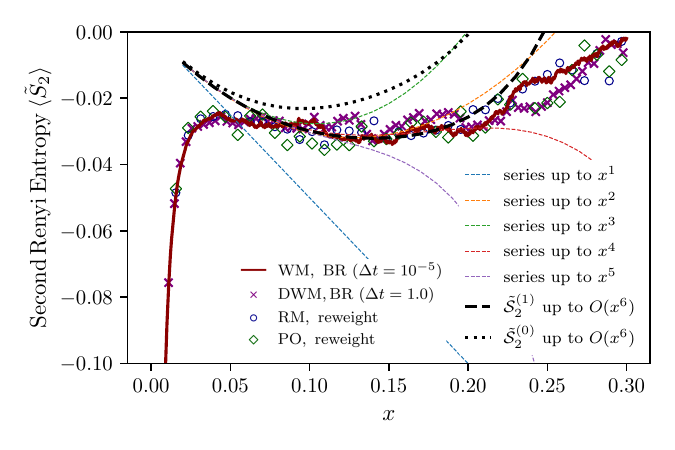}
}
\subfloat{
\includegraphics[width=0.5\linewidth]{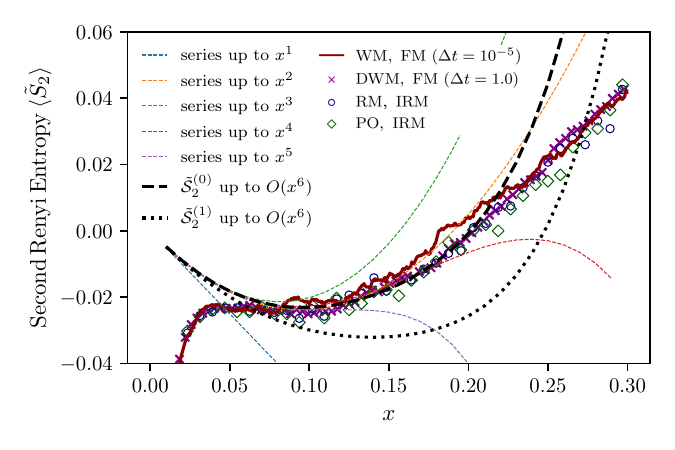}
}
\caption{Second Rényi entropy $\average{\tilde S_2}$ Eq.~\eqref{eq:shifted}, via the extrapolation formula \eqref{eq:extrapolation}. The convergence to the theoretical predictions $\tilde{\mathcal{S}}_2^{(1/0)} = \mathcal{S}_2^{(1/0)} + \ln x$ (dashed) of Eq.~\eqref{eq:S2_theo_N} is clear up to $x \approx 0.2$. Beyond that region, more terms in the small-$x$ expansion are needed to capture the empirical behavior, but data still collapse to single functions of $x$. 
\textit{Left:} BR - reweight - $N\to 1$ averages compared to $\tilde{\mathcal{S}}_2^{(1)}$ Eq.~\eqref{eq:S2_theo_N1}. \textit{Right:} FM - RM - $N\to 0$ averages compared to $\tilde{\mathcal{S}}_2^{(0)}$ Eq.~\eqref{eq:S2_theo_N0}.
}
\label{fig:S2_ortho_corrected}
\end{figure}

Furthermore, we employ the following stratagem in order to better compare the numerical results against the theoretical predictions, and to reduce the effects of non-universal corrections. By the assumption that the latter can be expanded perturbatively around $t\to \infty$, we consider the form
\begin{align}
    \langle S_n (x,t) \rangle^{(N)} = s_{n,N}^{(0)}(x) + \frac{1}{t} s_{n,N}^{(1)}(x) + \frac{1}{t^2} s_{n,N}^{(2)}(x) + \cdots,
\end{align}
where $s_{n,N}^{(0)}(x)$ is the leading universal part that survives in the scaling limit. 
Then, the approach to the universal part can be accelerated using the formula
\begin{equation} \label{eq:extrapolation}
    2 \langle S_n (x,2t)\rangle^{(N)}  - \langle S_n (x,t) \rangle^{(N)} = s_{n,N}^{(0)}(x) + O(t^{-2})\,,
\end{equation}
which is used to display data in Figs.~\ref{fig:S1_ortho_corrected} and \ref{fig:S2_ortho_corrected}. 
The convergence to the theoretical predictions is excellent when looking at the von Neumann entropy (see Fig.~\ref{fig:S1_ortho_corrected}) in the BR-$N\to 1$ case, from $x \approx 0.1$ in the considered region up to $x=0.3$. As shown in the figure, the expression for $\mathcal S_{\rm VN}^{(1)}(x)$ up to $O(x)$ (equivalent to the one for $\mathcal S_{\rm VN}^{(0)}(x)$) is sufficient to capture universal behavior only around $x\approx 0.05$. {For the FM-$N\to0$ case, only the $O(x^2)$ expression \eqref{eq:S1_theo_N0} is available, as higher-order coefficients could not be found (right panel).} 
Conversely, in the case of the second R{\'e}nyi entropy, both for BR and FM sample averages can be compared significantly with the predictions Eq.~\eqref{eq:S2_theo_N}, see Fig.~\ref{fig:S2_ortho_corrected}. The agreement between the small-$x$ expansions to $O(x^6)$ and the numerical results is excellent from $x \approx 0.05$ up to $x \approx 0.2$. For larger values of $x$, the theoretical curves detach from the numerical data, indicating that higher-order coefficients are needed, and that the perturbative expansion would need to be resummed at all orders. For the von Neumann entropy in the BR class this limitation is removed by the Borel--Pad\'e resummation of Sec.~\ref{sec:small_x}: the resummed scaling function, with its parameter-free confidence corridor, covers the whole window explored numerically and can be overlaid directly on the data {(see Fig.~\ref{fig:BP}, left panel, and Ref.~\cite{SM})}. We also test our method by comparing the purity $\Pi^{(\beta=1)}$ from data with the one obtained via approximate resummation, showing excellent agreement (right panel).

\subsection{Simulation Protocol for the Discrete Process} \label{subsec:simdiscrete}

Let us consider two classes of discrete random dynamics: $(i)$ products of matrices drawn from the real Ginibre ensemble, namely the model described in Sec.~\ref{sec:discrete_model}, and $(ii)$ projective measurements interleaved with dynamics generated by Haar-random orthogonal matrices. The calculation follows the methodology developed in Ref.~\cite{deluca2025universality} for the complex Ginibre ensemble and unitary circuits. For these two discrete ($\Delta t=1$) models, we simply have $\tp = q \ \Delta t$, therefore $x = \td/q$. 
The calculation of averages of $S_{\rm VN}$ and $S_2$ simply entails sample-averaging those quantities over multiple trajectories. However, two distinct ensemble averages are of particular interest: the reweighted average and the direct average, respectively
\begin{align}
    \langle S_{n}(t)  \rangle_{\rm BR} = \frac{\langle \Tr (\check{\rho}) \, S_n(t) \rangle}{\langle \Tr \check{\rho} \rangle}  \ ,\qquad 
    \langle S_{n}(t) \rangle_{\rm FM} = \langle S_n(t) \rangle \,,
\end{align}
which correspond to the Born's rule \eqref{eq:MaveBR}, $N \to 1$ replica limit, and the forced measurements \eqref{eq:MaveFM}, $N \to 0$ limit. Note, however, that we make no use of replicas for the numerical computation of these observables.

While the protocol $(i)$ is essentially the one described in detail in Sec.~\ref{sec:discrete_model}, we now turn to protocol $(ii)$ for purification, which involves measurements interleaved with time evolution generated by Haar-random orthogonal matrices. The setup is as follows. For a system of $L$ qubits with total Hilbert space dimension $q_{\mathrm{tot}} = 2^L$, we assume the initial state to be the maximally mixed state, $\rho_0 = q_{\mathrm{tot}}^{-1} \, \id$.
At each timestep $i \in \{1, 2, \dots, t\}$, we first apply a Haar-random orthogonal matrix $O \in \OO(q_{\mathrm{tot}})$ to the system, followed by a projective measurement of a single qubit in the computational basis. After this measurement, the remaining unmeasured qubits span a Hilbert space of dimension $q_{\mathrm{u}} = 2^{L-1}$.
The post-measurement state at time-step $\td$ is given by $\rho_\td = P_\td \rho_{\td-1} P_\td / \Tr(P_\td \rho_{\td-1} P_\td)$,
where $P_\td$ is a random orthogonal projector of rank $q_{\mathrm{u}}$, determined by the outcome of the single-qubit measurement. As each measurement projects onto one of the two computational-basis states of a single qubit, there are two possible outcomes at each timestep.
In this model, the relevant parameter $q$, which characterizes the effective dimension appearing in the averaged dynamics, is given by
\begin{align}
q = \left( \frac{1}{q_{\mathrm{u}}} - \frac{1}{q_{\mathrm{tot}}} \right)^{-1}.
\end{align}
This expression naturally arises in the analysis of the averaged purity or trace decay under repeated random measurements. In our setup with $q_{\mathrm{tot}} = 2^L$ and $q_{\mathrm{u}} = 2^{L-1}$, this simplifies to $q = 2^L$, although this equality is specific to the binary measurement structure.

\subsection{Simulation Protocol for the Continuous Process} \label{subsection:continuous_num}

The weak-measurement model defined in Sec.~\ref{sec:continuous_model} is simulated considering two distinct procedures: $(i)$ the explicit forward evolution of a discrete weak-measurement (DWM) protocol, as given by the Kraus operators Eq.~\eqref{eq:KaFull}; $(ii)$ the forward Euler evolution of the stochastic continuous-time Eq.~\eqref{eq:VNstochastic}. The advantage of the first method, which does not require a fine time-discretization, lies in its numerical efficiency. The equivalence of the two approaches provides a further proof of universality, as the two models collapse onto the same description after appropriate rescaling of the time variable through the function $\mathsf f$. 
In both approaches we compute the BR-averaged \eqref{eq:MaveBR} and the FM-averaged \eqref{eq:MaveFM} von Neumann and second R{\'e}nyi entropies. 

In particular, for the procedure $(i)$, we consider weak measurements at arbitrary $\Delta t$, via the Kraus operators $K_s(\hat O; \sqrt{\Gamma/\Delta t},\Delta t)$ \eqref{eq:KaFull}. The evolution of $\rho$ is dictated by 
\begin{equation} \label{eq:WMdiscrete}
    \rho + d\rho = \frac{\Kr_s \rho \Kr_s}{\Tr \left( \Kr_s \rho \Kr_s\right)} \,,
\end{equation}
conditioned on the selection of the random operator $\hat O$. Notice that for BR dynamics the choice of $K_{\pm1}$ is also conditioned on the ancilla measurement outcome $s=\pm 1$, which is realized with probability $\Tr \left( \Kr_s \rho \Kr_s\right)$ (meaning the outcomes are not equiprobable). In contrast, in the case of FM averaging, the choice $\Kr_{\pm 1}$ can be seen to be conditioned on random outcomes $s_\pm$ that are equiprobable. 
For model $(i)$, we use the general expression $\mathsf f(\Gamma \Delta t) = 1/2 - J_1(8\sqrt{\Gamma \Delta t})/(8\sqrt{\Gamma \Delta t})$~\cite{SM}. 

Turning to procedure $(ii)$, we consider the $\Delta t\to 0$ limit of the protocol Eq.~\eqref{eq:WMdiscrete}. For BR dynamics, we retrieve the
evolution Eq.~\eqref{eq:VNstochastic}, see Ref.~\cite{giachetti2023elusive}, which can be turned into an evolution equation for the normalized $\rho = \urho/\Tr \urho$. Also in the FM case, it is possible to retrieve a stochastic equation for the normalized density matrix, assuming equal probability of the outcomes $s_\td$, before the continuous-time limit. For the two cases, we have, respectively:
\begin{subequations}
\begin{equation}  \label{eq:stochastic_norm_BR}   d \rho = -i\ dt[H, \rho] - \Gamma dt \left( \rho - \frac{\id}{q} \right) + \sqrt{\Gamma} \left\{ dO-\average{dO}, \rho \right\} \qquad {\rm (BR)} \,,
\end{equation} 
\begin{equation}
    \label{eq:stochastic_norm_FM}   d \rho = -i\ dt [H, \rho] - \Gamma dt \left( \rho - \frac{\id}{q} + \frac 8 q (\rho^2 - \rho \ \Tr \rho^2)\right) + \sqrt{\Gamma} \left\{ dO-\average{dO}, \rho \right\}\qquad {\rm (FM)}\,.
\end{equation}
\end{subequations}
{where $\average{dO} := \Tr(\rho\, dO)$.}
The equations above can be easily used to numerically generate trajectories through the Euler method, for small $\Delta t \to 0$. 
However, the forward Euler approach is not easily viable to prove the $O(x)$ corrections beyond the leading behavior $\sim \ln x$. The numerical error introduced by the time-discretization is indeed quite relevant; in Figs.~\ref{fig:convergence}, \ref{fig:S1_ortho_corrected} and \ref{fig:S2_ortho_corrected}, the choice $\Delta t=10^{-5}$ has been used. Notice that $\mathsf f(\Delta t) \simeq 4\Delta t$ in this case.

\section{Conclusions}

In this work, we provide a comprehensive theoretical and numerical characterization of universal purification dynamics in monitored quantum processes, focusing on the scaling limit deep in the weak-measurement (volume-law) phase. In this regime, the exponential separation of timescales ensures that scrambling dominates over information extraction, enabling an effective $0+1$-dimensional replica statistical-mechanics description.

Specifically, we establish that restricting the system's effective symmetry group from the unitary group $\mathcal{U}(q)$ to the orthogonal group $\mathcal{O}(q)$---as realized in systems with {purely imaginary} Hamiltonians or circuits with real gates---qualitatively alters the scaling behavior of purification. To capture these universal features, we develop two complementary theoretical frameworks:
\begin{itemize}
    \item \textbf{Discrete-time random-matrix dynamics:} By mapping purification to products of matrices drawn from the real Ginibre ensemble ($\beta=1$), we use the combinatorial structure of the commutant space to characterize effective interactions between replicated degrees of freedom. In the real-matrix case, the basis of invariant states is enlarged to all pairings of $2N$ elements, generating a distinct universal transfer matrix.
    \item \textbf{Continuous-time weak monitoring:} We show that infinitesimal stochastic evolution of the density matrix maps eigenvalue dynamics to Dyson Brownian motion. Crucially, this mapping is formally equivalent to an imaginary-time Schrödinger equation for the integrable Calogero--Sutherland model, naturally accommodating different symmetry classes through the Dyson index $\beta$.
\end{itemize}

A central quantitative outcome of our analysis is the distinct scaling behavior of the Rényi entropies at short rescaled times $x = t/\tp$. In contrast to the unitary class ($\beta=2$), which exhibits an $O(x^2)$ leading-order correction, the orthogonal class ($\beta=1$) displays a linear $O(x)$ correction. This indicates that purification is inherently more effective in the orthogonal case, a direct consequence of the reduced parameter space of real matrices. These theoretical predictions were strongly corroborated by numerical simulations of both discrete and continuous protocols, which exhibited excellent data collapse onto our derived scaling functions.

Looking forward, this work opens several compelling avenues for future research. 
From an analytical standpoint, important subtleties of the replica formalism remain to be clarified. Although closed resummations of the $\beta=2$ type appear technically arduous at $\beta=1$ --- the natural Pfaffian analogue of the determinantal generating function resists the replica continuation --- the Borel--Pad\'e mechanism with the exponential decay at infinity enforced does work: it upgrades our perturbative expansions to a non-perturbative determination of the von Neumann scaling function, with a parameter-free confidence corridor (Sec.~\ref{sec:small_x} {and Ref.~\cite{SM}}).  Sharpening the deep-tail behaviour $x\gg1$, where the entropies are governed by a few dominant transfer-matrix singular values and the remainders decay exponentially with power-law prefactors, and taming the Pfaffian structure itself, remain rich analytical directions.

It is also natural to investigate additional dynamical universality classes generated by further constraints on the monitored dynamics. For example, monitored Clifford circuits
are associated with a different invariant-state structure~\cite{Leone2021quantumchaosis,gross2021schur,Turkeshi_2025,bittel2025completetheorycliffordcommutant,leone2025noncliffordcostrandomunitaries,magni2025anticoncentrationstatedesigndoped,aditya2025growthspreadingquantumresources,aditya2025mpembaeffectsquantumcomplexity,Magni2025quantumcomplexity,magni2025anti,Haug2025probingquantum,Leone2021quantumchaosis,dowling2025magic,dowling2025bridging}, while circuits with continuous symmetries (such as $U(1)$ conservation~\cite{PhysRevLett.133.140405,PRXQuantum.6.020343,PhysRevLett.129.120604,PhysRevX.8.031058,PhysRevLett.122.250602,m3np-p5xj,PhysRevD.110.L061901,gotta2026enhancingentanglementasymmetryfragmented}) introduce additional timescales such as charge sharpening before universal purification scaling sets in.

Finally, because the characteristic purification timescale $\tp$ grows exponentially with the {system size $L$}, the scaling regime can already emerge at modest system sizes. This makes direct observation of purification universality classes a realistic target for near-term quantum simulators and cross-entropy-benchmarking platforms.

\section*{acknowledgments}

{~}\\
We are indebted {to} Keiji Saito for illuminating discussions that motivated this project. ADL would also like to thank Adam Nahum and Pierre Le Doussal for the stimulating conversations on this topic, comparing the unitary and orthogonal cases and motivating a general formulation.
DK acknowledges support from the Hakubi Project at RIKEN.
X.T. acknowledges support from DFG under Germany's Excellence Strategy – Cluster of Excellence Matter and Light for Quantum Computing (ML4Q) EXC 2004/2 – 390534769, and DFG Collaborative Research Center (CRC) 183 Project No. 277101999 - project B01, and DFG Emmy Noether Programme proposal "Digital Quantum Matter Out-of-Equilibrium" No. 560726973. G.G. acknowledges the support of the European MSCA Grant 101152898 (DREAMS).

\putbib
\end{bibunit}


\clearpage
\onecolumngrid

\setcounter{page}{1}
\setcounter{equation}{0}
\setcounter{figure}{0}
\setcounter{table}{0}
\setcounter{section}{0}

\setcounter{secnumdepth}{3}

\renewcommand{\thesection}{\Roman{section}}
\renewcommand{\thesubsection}{\thesection.\arabic{subsection}}
\renewcommand{\thesubsubsection}{\thesubsection.\arabic{subsubsection}}

\makeatletter
\def\p@subsection{}
\def\p@subsubsection{}
\makeatother

\renewcommand{\theequation}{S\arabic{equation}}
\renewcommand{\thefigure}{S\arabic{figure}}
\renewcommand{\thetable}{S\arabic{table}}

\renewcommand{\theHequation}{S\arabic{equation}}
\renewcommand{\theHfigure}{S\arabic{figure}}
\renewcommand{\theHtable}{S\arabic{table}}

\begin{bibunit}

\begin{center}
    {\large \bfseries Supplemental Material:\\[0.4em]
    Universal purification dynamics in real non-unitary quantum processes \par}
    \vspace{1em}
    {\normalsize
    Federico Gerbino$^{1}$, Donghoon Kim$^{2}$, Guido Giachetti$^{3}$, Andrea De Luca$^{3}$ and 
    Xhek Turkeshi$^{4}$\par}
    \vspace{0.5em}
    {\small
    $^{1}$Laboratoire de Physique Th\'eorique et Mod\`eles Statistiques, Universit\'e Paris-Saclay, CNRS, 91405 Orsay, France\\ 
    $^{2}$Analytical Quantum Complexity RIKEN Hakubi Research Team,\\
    RIKEN Center for Quantum Computing (RQC), Wako, Saitama 351-0198, Japan\\
    $^{3}$Laboratoire de Physique de l'\'Ecole Normale Sup\'erieure, CNRS, ENS \& PSL University,\\
    Sorbonne Universit\'e, Universit\'e Paris Cit\'e, 75005 Paris, France\\
    $^{4}$Institut für Theoretische Physik, Universität zu Köln, Zülpicher Strasse 77, 50937 Köln, Germany\par}
    \vspace{1em}
\end{center}

\section{Weak measurement
\label{sec:weakmes}
}

In this Appendix, we provide a derivation of the weak measurement protocol introduced in Subsec.~\ref{subsec:weakmeastoeigen}, in order to show the origin of the Kraus operators $K_s(\hat O;\lambda)$ in Eq.~\eqref{eq:KaFull}. 
The basic observation for introducing this protocol, is that projective measurements do not naturally fit a continuous-time limit, as their effect on the system cannot be made arbitrarily small. Weak measurements solve this problem with an experimentally relevant protocol that allows for a continuous-time mathematical description: the idea is to couple the system with an ancilla, a spin $1/2$, for a small time $\Delta t$, so that the system and the ancilla become entangled. A projective measurement of the ancilla is then performed. The crucial aspect of this procedure is that the coupling between the system and the ancilla remains tunable, so that the effect on the system of the projective measurement on the ancilla can be made arbitrarily small. 

To be specific, we denote by $\hat X, \hat Y, \hat Z$ the Pauli matrices acting on the ancilla, by $\ket{\pm}$ the eigenstates of $\hat Z$ 
and assume that the ancilla is initially prepared in the eigenstate $\ket{\to} = (\ket{+} + \ket{-})/\sqrt{2}$ of $X$. Denoting as $\hat{O}$ a hermitian operator acting on the system, we assume that the system ($S$) and the ancilla ($A$) are coupled for a time $\Delta t$, according to the Hamiltonian
\begin{equation}
    H_{S+A} = \lambda \hat{O} \hat{Y} \;,
\end{equation}
with the coupling parameter $\lambda$.
Then, performing a projective measurement of the operator $\hat Z$ 
with outcome $s = \pm 1$
amounts to consider the Kraus operators (see also Appendix~(A) of \cite{giachetti2023elusive} for a treatment with similar notation)
\begin{equation}
\label{eq:KaCont}
    \Kr_{s = \pm}(\hat O; \lambda, \Delta t)  = \bra{\pm} e^{-i \lambda \hat O \hat{Y} \Delta t} \ket{\mathcal \to} 
    = \frac{1}{\sqrt 2}\begin{pmatrix}
        \frac{1+s}{2} \\
        \frac{1-s}{2}
    \end{pmatrix}^T \left( \cos(\lambda \hat O \Delta t) - i \sin (\lambda \hat O \Delta t) Y \right) \begin{pmatrix}
        1 \\1
    \end{pmatrix}  = \cos \left( \lambda \hat O \Delta t + s \frac{\pi}{4} \right) \,,
\end{equation}
where the specific choice $s=\pm 1$ with corresponding $\Kr_s$ is realized with probability $\Tr( \Kr_s \rho \Kr_s)$. One can check that the probability is conserved $\sum_{s=\pm 1} \Kr_s(\hat O; \lambda, \Delta t)^2=1$. The evolution, in a time-step $\Delta t$, for the non-normalized density matrix $\urho(t=\td \Delta t)$ such that $\rho(t) = \urho(t)/\Tr\urho(t)$ (see Eq.~\eqref{eq:rhoevol}) conditioned to the outcome $s$ is then simply
\begin{equation}
    \urho((\td+1)\Delta t) = \Kr_s \ \urho (\td\Delta t) \ \Kr_s \,.
\end{equation}
Scaling the coupling as $\lambda = \sqrt{\Gamma/\Delta t}$, the continuous-time limit $\Delta t\to 0$, $\lambda\to \infty$ with $\Gamma$ finite yields 
\begin{equation}
    \Kr_s\left(\hat O; \sqrt{\frac{\Gamma}{\Delta t}}, \Delta t\right) = \frac{1}{\sqrt 2} \left( \id - s \sqrt{\Gamma \Delta t}\ \hat O - \frac 1 2 \Gamma \Delta t \ \hat O^2 \right) +O(\Delta t^{3/2})\,,
\end{equation}
which leads, up to an overall multiplicative factor disappearing upon normalization $\urho \to \rho$, to the following evolution for the density matrix: 
\begin{equation}
     \urho((\td+1)\Delta t) - \urho(\td\Delta t) = \Gamma \Delta t \left( \hat O\  \urho(\td\Delta t) \  \hat O - \frac 1 2 \left\{ \hat O^2 , \urho(\td\Delta t) \right\} \right) -  \sqrt{\Gamma \Delta t} \left\{ \hat O, \urho(\td\Delta t)\right\} + O(\Delta t^{3/2})\,.
\end{equation}
If now we consider $\hat O = \sum_{\alpha,\alpha'=1}^q o_{\alpha\alpha'}\ket{\alpha}\bra{\alpha'}$ in the equation above, with $o$ a random matrix from a Gaussian ensemble with symmetry parameterized by a Dyson index $\beta$, upon the continuous-time limit $\Delta t \to dt$ , we effectively average the second moments over the extraction of $o$ at each time-step, as described in the main text,  we recover the continuous-time dynamics given in Eq.~\eqref{eq:VNstochastic}.

Beyond the continuous-time limit, every finite $\Delta t$ defines a different weak measurement protocol. As we discussed in the main text, in the considered scaling limit, all such models are expected to collapse onto the same universal curves only depending on the symmetry classes $\beta=1,2$, upon a correct choice of the purification time $t_P$. 
To compute the proper purification time as a function of the time step $\Delta t$, it is sufficient to compute the purity $\Pi(t=\Delta t)$ after one time-step. Note that this approach works in the current model, where we are using a dynamics that is already isotropic in the operator space. In a realistic model, Delta t must be chosen to be macroscopic, i.e., much larger than any scrambling timescale but smaller than the global purification timescale.

We start from the maximally mixed state $\rho(0) = \urho(0) = \id/q$, and we apply the weak measurement protocol as described above, so that 
\begin{equation}
    \Pi(t=\Delta t) = \average{\Tr \rho^2(t=\Delta t)} = \frac 1 q \sum_{s=\pm1} \aveE{\frac{\Tr \Kr_s^4 }{(\Tr \Kr_s^2)^2}  \Tr \Kr_s^2 }\,,
\end{equation}
where the Gaussian average $\aveE{\dots}$ is over the extraction of $\hat O$ from the chosen Gaussian ensemble ($\beta=1,2$), and the factor $\Tr \Kr_s^2$ is the probability for the corresponding outcome to be realized. 
It is convenient to turn the average over the matrix elements on $\hat O$ into an average over its eigenvalues, having 
\begin{equation}
    \Pi(\Delta t)  = \frac 1 q \sum_{s=\pm 1} \mathbb E \left[ \frac{\sum_{j=1}^q\cos^4(\sqrt{\Gamma\Delta t}\lambda_j + s \frac \pi 4)}{\sum_{j=1}^q\cos^2(\sqrt{\Gamma\Delta t}\lambda_j + s \frac \pi 4)} \right] \stackrel{q\to\infty}{=}
    \frac 2 q \frac{\int d\lambda\, n(\lambda) \cos^4(\sqrt{\Gamma\Delta t}\lambda +\frac \pi 4)}{\int d\lambda\, n(\lambda) \cos^2(\sqrt{\Gamma\Delta t}\lambda +\frac \pi 4)} \,,
\end{equation}
where in the second equality we exploited self-averagingness at $q\to\infty$, and introduced the spectral density $n(\lambda) = \aveE{q^{-1}\sum_{j=1}^q \delta(\lambda-\lambda_j)}$ for the Gaussian ensemble. The latter, in the limit of $q\to \infty$, converges to the semicircle law $n(\lambda) = \frac{1}{2\pi} \sqrt{4-\lambda^2}$, yielding
\begin{equation}
    \int d\lambda \ \frac{\sqrt{4-\lambda^2}}{2\pi} \left[ \cos\left( \sqrt{\Gamma\Delta t} \lambda + \frac \pi 4 \right)\right]^p = 2^{1-p}\sum_{l=0}^p \binom{p}{l} e^{i\frac \pi 4 (2l-p)} \frac{J_1(2\sqrt{\Gamma\Delta t}(2l-p))}{2\sqrt{\Gamma\Delta t}(2l-p)} = \begin{cases}
        \frac 1 2 \,, &  p=2 \\
        \frac 1 8 \left( 3-\frac{J_1(8\sqrt{\Gamma\Delta t})}{4\sqrt{\Gamma\Delta t}}\right) \,,& p=4
    \end{cases} \,,
\end{equation}
so that 
\begin{equation}
    \Pi(\Delta t) \stackrel{q\to \infty}{=} \frac 1 q \left( \frac 3 2-\frac{J_1(8\sqrt{\Gamma\Delta t})}{8\sqrt{\Gamma\Delta t}}\right) + O(q^{-2}) = \Pi(0) + \frac 1 q \left( \frac 1 2-\frac{J_1(8\sqrt{\Gamma\Delta t})}{8\sqrt{\Gamma\Delta t}}\right) + O(q^{-2})\,.
\end{equation}
Clearly, the function in round brackets defines the amount of ``gained purity" in one time-step, at the leading order in the size $q$, and we set $\mathsf f$ and the purification timescale $t_P$ for weak measurements as
\begin{equation}
    \mathsf f(\Gamma\Delta t) = \frac 1 2-\frac{J_1(8\sqrt{\Gamma\Delta t})}{8\sqrt{\Gamma\Delta t}} \,, \qquad t_P = \frac{q \ \Delta t}{\mathsf f(\Gamma \Delta t)} \,,
\end{equation}
so that $x = \td \Delta t / t_p=\td \ \mathsf f(\Gamma \Delta t) /q$. 
In particular, $\mathsf f(\Gamma\Delta) \simeq 4\Gamma\Delta t + O(\Delta t^2)$ and $t_P = q/(4\Gamma)$ at small $\Delta  t\to 0$ as in the continuous-time model, and $\mathsf f(\Gamma\Delta t\to \infty) = 1/2$, $t_P = 2q \ \Delta t$ at large $\Delta t$. Notice that this last limit can be also computed from the definition of $\Kr_s$ in Eq.~\eqref{eq:KaFull}, with $\Delta t\to \infty$. In that case, $\Kr_s \to P \ {\rm diag} \{\cos \theta_i \} \ P^T$, with $\theta_i$ distributed uniformly in $[0,2\pi)$, and $P\sim{\rm Haar}$ in $\mathcal S\OO(q)$.

\section{Jack Polynomials} \label{sec:jackpols}

In this Appendix, we better define the properties of the Jack symmetric polynomials, and give more details on the diagonalization of the hyperbolic Calogero-Sutherland Hamiltonian proving the main result of Sec.~\ref{sec:continuous_model}. 

Let us consider integer partitions $\lambda \vdash N$ in the notation $\lambda = \{\lambda_1, \dots, \lambda_{\ell}\}$ given above Eq.~\eqref{eq:partitionYoung} of the main text. Let us also recall that this notation is equivalent to the one used throughout the main text where one identifies an integer partition by means of the array of its multiplicities $\mathbf r$, i.e., $|\lambda| = \sum_j r_j$, $\sum_j r_j = \ell$. 
For a partition $\lambda$ of length $\ell$, we define the \textit{monomial symmetric polynomials} $m_\lambda(\vec y)$ of $q\geq \ell$ variables $\{y_j\}_{j=1}^q$ as 
\begin{equation} \label{eq:monomial}
m_\lambda(\vec y) =\sum_{\sigma\in \SS_q} \prod_{j=1}^q y_{\sigma(j)}^{\lambda_j}
\end{equation}
where $\lambda_{j>\ell}=0$ and the sum is over the symmetric group $\SS_q$ permuting the $y$'s variables.
Let us also recall the definition of the power-law symmetric polynomials $p_\mu(\vec y)$ of $q$ variables, indexed by $\mu\vdash N$, as given in Eq.~\eqref{eq:momfromys}. Both the set of monomial and power-law symmetric polynomials represent a basis of the space of symmetric polynomials, and as such there exists an invertible matrix encoding the change of basis from one set to the other. 
The basis of power-law symmetric polynomials is orthogonal with respect to the scalar product defined by 
\begin{equation} \label{eq:scalarP}
    \langle p_\lambda, p_\mu \rangle = \delta_{\lambda \mu} z_\lambda \ \alpha^{\ell(\lambda)} \,,\quad z_\lambda = \prod_j j^{r_j} \ r_j! \,,
\end{equation}
via the real parameter $\alpha$.

The Jack symmetric polynomials \cite{macdonald1998symmetric,stanley1989some} $J^{(\alpha)}_\lambda$ form a further family of bases for symmetric polynomials, labeled by the parameter $\alpha$. The linear change-of-basis relation is the one reported in Eq.~\eqref{eq:changeofbasis} of the main text. 
They are uniquely defined by the following two properties:
\begin{itemize}
    \item orthogonality with respect to the scalar product \eqref{eq:scalarP} of power-sum polynomials:
    \begin{equation} \label{eq:orthogonal}
        \langle J^{(\alpha)}_\lambda, J^{(\alpha)}_\mu \rangle  = c(\lambda,\alpha,1) c(\lambda,\alpha,\alpha) \delta_{\lambda \mu} \,,
    \end{equation}
    with 
    \begin{equation}
        c(\lambda,\alpha,t) = \prod_{i=1}^{\ell(\lambda)} \prod_{j=1}^{\lambda_i} (\alpha(\lambda_i-j) + (\lambda_j'-i) + t) \,,
    \end{equation}
    where $\lambda'_i$ is an element of the conjugated partition $\lambda'$, as defined in Eq.~\eqref{eq:conjPart} of the main text;
    \item the matrix of expansion coefficients on the basis of monomial polynomials is lower-triangular:
    \begin{equation}
        J^{(\alpha)}_\lambda = \sum_{\mu \leq \lambda} u_{\lambda\mu} \ m_\mu \,,
    \end{equation}
    where with $\mu \leq \lambda$ we intend that $|\lambda|=|\mu|$ and $\sum_{j=1}^i \mu_j \leq \sum_{j=1}^i \lambda_j$, $\forall i$. The diagonal coefficient $u_{\lambda\lambda}$ is fixed by the normalization choice imposed above, so that $u_{\lambda\lambda} = c(\lambda,\alpha,1)$. 
\end{itemize}
These two properties uniquely define the coefficients $u_{\lambda\mu}$. These can be computed algorithmically via Gram-Schmidt orthogonalization, or equivalently using the Cholesky decomposition of the Gram matrix associated to the scalar product Eq.~\eqref{eq:scalarP}. 
However, the analytical expressions are in general complicated, and direct extrapolations for arbitrary $N=|\lambda|$ are not possible, although simple relations exist $\forall N$ for some coefficients, following the normalization rule \eqref{eq:orthogonal}. For example $\theta_{(1^N)}^\lambda(\alpha)=1$, independently of $\lambda$ and $\alpha$, as stated in the main text. 
Notice that the change-of-basis rule Eq.~\eqref{eq:changeofbasis}, together with the orthogonality relation Eq.~\eqref{eq:orthogonal}, provides the following linear dependence:
\begin{equation} \label{eq:lineardep}
    \gamma^\lambda_\mu(\alpha) = \frac{\theta^\lambda_\mu(\alpha) \ z_\mu \ \alpha^{\ell(\mu)}}{c(\lambda,\alpha,\alpha) \ c(\lambda,\alpha,1)} \,.
\end{equation}

Even though these definitions are valid for arbitrary $\alpha\in \mathbb{R}$, a few special integer cases are of particular interest. Specifically, for $\alpha=1$, the Jack polynomials
\begin{equation} \label{eq:jacktoschur}
    J_\lambda^{(1)} = c(\lambda,1,1)\ s_\lambda
\end{equation}
are the Schur polynomials $s_\lambda$, with proportionality constant $c(\lambda,1,1)$ given by the product of hook-lengths of the partition $\lambda$. 
Given the known relations for Schur polynomials~\cite{macdonald1998symmetric}
\begin{equation} \label{eq:schurtop}
    s_\lambda = \sum_{\mu\vdash N} \frac{\chi^\lambda(\mu)}{z_\mu} p_\mu \,,\quad 
    p_\mu = \sum_{\lambda \vdash N} \chi^\lambda(\mu) s_\lambda 
\end{equation}
with $\chi^\lambda(\mu)$ the character of the Irrep of $\SS_N$ labeled by $\lambda$, and $z_\mu$ defined as in \eqref{eq:scalarP}, we retrieve the relation $\theta^\lambda_\mu(1) = \chi^\lambda(\mu) \ c(\lambda,1,1)/z_\mu$, and $\gamma_\mu^\lambda = \chi^\lambda(\mu)/c(\lambda,1,1)$. 
Moreover, as one has, for the Gelfand pair $(\SS_N\times \SS_N, \SS_N)$, the relation $\chi^\lambda(\mu) = d^\lambda \ \omega^{(\lambda,\lambda)}(\mu)$, in terms of the corresponding spherical function $\omega^{(\lambda,\lambda)}(\mu)$ (see Eq.~\eqref{eq:sphericalbeta2}) and of the Irrep dimension $d^\lambda = N!/c(\lambda,1,1)$, we can also identify 
\begin{equation}
    \theta^\lambda_\mu(1) = N! \ \frac{\chi^\lambda(\mu)}{d^\lambda z_\mu}= N! \frac{\omega^{(\lambda,\lambda)}(\mu)}{z_\mu} \,,\quad 
    \gamma_\mu^\lambda(1) = \frac{1}{N!} d^\lambda \chi^\lambda(\mu) = \frac{1}{N!} (d^\lambda)^2 \omega^{(\lambda,\lambda)}(\mu) \,,
\end{equation}
where \eqref{eq:lineardep} has been used. 
In particular, the first identity for $\gamma_\mu^\lambda(1)$ is the one used in order to write the moments Eq.~\eqref{eq:pmu_SL} in the shape of Eq.~\eqref{eq:momentsbeta2}.

For $\alpha=2$, instead, the Jack polynomials are the so-called Zonal polynomials $J^{(2)}_\lambda=Z_\lambda$ \footnote{
Macdonald~\cite{macdonald1998symmetric} defines the Jack polynomials $P^{(\alpha)}_\lambda$ such that they are orthogonal with respect to the scalar product for the symmetric power-sum polynomials, defined in Eq.~(1.4) Chapter VI, with the normalization convention that the leading coefficient for $m_\lambda$ is 1 (see Eq.~(10.13) Chapter VI). Then, one can introduce other equivalent polynomials $4J^{(\alpha)}_\lambda$ with a different normalization (Eq.~(10.22) in Chapter VI). The polynomials $J^{(\alpha)}_\lambda$ reduce precisely to the Zonal ones for $\alpha = 2$.
} 
(c.f. Eq.~(2.13) Chapter VII of \cite{macdonald1998symmetric}). 
Expressing the coefficients for the change of basis is in this case more complicated~\cite{Jiu2020}. We limit ourselves to reporting the following equivalent relation connecting Zonal and power-sum polynomials, using the nomenclature proper of the algebraic description of such objects~\cite{macdonald1998symmetric}: 
\begin{equation} \label{eq:zonal_omega}
    Z_\lambda = \sum_{\mu \vdash N} z^\lambda_\mu \ p_{\mu} \,, \quad
    p_\mu = |\SS_2^N \rtimes \SS_N | \sum_{\lambda \vdash N} \frac{\omega^{2\lambda}(\mu)}{c(2\lambda,1,1)} \ Z_\lambda \,,\quad 
    c(2\lambda,1,1) :=c(\lambda,2,2) c(\lambda,2,1) \,,
\end{equation}
where $\SS_2^N \rtimes \SS_N$ is the stabilizer group of $\SS_{2N}$, c.f. the discussion in Sec.~\ref{sec:general_G_invariant}, hence $|\SS_2^N \rtimes \SS_N| = 2^N N!$. Indeed, $(\SS_{2N}, \SS_2^N \rtimes \SS_N)$ is a Gelfand pair, and $\omega^{2\lambda}(\mu)$ is the corresponding zonal spherical function \eqref{eq:spherical}, labeled by the integer partition $2\lambda$ obtained by doubling each element of a given partition $\lambda\vdash N$, i.e., $2\lambda = \{2\lambda_1, \dots, 2\lambda_{\ell(\lambda)}\}$.
From the definition \eqref{eq:zonal_omega}, we immediately recover $z^\lambda_\mu= \theta^\lambda_\mu(2)$, whereas 
the matrix elements $\omega^{2\lambda}(\mu)$ are connected to the coefficients $\gamma_\mu^\lambda(2)$. 
Moreover, let us notice that the factor $c(2\lambda,1,1)$ corresponds to the hook-length product of $2\lambda$. Therefore, multiplying and dividing the second relation in \eqref{eq:zonal_omega} by $(2N)!$, we identify $|\Pa_N|^{-1}=2^NN!/(2N)!$, and $d^{2\lambda}=(2N)!/c(2\lambda,1,1)$ the dimension of the irreducible representation of $\SS_{2N}$ associated to $2\lambda$.
This yields, using \eqref{eq:lineardep}, the identification
\begin{equation}
    \theta^\lambda_\mu(2) = N!\  2^{N-\ell(\mu)} \frac{\omega^{2\lambda}(\mu)}{z_\mu}
    \,,\quad 
    \gamma_\mu^\lambda (2)= \frac{1}{{|\Pa_N|}}d^{2\lambda} \ \omega^{2\lambda}(\mu) \,,
\end{equation}
which we used to express the moments Eq.~\eqref{eq:pmu_SL} as Eq.~\eqref{eq:momentsbeta1}.

\section{Expressing the von Neumann entropy} \label{sec:VNentropy}

In this Appendix, we discuss the quantities
\begin{equation}
    \Omega_N^{(\beta)}(x) := \left[e^{x A^{(\beta)}}\right]_{\idp,\idp} \;, \qquad 
    M_N^{(\beta)}(x) := \left[e^{x A^{(\beta)}}\right]_{\sigma(r_N=1),\idp}
\end{equation} namely the quantities appearing in Eq.~\eqref{eq:S1_equiv}. Note that except for a normalization constant, these quantities can be reinterpreted as the moments
\begin{equation}
\Omega_N^{(\beta)}(x) \propto \aveE{\Tr[\check \rho]^N} = \aveE[t]{\left(\sum_{\alpha=1}^q y_\alpha\right)^N}\;, \qquad  M_N^{(\beta)}(x) \propto \aveE{\Tr[\check \rho^N]} = \aveE[t]{\sum_{\alpha=1}^q y_\alpha^N}
\end{equation}
where $\check\rho$ here can be seen a random matrix resulting from a discrete-time or continuous-time stochastic multiplicative process, in the usual scaling limit $x = t/t_P(L)$. Therefore, these two quantities also have an interpretation and a value in the context of pure random matrix theory.

In the case of $\beta = 2$, as we have already mentioned in the main text, there is a closed and simple expression for the second term
\begin{equation}
\label{eq:MNbeta2}
   M_N^{(\beta=2)}(x) = \frac{2^{N-1}}{N!}\,\sinh\!\left(\frac{Nx}{2}\right)^{N-1}.
\end{equation}
see Eq.~(97) in \cite{gerbino2024dyson}.
In contrast, 
\begin{equation}
    \label{eq:Omegabeta2}
    \Omega_N^{(\beta = 2)}(x) = \frac{1}{N!}\sum_{\lambda \vdash N} (d^\lambda)^2 e^{x \nu_{2}(\lambda)}
\end{equation}
does not have a simple closed expression. In~\cite{deluca2025universality}, a generating function $\sum_N y^N/N! \Omega_N(x)$ was introduced and, using the hook-length formula \eqref{eq:hooklengthformula},  was recast into a determinantal formula (see Appendix E in \cite{deluca2025universality}) which was computed numerically rather efficiently. 

In the case $\beta = 1$, one has to deal instead with
\begin{equation}
    \label{eq:Omegabeta1}
    \Omega_N^{(\beta = 1)}(x) = \frac{N! 2^N}{(2N)!}\sum_{\lambda \vdash N} d^{2\lambda} e^{x \nu_{1}(\lambda)}
\end{equation}
whose combinatorial structure is more involved. A closed resummation would require the Pfaffian analogue of that determinantal representation; this route proves technically arduous and we do not pursue it.  As we show in Sec.~\ref{subsec:borelpade}, however, a closed form is not needed: the exact small-$x$ coefficients, pushed to high order, admit a controlled Borel--Pad\'e resummation with the exponential approach to the asymptote enforced, which reconstructs the replica-limit scaling function on the whole physical window.

Moreover, as it was pointed out in the main text, an interesting consequence of the spectral representation of $\Omega_N^{(\beta)}$ is that the Born-rule purity can be obtained from its derivatives with respect to $x$ and $N$. For the two symmetry classes, one may write
\begin{equation}
\Omega_N^{(1)}(x)=\frac{2^N N!}{(2N)!}
\sum_{\lambda\vdash N}d_{2\lambda}e^{x\nu_1(\lambda)},
\qquad
\Omega_N^{(2)}(x)=\frac{1}{N!}
\sum_{\lambda\vdash N}d_\lambda^2e^{x\nu_2(\lambda)}.
\end{equation}
Differentiating the spectral sums (in the shape \eqref{eq:pmu_SL}) with respect to $x$ inserts the corresponding adjacency eigenvalue,
\begin{equation}
\partial_x\Omega_N^{(\beta)}(x)
=\sum_{\lambda\vdash N}\gamma_{(1^N)}^\lambda(2/\beta)
\nu_\beta(\lambda)e^{x\nu_\beta(\lambda)} \,.
\end{equation}
The eigenvalue is related to the spherical function evaluated on a representative $\tau$ of the swap class by
$\nu_1(\lambda) = N(N-1) \ \omega^{2\lambda}(\tau)$ and $\nu_2(\lambda) = N(N-1) \ \chi^\lambda(\tau)/2$. Therefore,
\begin{equation}
\partial_x\Omega_N^{(\beta)}(x)
=\frac{N(N-1)}{\beta}
\left[e^{xA^{(\beta)}}\right]_{\tau,1}.
\end{equation}
The matrix element on the right-hand side is, however, an unnormalized purity numerator. One must divide it by the return amplitude $\Omega_N^{(\beta)}(x)=\left[e^{xA_N^{(\beta)}}\right]_{1,1}$, obtaining
\begin{equation}
\Pi_N^{(\beta)}(x)
=\frac{\left[e^{xA_N^{(\beta)}}\right]_{\tau,1}}
{\Omega_N^{(\beta)}(x)}
=\frac{\beta}{N(N-1)}
\frac{\partial_x\Omega_N^{(\beta)}(x)}
{\Omega_N^{(\beta)}(x)}
=\frac{\beta}{N(N-1)}
\partial_x\ln\Omega_N^{(\beta)}(x).
\end{equation}
The apparent singularity at $N=1$ is harmless: since $\Omega_1^{(\beta)}(x)=1$, the regularity of the replica continuation gives
\begin{equation}
\Omega_N^{(\beta)}(x)
=1+(N-1)
\left.\partial_N\Omega_N^{(\beta)}(x)\right|_{N=1}
+O\left((N-1)^2\right).
\end{equation}
Consequently, $\partial_x\Omega_N^{(\beta)}$ contains the same factor $N-1$ as the denominator, with 
the normalization by $\Omega_N^{(\beta)}$ becoming automatic in the Born-rule limit,
and the replica limit becomes
\begin{equation} \label{eq:PifromS}
\Pi^{(\beta)}(x)
=\beta\partial_x
\left.
\partial_N\Omega_N^{(\beta)}(x)
\right|_{N=1}.
\end{equation}

We also observe that, as it happens in the case $\beta = 2$, the Taylor expansion in powers of $x$ has coefficients which are polynomials in $N$. This fact can be argued on the basis of combinatorial considerations on the flip graph of the pairings, and in Sec.~\ref{subsec:OmegaCoefficients} we present the detailed calculation of the first polynomial coefficients. Explicitly, we have, setting
\begin{equation}
    \Omega_N^{(\beta = 1)}(x) = \sum_{\ell=0}^\infty \frac{\omega_N^{(\beta = 1,\ell)}}{\ell!} \  x^\ell
\end{equation}
that $\omega_N^{(\beta = 1,\ell)}$ is a polynomial of degree $d\leq \ell$ and whose form for the first few values of $\ell$ is
\begin{equation}
    \label{eq:polyOmega}
    \omega_N^{(\beta = 1,0)} = 1\,, \quad 
    \omega_N^{(\beta = 1,1)} = 0\,, \quad 
    \omega_N^{(\beta = 1,2)} = \omega_N^{(\beta = 1,3)} = N(N-1)\,, \quad 
    \omega_N^{(\beta = 1,4)} = N(N-1)(3N^2+N-11) \,.
\end{equation}
For higher values of $\ell$, the polynomial assumption is enough to fix the exact form of $\omega_N^{(\beta = 1,\ell)}$ via polynomial interpolation on a finite set of values of $N$. 

\subsection{Small-$x$ expansion of $M_N^{(\beta = 1)}(x)$} \label{subsec:MCoefficients}

In order for this method to be useful to obtain the Taylor series in $x$ of the von Neumann entropy \eqref{eq:S1_equiv}, we also need a similar expansion for $M_N^{(\beta = 1)}(x)$. In contrast with the case $\beta = 2$, we do not have a simple closed expression akin to \eqref{eq:MNbeta2}. Explicitly, we have
from \eqref{eq:momentsbeta1}, in the shape \eqref{eq:pmu_SL}, recalling $\alpha = 2/\beta=2$ and $c(2\lambda,1,1) = c(\lambda,2,2)c(\lambda,2,1)$:
\begin{equation} \label{eq:momNeasy}
     M_N^{(\beta=1)}(x) = \sum_{\lambda\vdash N} \gamma^{\lambda}_{(N)}(2) \  e^{\nu_1(\lambda) x} = 2N\sum_{\lambda\vdash N } \frac{\theta^{\lambda}_{(N)}(2)} {c(2\lambda,1,1)}e^{\nu_1(\lambda) x} \,.
\end{equation}
Following Ref.~\cite{fyodorov2016moments}, we can use the closed expression 
\begin{equation}
\theta^{\lambda}_{(N)}(\alpha) = \alpha^{N-1} (\lambda_1 - 1)! \prod_{i=2}^{\ell(\lambda)} \left( -\frac 1 \alpha(i - 1) \right)_{\lambda_i} \,.
\end{equation}
Notice that the above equation implies $\theta^{\lambda}_{(N)}(1)=0$ if $\lambda_2>1$, and $\theta^{\lambda}_{(N)}(2)=0$ if $\lambda_3>1$. In other words, for $\alpha=2$, the coefficient $\theta^\lambda_{(N)}(2)$ is non-vanishing only for those partitions $\lambda=(\lambda_1,\lambda_2,1^{\ell-2})$ associated to Young tableaux composed of only $2$ rows longer than one, and $\ell-2$ rows of length 1. The length $\ell = \ell(\lambda)$ of the partition can then be written as $\ell=N-\lambda_1-\lambda_2+2$. Then:
\begin{equation}
   \theta^{\lambda}_{(N)}(\alpha) = (-1)^{N-\lambda_1-\lambda_2+1} \  2^{\lambda_1-1} \ (\lambda_1-1)! \ (2\lambda_2-3)!! \ (N-\lambda_1-\lambda_2+1)!\,,\quad \alpha=2
\end{equation}
Accordingly, the eigenvalue $\nu_{\beta=1}(\lambda)$ \eqref{eq:eigen_new} and the double-hook product associated to such $\lambda$'s can be written as
\begin{subequations}
\begin{equation}
    \nu_{\beta=1}(\lambda) = \frac 1 2 \left[ (\lambda_1 -\lambda_2 + 1)(\lambda_1 -\lambda_2 ) + 2N \left( \lambda_1+\lambda_2 - \frac{N+3}{2} \right) \right] \,,
\end{equation}
\begin{equation}
    c(2\lambda,1,1) = (N-\lambda_1-\lambda_2+1)! \ (N-\lambda_1-\lambda_2)! \
    (N+\lambda_1-\lambda_2+1) =  \frac{(2\lambda_2-2)!(2\lambda_1-1)!}{2(\lambda_1-\lambda_2)+1} \,.
\end{equation}
\end{subequations}
As can be seen, the sum over the partitions of $N$ is reduced in this specific case to the sum over only two integers, i.e., the measures of the first two rows $\lambda_1,\lambda_2$ of the corresponding Young diagram. A similar mechanism is at the origin of the simple expression \eqref{eq:MNbeta2}, in which the specific form of the characters on a cycle restricts the relevant partitions to $L$-shaped Young diagrams, i.e., to the sum over a single integer $\lambda_1$ (eventually involving a single geometric sum which leads to \eqref{eq:MNbeta2}).

Using the restriction $\lambda_3\le 1$, it is convenient to parametrize the relevant partitions by two nonnegative integers
\begin{equation}
\label{eq:param_app3}
m:=\lambda_2\in\{1,\dots,N-1\},
\qquad
k:=\lambda_1-\lambda_2\in\{0,\dots,m-1\},
\end{equation}
so that $\lambda_1=m+k$, $\lambda_2=m$ and the triangular constraint $\lambda_1+\lambda_2\le N$ becomes $2m+k\le N$.
With this parametrization, the weight
\(
\frac{2N\,\theta^\lambda_{(N)}(2)}{c(2\lambda,1,1)}
\)
and the eigenvalue $\nu_1(\lambda)$ become explicit elementary functions of $(m,k)$, and one finds
\begin{equation}
\label{eq:mk_sum_app3}
M_N^{(\beta=1)}(x)
=
\frac{2^N e^{N(N-1)x}\,N!}{(2N)!}
-\sum_{m=1}^{N-1}\sum_{k=0}^{m-1} \mathcal{W}_N(m,k)\,
\exp\!\Big(\tfrac{x}{2}A_{N,m}(k)\Big),
\end{equation}
with
\begin{align}
&A_{N,m}(k):=k^2-2Nk+k+N(4m-N-3). \\
&\mathcal{W}_N(m,k) :=    \frac{2^{k+1}(2k+1)\,N\,(-1)^{k+N}\,\Gamma(m)}
{(k-N)(k-N+1)(k+N)(k+N+1)\,\Gamma(2m)\,\Gamma(m-k)\,\Gamma(k-2m+N+1)}
\end{align}
We have not found an elegant way to summarize the expression in a closed formula~\cite{5127463}. However, the Taylor expansion at small $x$ provides non-polynomial coefficients but with a clear structure. Indeed, writing
\begin{equation}
\label{eq:MN_expansion_up_to_4_app}
M_N^{(\beta=1)}(x)= \frac{m_N^{(\beta=1,N-1)}}{(N-1)!} x^{N-1} + \frac{m_N^{(\beta=1,N)}}{N!} x^{N} + \frac{m_N^{(\beta=1,N+1)}}{(N+1)!} x^{N+1} +
\ldots \,,
\end{equation}
where the coefficients $m_N^{(\beta=1,\ell)}$ are the ones introduced in the discussion around Eq.~\eqref{eq:quasiminimal}, it is verified that the coefficients have an “almost” polynomial form. One sets
\begin{equation}
\label{eq:def_Cn_app}
C_N:= \frac{e^N\,\Gamma(N-1,N)}{N^{N-1}}
=\frac{(N-2)!}{N^{N-1}}\,S_0(N),
\qquad
S_0(N):=\sum_{j=0}^{N-2}\frac{N^j}{j!}.
\end{equation}
and we find the forms (c.f. the expressions given in Eqs.~(\ref{eq:coeff1}, \ref{eq:2quasiminimal})):
\begin{subequations}
\label{eq:qell_list_app}
\begin{align}
& m_N^{(\beta=1,N-1)} = N^{N-2} \,,\\[1mm]
& m_N^{(\beta=1,N)} = N^{N-2} \frac{N^2(N-1)}{2}\,C_N = \frac{N(N-1)}{2}e^N \ \Gamma(N-1,N) \,,\\[1mm]
& m_N^{(\beta=1,N+1)} = N^{N-2} \frac{N^3(N^2-1)}{3}\Big(\frac{7}{8}-C_N\Big) \,,\\[1mm]
& m_N^{(\beta=1,N+2)} = N^{N-1} (N+2) \frac{N^2(N^2-1)}{720}\Big[\big(60 N^2 + 40 N + 3\big) \,C_N -60N-4\Big],\\[1mm]
& m_N^{(\beta=1,N+3)} = N^{N-2} (N+3)(N+2) \frac{N^3(N^2-1)}{362880}\Big[
120+7N\big(200+9N(-67+185N)\big) \\
& \hspace{4cm}  -32\big(3+2N(16+21N(-4+15N))\big)\,C_N
\Big] \,.
\end{align}
\end{subequations}
We conjecture that the form of these coefficients is always of the type $N^{N-2}[P_\ell(N) + C_N Q_\ell(N)]$, where $P_\ell(N)$ and $Q_{\ell}(N)$ are polynomials in $N$.

\subsection{Small-$x$ expansion of $\Omega_N^{(\beta = 1)}(x)$} \label{subsec:OmegaCoefficients}

In this Subsection we focus on
\begin{equation}
\label{eq:easy_def_app}
\Omega^{(\beta=1)}_N(x)\equiv \left[e^{xA^{(\beta=1)}}\right]_{\idp,\idp}
=
2^N N! \sum_{\lambda\vdash N}\frac{1}{c(2\lambda,1,1)}\,e^{\nu_1(\lambda)x} \,.
\end{equation}
Notice that the trace of $e^{xA^{(\beta=1)}}$ is exactly
\begin{equation}
    \Tr \left( e^{xA^{(\beta=1)}} \right) = \sum_{p \in \Pa_N}\left[e^{xA^{(\beta=1)}}\right]_{p,p} = \sum_{\lambda\vdash N}\frac{(2N)!}{c(2\lambda,1,1)}\,e^{\nu_1(\lambda)x} = |\Pa_N| \ \Omega^{(\beta=1)}_N(x) \,,
\end{equation}
namely the contributions from each pairing $p\in \Pa_N$ are all equivalent. This can be understood expanding 
\begin{equation}
    \Omega^{(\beta=1)}_N(x) = \sum_{\ell=0}^\infty \frac{x^\ell}{\ell!} \left[\left(A^{(\beta=1)}\right)^\ell \right]_{\idp,\idp} =: \sum_{\ell=0}^\infty \frac{x^\ell}{\ell!} \omega_N^{(\beta = 1,\ell)}\,.
\end{equation}
The matrix element $\omega_N^{(\beta = 1,\ell)}= \left[\left(A^{(\beta=1)}\right)^\ell \right]_{\idp,\idp}$ counts the number of loops of $\ell$ steps starting and ending at the reference pairing $\idp$. Since the graph is invariant under global transformation of the pairing states, one has $\omega_N^{(\beta = 1,\ell)} = \left[\left(A^{(\beta=1)}\right)^\ell \right]_{p,p}$, $\forall p \in \Pa_N$, and the contributions to the trace are all equivalent. 
Our task is now to evaluate the coefficients $\omega_N^{(\beta = 1,\ell)}$ exactly in $N$ for the first few $\ell$'s. 
As a useful property used in the explicit calculation, we recall that the flip graph of perfect pairings in $\Pa_N$ is regular of degree $\rm d$.

\textbf{First coefficients. --- } First, the trivial walk of length $0$ is unique. Then, since the graph has no loops, we cannot come back to $\idp$ in only one step. Moreover, in two steps, we can instead choose one of the $\rm d$ neighbors of $\idp$ and then come back via the closed walk $\idp \to p \to \idp$. Therefore
\begin{equation}
\omega_N^{(\beta = 1,0)} = 1 \,,\quad 
\omega_N^{(\beta = 1,1)} = 0 \,,\quad 
\omega_N^{(\beta = 1,2)} = {\rm d}=N(N-1) \,.
\end{equation}

\textbf{Computation of $\omega_N^{(\beta = 1,3)}$. --- } To compute the third coefficient, fix an edge $\idp\sim p$ connecting vertex $\idp$ to a vertex $p$. Then $p$ differs from $\idp$ only on four indices
$\{a,b,c,d\}$, where the matching in $\idp$ is $(a,b)(c,d)$
and the matching in $p$ is $(a,c)(b,d)$ or $(a,d)(b,c)$.
But on the set $\{a,b,c,d\}$ there are exactly three perfect pairings:
\begin{equation}
(a,b)(c,d),\qquad (a,c)(b,d),\qquad (a,d)(b,c).
\end{equation}
Therefore, every edge $\idp\sim p$ lies in a unique triangle
\begin{equation}
    \idp \sim p \sim q \sim \idp \,,
\end{equation}
where $q$ is the third pairing on those same four indices. For the remaining $2N-4$ indices, the vertices $\idp$, $p$, $q$ are equal. 
Thus, for each first step $\idp\to p$, there is exactly one 2-step return path $p\to q\to \idp$, and
\begin{equation}
\omega_N^{(\beta = 1,3)} = {\rm d} =N(N-1) \,.
\end{equation}

\textbf{Computation of $\omega_N^{(\beta = 1,4)}$. --- }
We use the fact that 
\begin{equation}
\omega_N^{(\beta = 1,4)} = \left[\left(A^{(\beta=1)}\right)^4 \right]_{\idp,\idp}=\sum_{p\in \Pa_N} \left[\left(A^{(\beta=1)}\right)^2 \right]_{\idp,p}^2 \,.
\end{equation} 
So we classify the vertices $p$ that can be reached from $\idp$ in two steps.
\begin{itemize}
    \item \underline{Type I: $p=\idp$.} This means the second move undoes the first one. There are exactly ${\rm d}$ such 2-step paths (i.e., $L_2(N)$), so this contribution is ${\rm d}^2$.
    
    \item \underline{Type II: $p$ adjacent to $\idp$.} Fix $p\sim \idp$. Since the edge $\idp\sim p$ lies in a unique triangle, c.f. the discussion above, there is exactly one 2-step path from $\idp$ to $p$. Thus
    \begin{equation} \left[\left(A^{(\beta=1)}\right)^2 \right]_{\idp,p}=1\,, \quad  p \ {\rm neighboring\ to}\ \idp\,.
    \end{equation}
    Since there are $\rm d$ such vertices $p$, the contribution from this type is ${\rm d}\cdot 1^2={\rm d}$.    
    \item \underline{Type III: three reference pairs involved.} Choose $3$ pairs of $\idp$. After two flips, these three pairs may become fully mixed, so that $\idp\cup p$ is a single alternating 6-cycle. For a fixed triple of reference pairs, there are $8$ such final matchings $p$. Hence the total number of such vertices is
    \begin{equation}
    N_{(3)}=8\binom{N}{3}.
    \end{equation}
    For each such $p$, there are exactly $3$ possible intermediate neighbors of $\idp$, according to which two of the three pairs are flipped first. Therefore
    \begin{equation} \left[\left(A^{(\beta=1)}\right)^2 \right]_{\idp,p}=3\,, \quad  p \ {\rm has \ flipped}\ 3 \ \rm pairs \,.
    \end{equation}
    and 
    \begin{equation}
    N_{(3)}\cdot 3^2=8\binom{N}{3}\cdot 9.
    \end{equation}
    \item \underline{Type IV: four reference pairs involved.}
    Choose $4$ pairs of $\idp$. After two independent flips on two disjoint pairs of pairs,
    the union $\idp\cup p$ is a disjoint union of two alternating 4-cycles.
    For a fixed set of $4$ reference pairs: $(i)$ there are $3$ ways to partition them into $2$ unordered couples; $(ii)$ for each couple, there are $2$ possible flips. Hence the number of such vertices is
    \begin{equation}
    N_{(2,2)}=3\cdot 2^2 \binom{N}{4}=12\binom{N}{4}.
    \end{equation}
    For each such $p$, the two flips can be performed in either order, so
    \begin{equation} \left[\left(A^{(\beta=1)}\right)^2 \right]_{\idp,p}=2\,, \quad  p \ {\rm has \ flipped}\ 4 \ \rm pairs \,,
    \end{equation}
    and
    \begin{equation}
    N_{(2,2)}\cdot 2^2=12\binom{N}{4}\cdot 4.
    \end{equation}
    Summing all contributions,
    \begin{equation}
    \omega_N^{(\beta = 1,4)}={\rm d}^2+{\rm d}+9\cdot 8\binom{N}{3}+4\cdot 12\binom{N}{4} = N(N-1) \left(3N^2+N-11\right) \,.
    \end{equation}
\end{itemize}

\subsection{Borel--Pad\'e resummation of the replica limit} \label{subsec:borelpade}
We detail here the resummation announced in the main text, which turns the exact small-$x$ coefficients of the two replica derivatives
\begin{equation}\label{eq:FandD}
F(x):=\partial_N M^{(\beta=1)}_N(x)\Big|_{N=1},\qquad
D(x):=\partial_N \Omega^{(\beta=1)}_N(x)\Big|_{N=1},
\end{equation}
into a non-perturbative determination of $\mathcal S^{(1)}_{\rm VN}=D-F$, with a confidence corridor and no fitted parameter. The coefficients of both quantities are computable exactly to high order: we reached order $26$ for $F$ and order $24$ for $D$. The regular expansion coefficients of $F$ become increasingly small, and its series remains well behaved up to comparatively large $x$. By contrast, the series for $D$ is factorially divergent, but it can be resummed by a Borel--Pad\'e procedure in which the exponential approach to the large-$x$ asymptote is enforced explicitly. The same procedure can also be applied to $F$.

\textbf{Exact coefficients.} ---
The Taylor coefficients of $\Omega_N^{(\beta=1)}$ are polynomials in $N$, as discussed in Sec.~\ref{subsec:OmegaCoefficients}, so that $D(x)=\sum_{\ell\ge2}d_\ell x^\ell$ has exact rational coefficients. We computed them up to $\ell=24$:
\begin{equation}
d_2=\tfrac12,\quad d_3=\tfrac16,\quad d_4=-\tfrac7{24},\quad d_5=-\tfrac{43}{120},\quad
d_6=\tfrac{413}{720},\quad d_7=\tfrac{8257}{5040},\quad d_8=-\tfrac{11129}{5760},\ \dots \, .
\end{equation}
Similarly,
\begin{equation}
F(x)=\ln x+\gamma_{\rm E}-1+\sum_{\ell\ge1}f_\ell x^\ell,
\end{equation}
with $f_\ell\in\mathbb Q+C_1\mathbb Q$ and $C_1=e\,\Gamma(0,1)$. For example,
\begin{equation}
f_1=\tfrac{C_1}{2},\qquad
f_2=\tfrac7{24}-\tfrac{C_1}{3},\qquad
f_3=\tfrac{103}{720}C_1-\tfrac4{45}.
\end{equation}
These coefficients were computed up to $\ell=26$ by graded interpolation of the double sum defining $M_N$.

\textbf{The asymptote and the remainders.} ---
At large $x$, both replica derivatives approach the same linear asymptote,
\begin{equation}
F(x),D(x)\longrightarrow x+c_\infty,
\qquad
c_\infty=\ln 2-2+\gamma_{\rm E}.
\end{equation}
For $F$, this follows from the dominant contribution
\begin{equation}
B_N(x)=\frac{2^N\Gamma(N+1)}{\Gamma(2N+1)}e^{N(N-1)x}
\end{equation}
in the spectral sum for $M_N$. For $D$, it follows from $\mathcal S^{(1)}_{\rm VN}=D-F\to0$ under purification. All late-time information therefore resides in the remainders
\begin{equation}
r_M(x)=F(x)-x-c_\infty,\qquad
r(x)=D(x)-x-c_\infty.
\end{equation}
On the family of hook-shaped partitions $\lambda=(N-j,1^j)$, the analytically continued eigenvalues satisfy $\nu_j(1)=j(j-3)/2=0,-1,-1,0,+2,\dots$. The gap below the leading exponent is therefore equal to one, implying
\begin{equation}
r(x),r_M(x)\sim e^{-x}x^{-3/2}.
\end{equation}
The power-law prefactor reflects the double degeneracy of the subleading exponent and the turnaround in the ordering of the spectrum upon the analytical continuation.

\textbf{The Borel--Pad\'e algorithm.} ---
A root test on the coefficients $d_\ell$ gives $|d_\ell|^{1/\ell}/\ell\to0.12$, corresponding to $d_\ell\sim \ell!/A^\ell$ with $A\simeq3.3$. Hence the series for $D$ is factorially divergent and has zero radius of convergence. Moreover, the Borel coefficients $d_\ell/\ell!$ exhibit the period-four sign pattern $({+}{+}{-}{-})$, indicating that the leading Borel singularities lie away from the positive real axis. The Laplace contour is therefore unobstructed and the series is Borel-summable.

To incorporate the known exponential decay, we first define
\begin{equation}
u(x):=e^x r(x)=\sum_{\ell=0}^{\ell_{\rm max}} u_\ell x^\ell\,,
\end{equation}
where $\ell_{\rm max}$ determines the number of available Taylor coefficients, 
and then form its Borel transform $\hat u$, which is approximated by its a Pad\'e approximant: 
\begin{equation}\label{eq:uBPmain}
\hat u(t):=\sum_{\ell=0}^{\infty}\frac{u_\ell}{\ell!}t^\ell
=[m/n]_{\hat u}(t)+O\!\left(t^{m+n+1}\right),
\qquad
[m/n]_{\hat u}(t):=\frac{P_m(t)}{Q_n(t)},
\end{equation}
with $P_m$ and $Q_n$ polynomials of order $m$ and $n$ respectively. 
The Pad\'e approximation is uncontrolled in the original $x$-plane, where the series has zero radius of convergence, but it is controlled in the Borel plane, where $\hat u$ has a finite convergence radius. The nearest singularities are found near $t\simeq1.0\pm3.6\,i$; the real shift by one originates from stripping the factor $e^{-x}$ associated with the spectral gap. The Borel sum is reconstructed by a Laplace transform and the exponential factor is restored at the end: we define the resummed Borel-Pad\'e remainder 
\begin{equation}\label{eq:uBPmain2}
r_{\rm BP}^{(m,n)}(x)=e^{-x}\int_0^\infty ds\,e^{-s}\,
\frac{P_m(xs)}{Q_n(xs)} \,.
\end{equation}
The finite-order result depends on the chosen Pad\'e orders. Since the exact asymptotic form contains the non-rational factor $x^{-3/2}$, no single ratio of polynomials can reproduce it exactly. We therefore use the neighboring choices $n=m+1$ and $n=m+2$ to define a parameter-free confidence corridor, taking the mean as the central value and their envelope as the uncertainty. The value of $m$ is fixed by the number of available Taylor coefficients. With the $24$ coefficients of $D$, the relevant neighboring approximants are $[10/11]_{\hat u}$, $[11/12]_{\hat u}$, and $[10/12]_{\hat u}$. The same construction is applied to $r_M$.

\textbf{Purity from the resummed return amplitude.} ---
At finite replica number, the swap matrix element must be normalized by the ``return amplitude'' $\Omega_N^{(\beta)}$. This gives
\begin{equation}
\Pi_N^{(\beta)}(x)=\frac{\beta}{N(N-1)}\,\partial_x\ln\Omega_N^{(\beta)}(x) \,.
\end{equation}
Since $\Omega_1^{(\beta)}(x)=1$, its regular expansion around $N=1$ reads
\begin{equation}
\Omega_N^{(\beta)}(x)=1+(N-1)
\left.\partial_N\Omega_N^{(\beta)}(x)\right|_{N=1}
+O\!\left((N-1)^2\right) \,.
\end{equation}
The factor $N-1$ in the denominator is therefore cancelled by the same factor in $\partial_x\ln\Omega_N^{(\beta)}$, and the Born-rule purity is recovered by Eq.~\eqref{eq:PifromS}.
In the present orthogonal case, $D(x)=\left.\partial_N\Omega_N^{(1)}(x)\right|_{N=1}$, so that the resummed purity follows directly from
\begin{equation}\label{eq:purityFromD}
P_{\rm BR}^{(1)}(x)=\partial_xD(x)=1+\partial_x r(x).
\end{equation}
Thus, once $r_{\rm BP}^{(m,n)}$ has been constructed, no additional resummation is required: one differentiates $D_{\rm BP}^{(m,n)}(x)=x+c_\infty+r_{\rm BP}^{(m,n)}(x)$. The spread among neighbouring Pad\'e orders can be propagated in the same way to obtain a confidence corridor for the purity.

\textbf{Benchmark.} ---
For $\beta=2$, the identical procedure reproduces the exact resummation formula available from the determinantal structure of Ref.~\cite{deluca2025universality} with relative accuracy better than $10^{-3}$ up to $x\simeq6$, while the spread among neighbouring Pad\'e orders contains the exact value throughout this interval. This validates the order spread as a parameter-free estimate of the resummation uncertainty.

\textbf{Result.} ---
Combining the two resummed remainders gives
\begin{equation}\label{eq:SVNresummed}
\mathcal S^{(1)}_{\rm VN}(x)=r(x)-r_M(x).
\end{equation}
Using worst-case propagation of the two confidence corridors, we obtain
\begin{center}
\begin{tabular}{llllllll}
$x$ & $0.5$ & $1.0$ & $1.5$ & $2.0$ & $2.5$ & $3.0$ & $4.0$\\
$\mathcal S^{(1)}_{\rm VN}(x)$ & $1.0709(11)$ & $0.4806(5)$ & $0.2351(3)$ & $0.1194(4)$ & $0.0618(5)$ & $0.0324(5)$ & $0.0090(5)$
\end{tabular}
\end{center}
At $x\le1.5$, the raw corridor is narrower than the $10^{-3}$ relative level at which the construction was independently validated, and this value is therefore used as an uncertainty floor. The results agree, within errors, with the small-$x$ expansion in Eq.~\eqref{eq:S1_theo_N1} over its range of validity and with the numerical protocols of the main text. 

\putbib
\end{bibunit}

\end{document}